\documentclass[11pt,a4paper]{article}
\pdfoutput=1
\usepackage{amssymb,amsmath,amsfonts, mathtools, mathrsfs}
\usepackage[utf8]{inputenc} % per usare le lettere accentate come à,è etc...
\usepackage[dvipsnames]{xcolor}

\newif\ifnatbibsort\natbibsorttrue

\relax

\ifnatbibsort\RequirePackage[numbers,sort&compress]{natbib}\else\RequirePackage[numbers,compress]{natbib}\fi
\RequirePackage[colorlinks=true
,urlcolor=blue
,anchorcolor=blue
,citecolor=blue
,filecolor=blue
,linkcolor=blue
,menucolor=blue
,pagecolor=blue
,linktocpage=true
,pdfproducer=medialab
,pdfa=true
]{hyperref}

\usepackage{graphicx}
\usepackage{caption}
\usepackage{tensor}
\usepackage{subfigure}
\usepackage{enumerate}
\usepackage{dsfont}
\usepackage{verbatim}
\usepackage{amsfonts,euscript,amssymb,stmaryrd,braket}
\usepackage{tikz}
\usetikzlibrary{arrows,decorations.markings,patterns}
\usepackage{slashed}

\def\clock{{\count0=\time
		\divide\count0 60
		\ifnum\count0<10 0\fi\the\count0
		\multiply\count0 -60 \advance\count0 \time
		:\ifnum\count0<10 0\fi \the\count0
}}
\newcommand{\timestamp}{{\small\vbox{\hbox{\tt\jobname.tex}
			\hbox{\the\day/\the\month/\the\year, \clock}}}}

\newcommand{\bea}{\begin{eqnarray}}
\newcommand{\eea}{\end{eqnarray}}

\newcommand{\be}{\begin{equation}}
\newcommand{\ee}{\end{equation}}

\makeatletter
\let\old@startsection=\@startsection
\let\oldl@section=\l@section
\renewcommand{\@startsection}[6]{\old@startsection{#1}{#2}{#3}{#4}{#5}{#6\mathversion{bold}}}
\renewcommand{\l@section}[2]{\oldl@section{\mathversion{bold}#1}{#2}}
\makeatother

\numberwithin{equation}{section}
\usepackage{color}

\def \RR {{\mathbb R}}

\def\ri {{\rm i}}
\def\rd {{\rm d}}
\def\e {{\rm e}}

\begin{document}
	\renewcommand{\thefootnote}{\arabic{footnote}}

	\overfullrule=0pt
	\parskip=2pt
	\parindent=12pt
	\headheight=0in \headsep=0in \topmargin=0in \oddsidemargin=0in

	\vspace{ -3cm} \thispagestyle{empty} \vspace{-1cm}
	\begin{flushright} 
		\footnotesize
		\textcolor{red}{\phantom{print-report}}
	\end{flushright}

\begin{center}
	\vspace{.0cm}

	{\Large\bf \mathversion{bold}
	%Notes on the 
	%\textcolor{blue}{(Geometric action of)} 
	Modular conjugations 
	in 2D conformal field theory
	%in two-dimensional conformal field theories
	%and holographic bit threads
	}
	\\
	\vspace{.25cm}
	\noindent
	{\Large\bf \mathversion{bold}
	and holographic 
	%\textcolor{blue}{(geodesic?)} 
	bit threads}

	\vspace{0.8cm} {
		Mihail Mintchev$^{\,a}$
		and Erik Tonni$^{\,b}$
	}
	\vskip  0.7cm
	
	\small
	{\em
		$^{a}\,$Dipartimento di Fisica, Universit\'a di Pisa and INFN Sezione di Pisa, \\
		largo Bruno Pontecorvo 3, 56127 Pisa, Italy
		\vskip 0.1cm
		$^{b}\,$SISSA and INFN Sezione di Trieste, via Bonomea 265, 34136, Trieste, Italy 
	}
	\normalsize

\end{center}

\vspace{0.3cm}
\begin{abstract} 

We study the geometric action of some modular conjugations in two dimensional (2D) conformal field theories (CFT).
We investigate the bipartition given by an interval
when the system is in the ground state, either on the line or on the circle,
and in the thermal Gibbs state on the line. 
We find that the restriction of the corresponding inversion maps to a spatial slice
is obtained also in the gauge/gravity correspondence
through the geodesic bit threads in a constant time slice 
of the dual static asymptotically AdS background. 
For a CFT in the thermal state on the line, 
the modular conjugation suggests the occurrence of a second world 
which can be related through the geodesic bit threads to the horizon 
of the BTZ black brane background. 
An inversion map is constructed 
also for the massless Dirac fermion in the ground state and on the line bipartite by the union of two disjoint intervals. 

\end{abstract}

%
%\begin{center}
%{\it Dedicated to the loving memory of Ios\'e Scalfi}
%\end{center}

\newpage
%%%%%%%%%%%%%%%%%%%%%%%%%%%%%%%%%%%%%
\tableofcontents
%%%%%%%%%%%%%%%%%%%%%%%%%%%%%%%%%%%%%

%\newpage
%%%%%%%%%%%%%%%%%%%%%%%%%%%%%%%%%%%%%%%%%%%
\section{Introduction}
\label{sec_intro}
%%%%%%%%%%%%%%%%%%%%%%%%%%%%%%%%%%%%%%%%%%%

The modular theory of Tomita and Takesaki \cite{takesaki-book-70, Haag:1992hx, Brattelli2, Borchers:2000pv, takesaki-book-03},
developed within the context of the algebraic quantum field theory,
provides important rigorous tools to understand various aspects of the bipartite entanglement 
in quantum field theories.

Considering a von Neumann algebra $\mathcal{A}$ of observables localised in a spacetime region
and in standard form (meaning that it acts on a Hilbert space $\mathcal{H} $ containing a cyclic and separating vector $\Omega$),
the conjugate linear operator $S : \mathcal{A} \Omega \to \mathcal{A} \Omega$
is such that $S\, \mathcal{O} \Omega =   \mathcal{O}^\ast \Omega$, for any operator $  \mathcal{O}\ \in \mathcal{A}$.
The unique polar decomposition of $S$ reads $S= J\, \Delta^{1/2}$, where
$\Delta$  is a self adjoint, positive (unbounded in general) operator called modular operator of $( \mathcal{A} , \Omega)$
and $J$ is an antiunitary operator called modular conjugation of $( \mathcal{A} , \Omega)$.
It turns out that the vector $\Omega$ is invariant, 
i.e. $\Delta \Omega = \Omega$ and $J \Omega = \Omega$,
and also that $J = J^\ast = J^{-1}$.
The modular operator $\Delta$ can be written as $\Delta = \e^{- K}$ where 
$K$ is the (full) modular Hamiltonian, a self adjoint operator whose real spectrum extends 
in general over $\RR$ and such that $\Omega$ is one of its eigenvectors 
with eigenvalue equal to zero. 
A crucial role is played by the
one-parameter group made by the unitary operators $\Delta^{\ri \tau}$ with $\tau \in \RR$.
The theorem of Tomita and Takesaki tells us that
$\Delta^{-\ri \tau} \mathcal{A} \,\Delta^{\ri \tau} = \mathcal{A}$
and  $\Delta^{-\ri \tau} \mathcal{A}' \, \Delta^{\ri \tau} = \mathcal{A}'$
for $\mathcal{A} $  and its commutant $\mathcal{A}'$ for any $\tau \in \RR$,
and that $J \mathcal{A}J = \mathcal{A}'$.
This theorem leads to introduce the group of the modular automorphisms of the state $\Omega$ on $\mathcal{A}$,
whose generic element is the map $\sigma_\tau \mathcal{O} \equiv \Delta^{\ri \tau} \mathcal{O} \,\Delta^{-\ri \tau} $, 
for any $\mathcal{O} \in \mathcal{A}$,
which determines the modular evolution (flow) of $\mathcal{O}$ generated by $K$.

For a local relativistic quantum field theory in its ground state whose space is bipartite 
by a region $A$ and its complement $B$ in a non trivial way, 
the theorem of Tomita and Takesaki can be applied
and $K = K_A \otimes \boldsymbol{1}_B - \boldsymbol{1}_A \otimes K_B$,
where $K_A$ and $K_B$ are the modular Hamiltonians of $A$ and $B$ respectively. 
Typically, $\sigma_\tau$ and $J$ do not have a geometric action;
indeed, a direct geometric meaning for $\sigma_\tau$ and $J$
has been found only in very few special cases \cite{Haag:1992hx}.
The most important one corresponds to the 
theorem of Bisognano and Wichmann \cite{Bisognano:1975ih, Bisognano:1976za},
which considers the bipartition associated to half space 
for a local relativistic quantum field theory in its ground state.
In this case the geometric action of $\sigma_\tau$ is given by the Lorentz boosts
in the direction preserving the wedge
and the geometric action of $J$ by a reflection of the time coordinate
and of the spatial coordinate orthogonal to the entangling hyperplane.
Another important example is described in 
the theorem of Hislop and Longo \cite{Hislop:1981uh, Brunetti:1992zf},
which considers a conformal field theory (CFT) in its ground state
and the bipartition determined by a sphere, 
providing also the geometric actions for the corresponding $\sigma_\tau$ and $J$.
Instead, the action of $\sigma_\tau$ and $J$ is non geometric e.g.
for a free massive field theory in its ground state when the space bipartite by a sphere
\cite{Borchers:2000pv,Saffary:2006ci,Saffary:2005mz}.

%\noindent
%$\bullet$ 
%{\bf 2D CFT}

We consider a local CFT in $1+1$ spacetime dimensions (2D).
The modular operator $\Delta$ in these models has been largely studied
and various explicit expressions of modular Hamiltonians have been found,
both when the entire system is in its ground state 
\cite{Bisognano:1975ih, Bisognano:1976za, Hislop:1981uh, Brunetti:1992zf, Casini:2009vk, Longo:2009mn, Rehren:2012wa,
Casini:2011kv, Wong:2013gua, Cardy:2016fqc, Arias:2016nip, Tonni:2017jom, Arias:2018tmw, Mintchev:2020uom, Mintchev:2020jhc}
and in the thermal case  
\cite{Borchers:1998ye, Borchers:1999short, Schroer:1998pj, Longo:2000ui, Buchholz:2006hp, Camassa:2011te, Camassa:2011wk,
Wong:2013gua, Cardy:2016fqc, Hollands:2019hje, Blanco:2019xwi, Fries:2019ozf}.
Instead, the modular conjugation $J$ and its geometrical action have been less explored \cite{Hislop:1981uh, Haag:1992hx}.
It is worth remarking that the Tomita-Takesaki modular theory 
provides insightful results when the 2D CFT is in a thermal state, 
already when the space is the real line. 
Indeed, the occurrence of a non trivial commutant naturally leads to consider
also a {\it second world}
which inherits the structure of the real word through the modular conjugation,
as highlighted by Borchers \cite{Borchers:1999short},
by Schroer and Wiesbrock \cite{Schroer:1998pj} (who called it virgin spacetime or thermal shadow world)
and by Longo et al. \cite{Longo:2000ui, Camassa:2011te, Camassa:2011wk}
(who formalise this idea through the thermal completion).

In this manuscript we study the geometric action of some modular conjugations in 2D CFT
by employing the results in the Euclidean spacetime discussed in \cite{Cardy:2016fqc}.
They include the case of a 2D CFT in a thermal state on the line bipartite by an interval
and the massless Dirac fermion in the ground state on the line bipartite by the union of two 
disjoint intervals.

In the context of the gauge/gravity (holographic) correspondence, 
the prescription proposed by Ryu an Takayanagi (RT) \cite{Ryu:2006bv, Ryu:2006ef}
to evaluate the holographic entanglement entropy 
in asymptotically Anti de Sitter (AdS) gravitational backgrounds
has provided a very important tool
to explore quantum gravity through quantum information concepts in certain models. 
Focussing on static gravitational backgrounds, 
this formula claims that the entanglement entropy $S_A$ of a region $A$ in the dual CFT on the boundary
of the asymptotically AdS spacetime
can be computed in such dual gravitational background as $S_A = \textrm{Area}(\gamma_A)/(4G_{\textrm{\tiny N}})$,
where $G_{\textrm{\tiny N}}$ is the Newton constant
and $\gamma_A$ (often called RT hypersurface)
is the codimension-two minimal area hypersurface 
belonging to the same constant time slice of $A$,
anchored to the boundary of $A$ and homologous to $A$.

An insightful reformulation of the RT prescription has been suggested by Freedman and Headrick 
in terms of the holographic bit threads \cite{Freedman:2016zud}.
Consider the bulk vector fields $V$ which are divergenceless $\nabla \cdot V = 0$ and norm bounded as $|V| \leqslant 1$.
By applying the Riemannian geometry version of the max-flow/min-cut theorem,
it turns out that $\textrm{Area}(\gamma_A)$ is equal to the maximum flux of one of these vector fields through 
the spatial region $A$ in the boundary of the asymptotically AdS background. 
The integral curves (or flow lines) of this vector field are called (holographic) bit threads.
In this setting, the RT hypersurface $\gamma_A$ is the bottleneck of the flow of a vector field satisfying this extremization condition.
It is important to remark that, while such vector field is uniquely fixed on $\gamma_A$,
it is infinitely degenerate on $A$.
Further properties of the holographic bit threads at some fixed time slice have been discussed in 
\cite{Headrick:2017ucz, Hubeny:2018bri, Agon:2018lwq, Headrick:2020gyq, Harper:2019lff}
and also a covariant generalisation has been recently proposed \cite{Headrick:2022nbe}.
Some holographic bit threads configurations have been explicitly constructed in  \cite{Agon:2018lwq}
by imposing further reasonable requirements 
in order to reduce the above mentioned degeneracy.

Among the explicit holographic bit threads constructions proposed in \cite{Agon:2018lwq}, 
in this manuscript we consider the geodesic flows
(where the integral lines of the holographic bit threads are also geodesics),
showing that, 
for simple three dimensional gravitational backgrounds 
and for some bipartitions given by an interval, 
they provide the same inversion relation 
obtained from the geometric action of the modular conjugation 
for the corresponding setup in the dual CFT.

The outline of the paper is as follows. 
In Section\;\ref{sec_mod_flow} the main quantities considered throughout our analyses are introduced. 
The CFT analysis for the simplest case,
which involves the bipartition of the line given by a single interval 
when the entire system is in the ground state,
is carried out in Section\;\ref{sec-line-single-interval}.
In Section\;\ref{sec-AdS} the results of the previous section 
are related to the geodesic bit threads in a constant time slice of Poincar\'e AdS$_3$.
In Section\;\ref{sec-thermal} we discuss 
the geometric action of the modular conjugation 
for a CFT in a thermal state and on the line bipartite by an interval.
In Section\;\ref{sec-BTZ}, the resulting inversion map is related to the geodesic bit threads 
in the BTZ black brane geometry. 
In Section\;\ref{sec-line-two-interval} we propose an inversion map 
for the free massless Dirac field in the ground state and on the line
bipartite by the union of two disjoint intervals.
Some conclusions are drawn in Section\;\ref{sec_conclusions}.
In Appendix\;\ref{sec_app_circle} we explore the inversion map 
for a CFT on a circle bipartite by an interval and in its ground state,
and in Appendix\;\ref{sec_app_global_AdS} we discuss 
the corresponding holographic construction 
through geodesic bit threads in a constant time slice of  global AdS$_3$.

%\newpage
%%%%%%%%%%%%%%%%%%%%%%%%%%%%%%%%%%%%%%%%%%%%%%%%
\section{Single interval: Modular flow of a chiral field}
\label{sec_mod_flow}
%%%%%%%%%%%%%%%%%%%%%%%%%%%%%%%%%%%%%%%%%%%%%%%%

In this section we summarise some basic results 
about the modular Hamiltonians in 2D CFT and their modular flows 
employed in the rest of the paper
\cite{Hislop:1981uh, Brunetti:1992zf, Casini:2011kv, Wong:2013gua, Cardy:2016fqc, Casini:2009vk, Hollands:2019hje, Mintchev:2020uom}.

In the $1+1$ dimensional Minkowski spacetime $\mathcal{M}$ described by the coordinates $(x,t)$,
we consider the light ray coordinates 
\be
\label{lc1}
u_\pm \equiv x\pm t
\ee
and a right ($+$) and a left ($-$) primary chiral field $\phi_\pm(u_\pm)$ of dimension $h$. 
Such fields satisfy the following commutation relation \cite{luscher}  
\be
\big[\,T_\pm(u_\pm)\, ,\, \phi_\pm(v_\pm)\,\big]
\,=\,
\ri \, \Big( \delta(u_\pm-v_\pm) \,(\partial\phi_\pm)(v_\pm) - h\, \phi_\pm (v_+) \, \delta^\prime (u_\pm -v_\pm ) \Big)
\ee
where $T_\pm$ are the chiral components of the energy-momentum tensor. 
For the cases we are interested in,
the right and left modular Hamiltonians  of a single interval $A = [a,b]$ 
can be written in terms of $T_\pm$ as follows
\be
\label{K-pm-A-def}
K^\pm_A \,=\, \pm\,2\pi \int_a^b \beta_0(u_\pm) \, T_\pm (u_\pm )\, \rd u_\pm  
\ee
with 
\be
\label{beta0-def-local}
 \beta_0(u) \equiv \frac{1}{w'(u)}
\ee
where the explicit form of the function $w$ depends on the state of the entire system. 
%on the representation of $\phi_\pm$. 
For instance, in the ground state we have
\be
\label{wfund}
 w(u) =  \log\!\bigg( \frac{u-a}{b-u} \bigg)
 \;\;\;\qquad\;\;\;
  \beta_0(u) \equiv \frac{(b-u)(u-a)}{b-a}\,.
 \ee
At finite temperature $1/\beta$, we consider a thermal state for which the mean energy density satisfies 
the Stefan-Boltzmann law \cite{Cardy:2010fa}
\be
\label{T}
\langle T_{tt}(t,x) \rangle 
\equiv
\frac{1}{2} \, \langle T_+(x+t) + T_-(x-t) \rangle 
\,= \,
\frac{c\, \pi}{12 \, \beta^2} 
\ee 
where $c$ is the central charge of the 2D CFT. 
According to \cite{Camassa:2011te, Camassa:2011wk},
for $0<c<1$ this state is unique,
while for $c\geqslant 1$ the condition (\ref{T}) selects a specific (geometric) thermal state. In this case 
\be
\label{wtherm}
w(u) =  \log\!\bigg( \frac{\sinh[\pi (u-a)/\beta]}{\sinh [\pi (b-u)/\beta ]} \bigg)
 \;\;\qquad\;\;
  \beta_0(u) \equiv  \frac{\beta}{\pi}\; \frac{\sinh[\pi (b-u)/\beta]\, \sinh[\pi(u-a)/\beta]}{\sinh [\pi(b-a)/\beta]}\,.
\ee

The modular evolution (flow) of $\phi_\pm$, generated by (\ref{K-pm-A-def}), reads
\bea
\label{evolution}
\phi_\pm (\tau,u_\pm) 
&\equiv  &
\e^{\ri \tau K^\pm_A} \, \phi_\pm (u_\pm) \; \e^{-\ri \tau K^\pm_A}
\\
\rule{0pt}{.8cm}
&=&
\bigg[\, \frac{\beta_0\big( \xi(\pm \tau,u_\pm) \big)}{\beta_0(u_\pm)} \,\bigg]^h  \phi_\pm \big( \xi(\pm\tau,u_\pm) \big) 
=
\left [\partial_{u_\pm} \xi(\pm\tau,u_\pm) \right ]^h \phi_\pm \big(\xi(\pm \tau,u_\pm) \big) 
\nonumber
\eea
(see e.g. the Appendix\;B of \cite{Mintchev:2020uom}),
where $\tau \in \RR $ is the modular parameter and 
the function $ \xi(\tau,u) $ is  fully determined by the function $w$ as follows
\be
\label{xi-def-gen}
 \xi(\tau,u) 
=
w^{-1}\big( w(u)+ 2\pi\,\tau\big)
\ee
where $u \in A$.
Since  $\xi(0,u)  = u$ for any $u\in A$ by construction,
the fields in (\ref{evolution}) satisfy the initial condition $\phi_\pm (0,u_\pm)  = \phi_\pm (u_\pm)$.

In our analyses we also employ the modular evolution in Euclidean spacetimes described in \cite{Cardy:2016fqc},
which is given by a complex function related to (\ref{xi-def-gen}) as follows
\be
\label{z-euclid-def-gen}
 z(\theta,x)  
 =  \xi(\tfrac{\ri \theta}{2\pi},x)
 = w^{-1}\big( w(x)+ \ri \theta \big)
\ee
where $x\in A$ and $\theta \in [0,2\pi)$.

A fundamental characteristic feature of any modular flow is the Kubo-Martin-Schwinger (KMS) condition \cite{Haag:1992hx}. 
In order to check that the flow (\ref{evolution}) satisfies this condition, 
consider the two-point functions 
\be 
\label{Wightman}
W_\pm (\tau_1,t_1,x_1;\tau_2,t_2,x_2) \equiv 
\langle 
\phi_\pm(\tau_1 , x_1\pm t_1)   \,
\phi_\pm(\tau_2 , x_2\pm t_2)  
\rangle
\ee
where $x_1$ and $x_2$ belong to $A$.
The KMS condition implies 
\be
\label{KMStau}
W_\pm (\tau_1+ \textrm{i}, t_1,x_1\,;  \tau_2, t_2, x_2)  
\,=\,
 W_\pm (\tau_1,t_1,x_1 ; \tau_2+ \textrm{i}, t_2,  x_2, t_2 ) 
\,=\,
 W_\pm (\tau_2,t_2,x_2 ;  \tau_1,t_1,x_1) 
\ee 
which can be verified deriving the expectation values (\ref{Wightman}) in explicit form. 
To deal with the left and right movers simultaneously, let us adopt the notation 
\be 
\label{notation}
\phi (u) = 
\left\{ 
\begin{array}{ll}
\phi_+(u) \hspace{.8cm} &  u = u_+ 
\\
\rule{0pt}{.7cm}
\phi_-(u) & u=u_- \,.
\end{array}
\right.
\ee
From (\ref{evolution}), 
for the two-point function along the modular flow one obtains 
\be
\label{2-point-primary-mod-gen}
\langle 
\phi (\tau_1 , u_1)   \,
\phi (\tau_2 , u_2)  
\rangle
=
\big[
\partial_{u_1} \xi (\tau_1 , u_1) \,
\partial_{u_2} \xi (\tau_2 , u_2)
\big]^h 
\,\langle 
\phi \big(\xi(\tau_1 , u_1) \big) \,
\phi \big(\xi(\tau_2 , u_2) \big)
\rangle
\ee
where the expectation value  in the r.h.s. is determined by the choice of representation for $\phi $. 
In the ground state and in the thermal state, we have respectively
\be 
\langle \phi (u_1 ) \,\phi (u_2)  \rangle 
= 
\frac{1}{2\pi\ri \, (u_1-u_2-\ri \varepsilon)^{2h}} 
\;\;\qquad\;\;
\langle \phi (u_1 ) \,\phi (u_2)  \rangle 
= 
\frac{1}{2\pi\ri \, \big[\frac{\beta}{\pi} \,\sinh \! \big(\frac{\pi}{\beta} (u_1 - u_2 -\ri \varepsilon ) \big)\big]^{2h}}\,.
\ee
Combining these expressions with (\ref{evolution}), in the ground state one finds  
\be
\label{Wfunc-gs}
\langle 
\phi (\tau_1 , u_1)   \,
\phi (\tau_2 , u_2)  
\rangle
= 
\frac{1}{2\pi \ri} \left [\frac{1}{u_1 - u_2 - \textrm{i} \varepsilon }\;\,
\frac{e^{w(u_1)} - e^{w(u_2)}}{e^{w(u_1)+\pi \tau_{12}} - e^{w(u_2)-\pi \tau_{12}} - \textrm{i} \varepsilon} \right ]^{2h}
\ee
with $\tau_{12} \equiv \tau_1 - \tau_2$ and $w$ given by (\ref{wfund}). 
Instead, at finite temperature one has 
\be
\label{Wfunc-thermal}
\langle 
\phi (\tau_1 , u_1)   \,
\phi (\tau_2 , u_2)  
\rangle
=
\frac{1}{2\pi \ri} \left [ \frac{1}{ \frac{\beta}{\pi} \,\sinh \! \big(\frac{\pi}{\beta} (u_1 - u_2 -\ri \varepsilon ) \big) }\;\,
\frac{e^{w(u_1)} - e^{w(u_2)}}{e^{w(u_1)+\pi \tau_{12}} - e^{w(u_2)-\pi \tau_{12}} - \textrm{i} \varepsilon} \right ]^{2h}
\ee
with $w$ defined by (\ref{wtherm}).  
By using the explicit expressions (\ref{Wfunc-gs}) and  (\ref{Wfunc-thermal}),
one can directly verify the KMS condition (\ref{KMStau}). 

We observe also that,  in addition to (\ref{KMStau}),
the thermal two-point function (\ref{Wfunc-thermal}) 
satisfies the KMS condition in the physical time 
\be
\label{KMSt}
W_\pm(\tau_1, t_1 + \textrm{i}\beta , x_1;  \tau_2,t_2,x_2)  
\,=\,
 W_\pm (\tau_1, t_1, x_1;  \tau_2, t_2 + \textrm{i}\beta, x_2) 
\,=\,
 W_\pm (\tau_2, t_2, x_2;  \tau_1, t_1, x_1) 
\ee
which involves $\beta$.

In the $1+1$ dimensional Minkowski spacetime in  the coordinates $(x,t)$,
let us consider the domain of dependence $\mathcal{D}_A$ of the interval $A=[a,b]\in \RR$
(also known as diamond or double cone). A modular trajectory in $\mathcal{D}_A$ 
is made by the points whose spacetime coordinates are
\be
\label{mod-trajec-1int-u}
x(\tau) = \frac{\xi(\tau, u_{+,0}) + \xi(-\tau, u_{-,0})}{2}
\;\;\qquad\;\;
t(\tau) = \frac{\xi(\tau, u_{+,0}) - \xi(-\tau, u_{-,0})}{2}
%\;\;\qquad\;\;
%x \in A
\ee
in terms of (\ref{xi-def-gen}),
where $\tau \in \RR$ and $u_{\pm,0}$ are the light ray coordinates of the point at  $\tau=0$.
It is often convenient to describe a modular trajectory in terms of the point $x\in A$ at $t=0$. 
Setting $\tau=0$ at this point, this modular trajectory is given by $(x(\tau), t(\tau)) \in \mathcal{D}_A$ with 
\be
\label{mod-trajec-1int}
x(\tau) = \frac{\xi(\tau, x) + \xi(-\tau, x)}{2}
\;\;\;\;\;\qquad\;\;\;\;\;
t(\tau) = \frac{\xi(\tau, x) - \xi(-\tau, x)}{2}
%\;\;\qquad\;\;
%x \in A
\ee
in terms of (\ref{xi-def-gen}), where $\tau \in \RR$.

%\newpage
%%%%%%%%%%%%%%%%%%%%%%%%%%%%%%%%%%%%%%%%%%%%%%%%
\section{Single interval: Ground state}
\label{sec-line-single-interval}
%%%%%%%%%%%%%%%%%%%%%%%%%%%%%%%%%%%%%%%%%%%%%%%%

In this section we focus on a CFT in the ground state and on the line
bipartite by an interval $A=(a,b) \subset \RR$ and its complement.

\subsection{Internal modular evolution}
\label{sec-mod-evo-gs}

The function $w$ to consider in this case is given in (\ref{wfund})
\cite{Hislop:1981uh, Casini:2009vk, Casini:2011kv, Wong:2013gua, Cardy:2016fqc}.
Specifying (\ref{xi-def-gen}) to this function,
for each chirality  we obtain 
the following geometric action of the modular operator 
\cite{Hislop:1981uh, Brunetti:1992zf, Haag:1992hx}\footnote{The Eq.\,(V.4.33) of \cite{Haag:1992hx} is obtained by setting $b=-\,a=1$ and $\tau_{\textrm{\tiny here}} = - \,\tau_{\textrm{\tiny there}}$ in (\ref{xi-interval-line-gs}).}
\be
\label{xi-interval-line-gs}
 \xi(\tau,u) 
%%=
%%w^{-1}\big( w(x)+ 2\pi\,\tau\big)
 =
 \frac{a  + b\, \e^{w(u)}\, \e^{2\pi \tau}}{1 + \e^{w(u)} \, \e^{2\pi \tau}}
 =
 \frac{(b-u) \,a+(u-a)\, b\, \e^{2\pi \tau}}{b-u +(u-a)\, \e^{2\pi \tau}}
 \;\;\qquad\;\;\;
  \xi(\tau=0,u) = u
\ee
which is also known as 
the geometric action of the modular automorphism group of the diamond $\mathcal{D}_A$ 
induced by the vacuum state.
The expression (\ref{xi-interval-line-gs}) must be employed for both the chiralities 
and it provides the modular trajectories in the diamond $\mathcal{D}_A$ 
through (\ref{mod-trajec-1int-u}), or (\ref{mod-trajec-1int}) equivalently. 

\begin{figure}[t!]
\vspace{-.6cm}
\hspace{-1.2cm}
%\begin{center}
\includegraphics[width=1.17\textwidth]{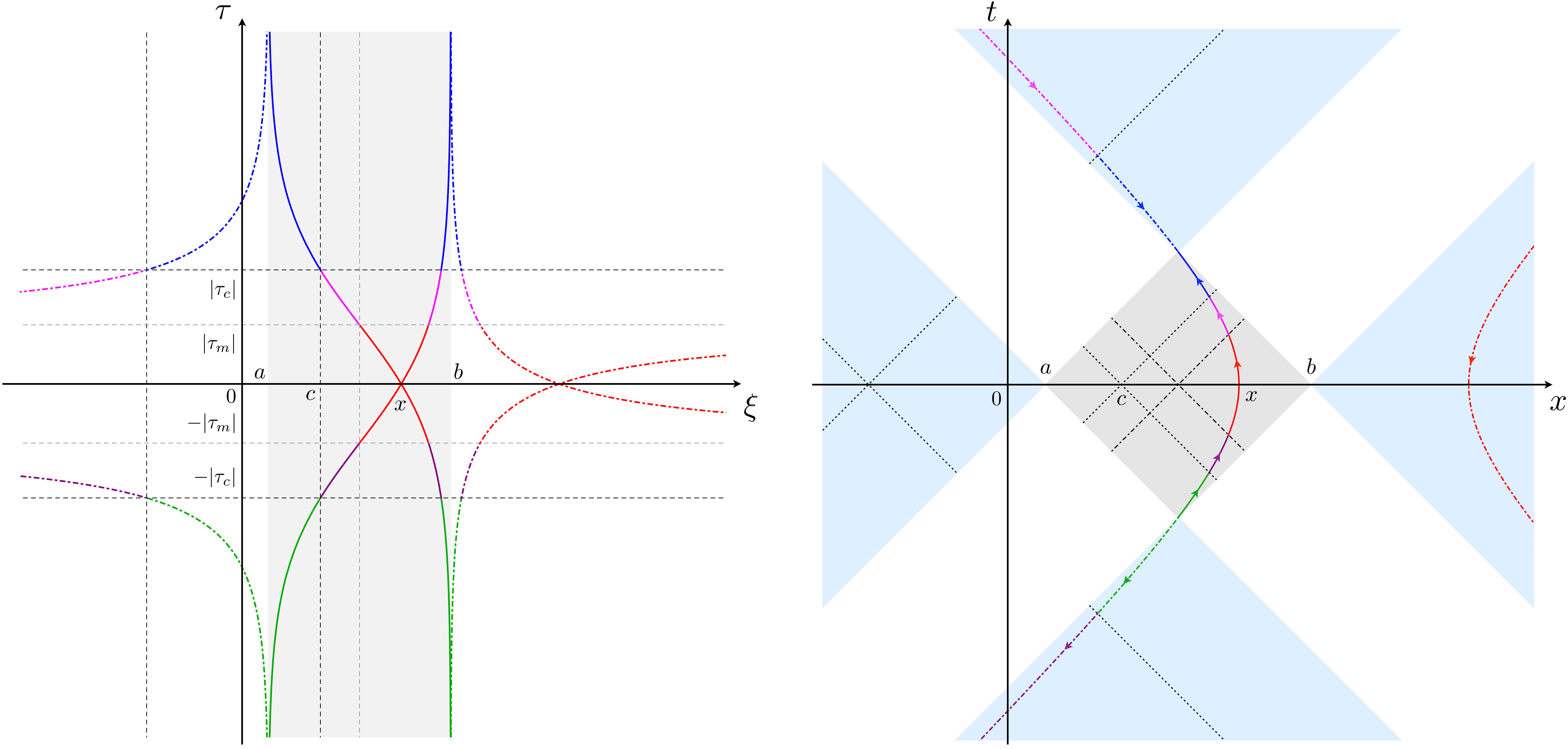}
% \end{center}
\vspace{-.3cm}
\caption{Left: 
The curves for $\xi(\tau, x)$ and $\xi(-\tau, x)$ (solid lines) 
and for $\xi(\tau, z_{\textrm{\tiny inv}}(x) )$ and $\xi(-\tau, z_{\textrm{\tiny inv}}(x) )$ (dot-dashed lines),
obtained from (\ref{xi-interval-line-gs}), (\ref{x-inv = z-inv t=0}) and (\ref{Haag-inversion-map-upm}).
Right: Modular trajectory (solid line) in the diamond $\mathcal{D}_A$ (grey region) 
and its image under the inversion (\ref{Haag-inversion-map}) (dot-dashed line).
The dot-dashed segment in $\mathcal{D}_A$ provide the partition (\ref{DA-dec-gs}).
The modular parameter $\tau$ grows along the curves as indicated by the arrows.
}
\label{figure-xi-double-cone}
\end{figure}

For any assigned $x\in (a,b)$, from (\ref{xi-interval-line-gs}) we have that 
the functions in (\ref{mod-trajec-1int}) satisfy
$ x(\tau=0) = x$ and $ t(\tau=0) = 0$, as expected.
Notice that $\xi(\pm\tau,x) $ take all real values for $\tau \in \RR$, 
with $\xi(\tau,x) \to a$ as $\tau\to -\infty$ 
and $ \xi(\tau,x) \to b$ as $\tau\to +\infty$.
Moreover, $\xi(\tau,a) = a$ and $\xi(\tau,b) =b$ for any $\tau \in \RR$.
In the left panel of Fig.\,\ref{figure-xi-double-cone}, the solid curves in the grey region 
correspond to $\xi(\tau,x) $ and $ \xi(-\tau,x) $, with $\tau\in \RR$.
These curves intersect at $x\in (a,b)$, when $\tau=0$.

Given a generic $c\in (a,b)$, from (\ref{xi-def-gen}) we have that
$ \xi(\tau_c ,x) = c$ when $\tau_c = \tfrac{1}{2\pi} [w(c) - w(x)]$.
This value splits the range of the modular parameter $\tau \in \RR$ into three sets $(-\infty, -|\tau_c|)$, $(-|\tau_c| , |\tau_c|)$ and $(|\tau_c|, +\infty)$,
as highlighted by the darker grey horizontal dashed lines in the left panel of Fig.\,\ref{figure-xi-double-cone}.
A special role is played by the mid-point of the interval $\tfrac{a+b}{2}$ (see Sec.\,\ref{sec-inversion-gs-lorentz}))
which corresponds to the value $\tau_m= \tfrac{1}{2\pi} [w(\tfrac{a+b}{2}) - w(x)]$ of the modular parameter,
that provides the lighter grey horizontal lines in the left panel of Fig.\,\ref{figure-xi-double-cone}.

The important case considered by Bisognano and Wichmann \cite{Bisognano:1975ih,Bisognano:1976za}
corresponds to the limiting regime where the subsystem $A$ becomes the semi-infinite line $x>0$.
Setting $a=0$ first and then taking $b\to +\infty$ in (\ref{xi-interval-line-gs}), 
one finds $ \xi(\tau,x) = x\, \e^{2\pi \tau}$ with $x>0$,
i.e.  the dilations of the semi-infinite line parameterised by $\tau$.

The action of the modular operator and of the modular conjugation  in the Minkowski space 
can be found by combining the corresponding actions for the two chiralities. 
In the right panel of Fig.\,\ref{figure-xi-double-cone},
the grey region is the domain of dependence $\mathcal{D}_A$ of the interval $A=[a,b]\in \RR$,
which is also called diamond or double cone.
The null rays departing from its central point $(\tfrac{a+b}{2},0)$ (dot-dashed segments in $\mathcal{D}_A$) 
provide the following partition of $\mathcal{D}_A$ into four smaller diamonds
\be
\label{DA-dec-gs}
\mathcal{D}_A \,\equiv\, 
\mathcal{D}_{\textrm{\tiny R}} \cup \mathcal{D}_{\textrm{\tiny L}} \cup 
\mathcal{D}_{\textrm{\tiny F}} \cup \mathcal{D}_{\textrm{\tiny P}} 
\ee
where in the r.h.s. the letter in the subindex stands for either 
right (R) or left (L) or future (F) or past (P),
depending on the position of the corresponding diamond with respect to the central point $(\tfrac{a+b}{2},0)$ of $\mathcal{D}_A$.
The latter point is the only one which provides a partition of $\mathcal{D}_A$ in four diamonds;
indeed, the partitions of $\mathcal{D}_A$ induced by the other points include at least two rectangular regions. 
The solid line inside $\mathcal{D}_A$ 
is the modular trajectory obtained from (\ref{mod-trajec-1int}),
with $\xi(\tau, x)$ and $\xi(-\tau, x)$ given by the solid lines shown in left panel of the same figure
(an arc having certain colour in the left panel provides the arc with the same colour in the right panel). 
The different arcs of the modular trajectory in $\mathcal{D}_A$
can be identified through their intersections with the light rays departing 
from $(c,0)$ (dotted segments in $\mathcal{D}_A$)
and from the mid-point of the interval at $t=0$ (dot-dashed segments in $\mathcal{D}_A$).
The modular parameter $\tau$ grows along the modular trajectory as indicated by the arrows.

When $A$ becomes the semi-infinite line $x>0$ (i.e. for $a=0$ and $b\to +\infty$),
the modular trajectory obtained from (\ref{xi-interval-line-gs}) and (\ref{mod-trajec-1int})
simplifies to $(x(\tau), t(\tau)) = (x  \cosh(\tau) , \, x \sinh(\tau))$.

%\newpage
\subsection{Geometric action in Minkowski spacetime}
\label{sec-inversion-gs-lorentz}

For a CFT in the ground state and on the line bipartite through an interval, 
the geometric action of the modular conjugation
is the following inversion map 
$(x,t) \to (x_{\textrm{\tiny inv}}, t_{\textrm{\tiny inv}}) $ 
\cite{Hislop:1981uh, Haag:1992hx}\footnote{The two dimensional case of Eq.\,(V.4.34) in \cite{Haag:1992hx}
corresponds to (\ref{Haag-inversion-map}) for $b=-a=1$.}
\be
\label{Haag-inversion-map}
x_{\textrm{\tiny inv}}(x,t) 
\equiv
\frac{a+b}{2} - \,\bigg( \frac{b-a}{2} \bigg)^2 \,\frac{ x - \tfrac{a+b}{2}  }{t^2 - \big(x - \tfrac{a+b}{2} \big)^2}
\;\;\qquad\;\;
t_{\textrm{\tiny inv}}(x,t) 
\equiv
\bigg( \frac{b-a}{2} \bigg)^2 \,\frac{ t }{t^2 - \big(x - \tfrac{a+b}{2} \big)^2}
\ee
which is idempotent, as expected.

In terms of the light ray coordinates (\ref{lc1}),
the geometric action (\ref{Haag-inversion-map})  leads to define the inversion map $z_{\textrm{\tiny inv}} $ as follows
%gives $u_{\pm,\textrm{\tiny inv}} \equiv x_{\textrm{\tiny inv}}\pm t_{\textrm{\tiny inv}}$ given by
\be
\label{Haag-inversion-map-upm}
u_{\pm,\textrm{\tiny inv}}(u_\pm) 
\equiv
x_{\textrm{\tiny inv}}(x,t) \pm t_{\textrm{\tiny inv}}(x,t) 
= 
%\frac{a+b}{2}\; \frac{u_\pm -\tfrac{2 a b}{a+b}}{u_\pm -\tfrac{a+b}{2}}
%= 
\frac{a+b}{2} +  \frac{(\tfrac{b-a}{2})^2}{u_\pm -\tfrac{a+b}{2}}
\equiv 
z_{\textrm{\tiny inv}}(u_\pm)
\ee
which allows to write (\ref{Haag-inversion-map}) as
\be
\label{x-t-inv-z-one-int}
x_{\textrm{\tiny inv}} (x, t)
=
\frac{z_{\textrm{\tiny inv}}(x+t) + z_{\textrm{\tiny inv}}(x-t) }{2} 
\;\;\qquad\;\;
t_{\textrm{\tiny inv}} (x, t)
=
\frac{z_{\textrm{\tiny inv}}(x+t) - z_{\textrm{\tiny inv}}(x-t) }{2} \,.
\ee

Another insightful form of the geometric action of the modular conjugation
can be written in terms of the distance from the mid-point of $A$.
Indeed, by introducing 
\be
\tilde{x} \,\equiv\, x - \frac{a+b}{2} 
\;\;\;\qquad\;\;\;
\tilde{x}_{\textrm{\tiny inv}} \,\equiv\, x_{\textrm{\tiny inv}} - \frac{a+b}{2} 
\ee
from (\ref{Haag-inversion-map}) one finds that
\be
\tilde{x}_{\textrm{\tiny inv}} \pm t_{\textrm{\tiny inv}}
\,=\,
\frac{(\ell/2)^2}{\tilde{x} \pm t}
\ee
where $\ell \equiv b-a$ is the length of the interval $A$, 
whose endpoints are the entangling points.

When $t=0$, the inversion map (\ref{Haag-inversion-map}) simplifies to
\be
\label{x-inv = z-inv t=0}
x_{\textrm{\tiny inv}}(x,t=0)  = z_{\textrm{\tiny inv}}(x) 
\;\;\;\qquad\;\;\;
t_{\textrm{\tiny inv}}(x,t=0)  = 0
\ee
which sends $A$ into its complement on the line, and viceversa. 
In particular, $z_{\textrm{\tiny inv}}(x) < a$ when $x\in (a, \tfrac{a+b}{2})$
and $z_{\textrm{\tiny inv}}(x) > b$ when $x\in ( \tfrac{a+b}{2}, b)$.
%Since the endpoints of $A$ are shared with its complement on the line, they are also called entangling points. 
We remark that $z_{\textrm{\tiny inv}}(x) $ in (\ref{x-inv = z-inv t=0}) sends a point in $A$ close to an entangling point 
into a point in $B$ close to the same entangling point, and viceversa.

In the left panel of Fig.\,\ref{figure-xi-double-cone} the dot-dashed curves 
correspond to $\xi(\tau, z_{\textrm{\tiny inv}}(x) )$ and $\xi(-\tau, z_{\textrm{\tiny inv}}(x) )$,
with $\tau \in \RR$, which intersect at $\tau=0$ and $z_{\textrm{\tiny inv}}(x) \in B$.
These curves and the corresponding ones for $\xi(\tau, x)$ and $\xi(-\tau, x)$ in the grey region
are partitioned into five arcs (indicated with different colours)
at the values $\tau = \mp |\tau_c|$ and $\tau = \mp |\tau_m|$ of the modular parameter.
Notice that,
while $|\xi(\tau, x)|$ remains finite, 
we have that $\big| \xi(\tau, z_{\textrm{\tiny inv}}(x)  ) | \to  +\infty$
as $|\tau| \to |\tau_m|$.

In the right panel of Fig.\,\ref{figure-xi-double-cone}, 
the light blue region $\mathcal{W}_A$ 
can be partitioned into four infinite triangular domains as follows
\be
\label{WA-dec-gs}
\mathcal{W}_A \,\equiv\, 
\mathcal{W}_{\textrm{\tiny R}} \cup \mathcal{W}_{\textrm{\tiny L}} \cup 
\mathcal{V}_{\textrm{\tiny F}} \cup \mathcal{V}_{\textrm{\tiny P}} 
\ee
(the notation introduced in (\ref{DA-dec-gs}) has been adopted also here)
where the subindex indicates the position of the vertex of the corresponding infinite triangle 
with respect to the center of $\mathcal{D}_A$.
The vertices of the left spacelike wedge $\mathcal{W}_{\textrm{\tiny L}}$ and the right spacelike wedge $\mathcal{W}_{\textrm{\tiny R}}$ 
are $(a,0)$ and $(b,0)$ respectively, 
while the forward light cone $\mathcal{V}_{\textrm{\tiny F}}$ and backward light cone $\mathcal{V}_{\textrm{\tiny P}}$ 
have vertices in $(\tfrac{a+b}{2}, \tfrac{b-a}{2})$ and $(\tfrac{a+b}{2}, -\tfrac{b-a}{2})$ respectively. 
The geometric action of the modular conjugation given by (\ref{Haag-inversion-map}),
or equivalently by (\ref{Haag-inversion-map-upm}) in the light ray coordinates,
maps $\mathcal{D}_A$ into the light blue region and viceversa
(see also Fig.\,V.4.1 of \cite{Haag:1992hx}).
The mid-point $(\tfrac{a+b}{2}, 0)$ is sent to infinity by this map.
Considering the geometric action (\ref{Haag-inversion-map}) 
separately on each element of the partitions of $\mathcal{D}_A$ and $\mathcal{W}_A$ 
introduced in (\ref{DA-dec-gs}) and (\ref{WA-dec-gs}),
one finds that it relates domains that share a vertex (hence with the same subindex).

The modular trajectory corresponding to the solid curve in $\mathcal{D}_A$ 
passes through $(x,0)$ when $\tau=0$
and is mapped into the dot-dashed curve in the light blue region.
The modular parameter $\tau$ grows along these curves as indicated by the arrows.
Hence, both the modular trajectory in $\mathcal{D}_A$ and its dot-dashed image in $\mathcal{W}_A$
originate in the lower vertex of $\mathcal{D}_A$ 
as $\tau \to -\infty$ and end in the upper vertex of $\mathcal{D}_A$  as $\tau \to +\infty$.
Each arc in the modular trajectory having a certain colour is sent by (\ref{Haag-inversion-map})
to the arc on the dot-dashed curve having the same colour
and they both correspond to the same range of the modular parameter $\tau$.
As already remarked, the partition of the modular trajectory 
into the five arcs with different colours is determined by its intersections 
with the light rays departing from the mid-point of the interval $P_m =(\tfrac{a+b}{2}, 0)$
(black dot-dashed segments in $\mathcal{D}_A$)
and from the generic point $P_c =(c, 0)$ in the interval (black dotted segments in $\mathcal{D}_A$) at $t=0$.
The same holds for the image of the modular trajectory 
under (\ref{Haag-inversion-map}) in the light blue region, 
with the obvious difference that its partition is determined
by the light rays departing from the images of $P_m$ and $P_c$ under the same map. 
Since this map sends $P_m$ to infinity,
the red, purple and magenta dot-dashed arcs in the light blue region
reach infinity in $\mathcal{W}_{\textrm{\tiny R}}$, $\mathcal{V}_{\textrm{\tiny P}}$ and $\mathcal{V}_{\textrm{\tiny F}}$
respectively. 
In particular, the red dot-dashed arc in $\mathcal{W}_A$
entirely belongs either to  $\mathcal{W}_{\textrm{\tiny R}}$ or to  $\mathcal{W}_{\textrm{\tiny L}}$,
depending on whether $x>\tfrac{a+b}{2}$ or $x<\tfrac{a+b}{2}$ respectively,
and  (\ref{x-inv = z-inv t=0}) tells us that
it passes through the point $(z_{\textrm{\tiny inv}}(x) ,0)$ when $\tau=0$.
Any arc in the right panel of Fig.\,\ref{figure-xi-double-cone}
is obtained through (\ref{mod-trajec-1int}), 
by using the arcs denoted by the same kind of line and the same colour 
in the left panel of the same figure. 

It is worth considering two special bipartitions of the line.
When $a=-\,b$, the interval $A$ is centered in the origin of the $x$-axis
and the inversion map (\ref{Haag-inversion-map}) becomes
\cite{Haag:1992hx}\footnote{The Eq.\,(V.4.34) of \cite{Haag:1992hx} corresponds to (\ref{xt-inv-centered}) for $b=1$.}
\be
\label{xt-inv-centered}
x_{\textrm{\tiny inv}} = - \, \frac{b^2\,x}{t^2 - x^2}
\;\;\qquad\;\;
t_{\textrm{\tiny inv}} =   \frac{b^2\, t}{t^2 - x^2}
\ee
which further simplifies for $t=0$ to the following suggestive form 
\be
\label{xt-inv-centered-t=0}
x_{\textrm{\tiny inv}} = \frac{b^2}{x}
\;\;\;\;\qquad\;\;\;\;
t_{\textrm{\tiny inv}} =  0\,.
\ee

When $b \to +\infty$, the interval $A$ and its diamond $\mathcal{D}_A$
become respectively the half line $x\geqslant a$ and the right Rindler wedge with vertex in $(a,0)$.
The inversion map (\ref{Haag-inversion-map}) simplifies to 
the modular conjugation for such Rindler wedge, which reads
\cite{Haag:1992hx}\footnote{The case $a=0$ of (\ref{mod-conj-wedge}) is discussed in Sec.\,V.4.2 of \cite{Haag:1992hx}.}
\be
\label{mod-conj-wedge}
x_{\textrm{\tiny inv}}(x,t)  -a \,=\,  - \,( x - a )
\;\;\qquad\;\;
t_{\textrm{\tiny inv}}(x,t)  =  - \,t\,.
\ee
In terms of the light ray coordinates (\ref{lc1}), taking $b \to +\infty$ in (\ref{Haag-inversion-map-upm}), for this map one obtains
\be
\label{Haag-inversion-map-upm-half-line}
u_{\pm,\textrm{\tiny inv}} - a \,=\,  -\big(  u_\pm - a \big)
\ee
which is consistent with (\ref{mod-conj-wedge}), as expected.

The above results can be extended in a straightforward way 
to the case in higher dimensions
where the subsystem is a spatial sphere in the Minkowski spacetime \cite{Haag:1992hx, Hislop:1981uh},
by employing the light cone coordinates $r \pm t$, 
being $r$ the radial coordinate when the origin coincides with the center of the sphere.

%\newpage
\subsection{Geometric action in the Euclidean spacetime}
\label{sec-euclid-gs}

In \cite{Cardy:2016fqc}
the modular evolution in the Euclidean spacetime 
for a 2D CFT in its ground state and on a line bipartite by an interval $A=(a,b)$
has been described in terms of the function $w(x)$ given in (\ref{wfund}) 
through the following complex function
\be
\label{z-inv-1int}
z(\theta, x) 
\equiv 
w^{-1} \big( w(x) + \textrm{i} \theta \big)
=\frac{a\,(b-x) + b\, (x-a) \, \e^{\ri \theta}}{b-x + (x-a) \, \e^{\ri \theta}}
\;\;\qquad\;\;
\theta \in [0,2\pi)
\qquad
x \in A
%\;\;\qquad\;\;
%x\in A
\ee
which can be obtained also by replacing $2\pi \tau$ with $\ri \theta$ in (\ref{xi-interval-line-gs}).
This complex function can be equivalently written in the form
\bea
\label{z-theta-circle}
z(\theta, x) 
%\,=\, \frac{a\,(b-x) + b\, (x-a) \, \e^{\ri \theta}}{b-x + (x-a) \, \e^{\ri \theta}}
\,=\,
\mathcal{C} + \mathcal{R} \, \e^{\ri \gamma}
\eea
where
\be
\mathcal{C}  \equiv \frac{x^2 - a\, b}{2\big( x - \frac{a+b}{2} \big)}
\;\;\qquad \;\;
\mathcal{R}  \equiv \frac{(b-x)(x-a)}{2\big( x - \frac{a+b}{2} \big)}
\;\; \qquad \;\;
\gamma \equiv
\pi - \theta -\, \ri  \log\! \bigg( \frac{ x-a + \e^{\ri \theta}(b-x) }{ x-a + \e^{-\ri \theta}(b-x) } \bigg) \,.
\ee
Hence, in the complex plane described by the complex coordinate $z$,
the curve $z(\theta, x) $ corresponding to an assigned value $x\in A$ and parameterised by $\theta \in (0, 2\pi)$ is a circle
with radius $\mathcal{R}$  centered in the  point $(\mathcal{C}, 0)$ on the real axis.
From (\ref{z-inv-1int}), notice that $z(\theta, a) = a$ and $z(\theta, b)  = b$ for any allowed value of $\theta$,
despite the fact that $w(x)$ in (\ref{wfund}) is not defined when either $x=a$ or $x=b$.

\begin{figure}[t!]
\vspace{-.6cm}
\hspace{.3cm}
%\begin{center}
\includegraphics[width=.95\textwidth]{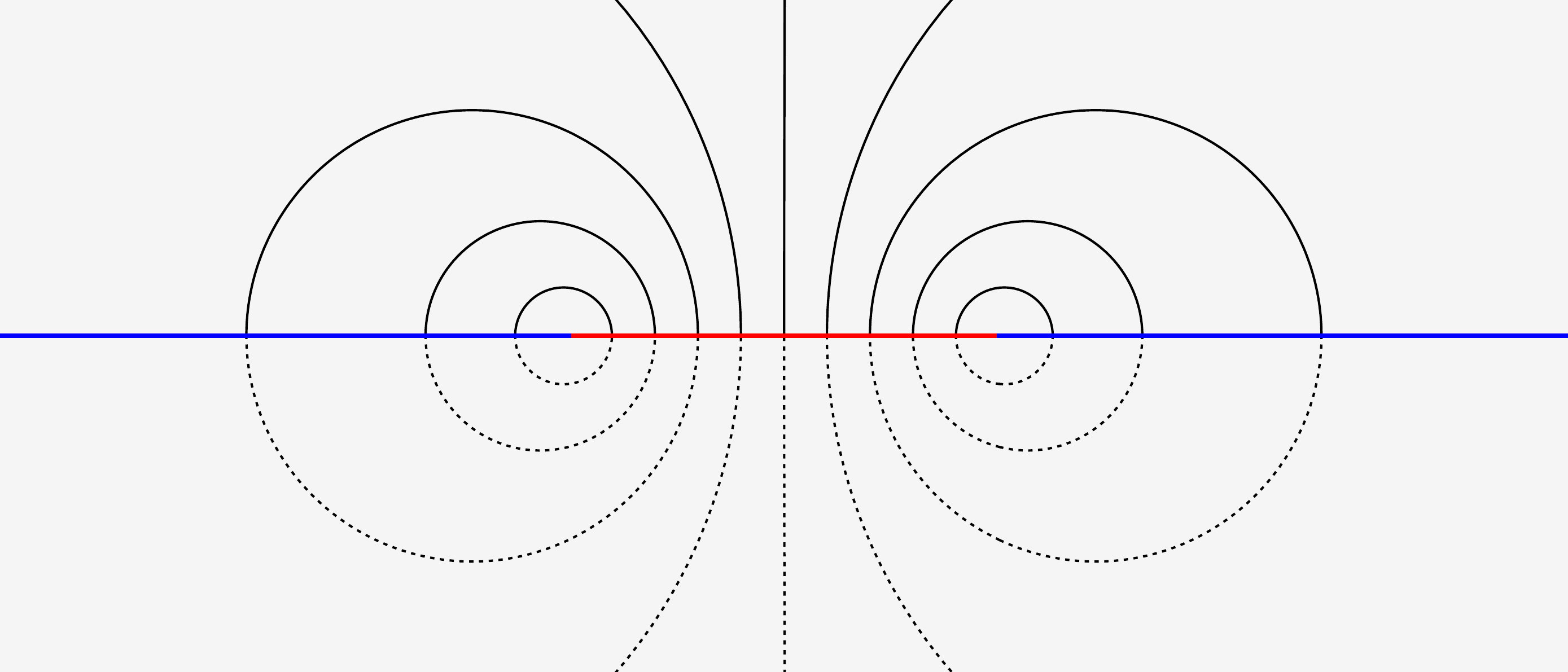}
% \end{center}
\vspace{.2cm}
\caption{
%Inversion map  (\ref{z-inv-1int-pi}) on the real axis 
Euclidean modular evolution in the complex plane:
The curves are half circles obtained from (\ref{z-inv-1int}) (or (\ref{z-theta-circle}) equivalently)
with $\theta \in (0,\pi)$ (solid arcs) or $\theta \in (\pi,2\pi)$ (dotted arcs), for some $x\in A$.
They map a point in $A$ (red segment) to a point in its complement (blue half lines) and viceversa,
which are related by (\ref{z-inv-1int-pi}).
}
\label{figure-inversion-Euclid-1int}
\end{figure}

In Fig.\,\ref{figure-inversion-Euclid-1int},
considering the complex plane described by the complex variable $z$ (grey plane),
we show the half circles obtained from (\ref{z-inv-1int}) (or (\ref{z-theta-circle}) equivalently) 
for various assigned  $x\in A$ and parameterised either by $\theta \in (0,\pi)$ (solid arcs) or by $\theta \in (\pi,2\pi)$ (dot-dashed  arcs).
In the same complex plane, it is useful to explore also the curves 
given by (\ref{z-inv-1int}) (or (\ref{z-theta-circle}) equivalently) parameterised by $x\in A$ for some fixed $\theta \in (0,2\pi)$.
For these curves we have that $\partial_x z(\theta, x) \to \e^{\ri \theta}$ as $x \to a$
and $\partial_x z(\theta, x) \to \e^{-\ri \theta}$ as $x \to b$;
hence, the curve that  intersects orthogonally the real axis at $x=a$ and $x=b$ corresponds to $\theta =\pi/2$.
From  (\ref{z-inv-1int}) one finds that such curve is the half circle centered in the mid-point of $A$
(see also Sec.\,\ref{sec-AdS} and Fig.\,\ref{figure-1int-AdS}).

The inversion map (\ref{Haag-inversion-map-upm}) 
naturally provides the complex map $z_{\textrm{\tiny inv}} : \mathbb{C} \to \mathbb{C}$ 
given by
\be
\label{Haag-inversion-map-z-map}
z_{\textrm{\tiny inv}}(\zeta) = \frac{a+b}{2} + \frac{(\tfrac{b-a}{2})^2}{\zeta - \tfrac{a+b}{2}}
\ee
which is a rational function with real parameters. 
By introducing the real variable $t_{\textrm{\tiny E}} \in \RR$ and exploiting (\ref{x-t-inv-z-one-int}),
we can relate (\ref{Haag-inversion-map-z-map})  to the inversion map (\ref{Haag-inversion-map}) 
by replacing $t$ with $\ri t_{\textrm{\tiny E}}$, finding 
\bea
\label{Re-z-inv-1int}
\textrm{Re}\big[ z_{\textrm{\tiny inv}}(x+\textrm{i} t_{\textrm{\tiny E}})  \big] 
&=& 
\frac{z_{\textrm{\tiny inv}}(x+\textrm{i} t_{\textrm{\tiny E}}) + z_{\textrm{\tiny inv}}(x-\textrm{i} t_{\textrm{\tiny E}}) }{2} 
= x_{\textrm{\tiny inv}} (x, \textrm{i} t_{\textrm{\tiny E}})
\\
\label{Im-z-inv-1int}
\rule{0pt}{.7cm}
\textrm{i}\,\textrm{Im}\big[ z_{\textrm{\tiny inv}}(x+\textrm{i} t_{\textrm{\tiny E}})  \big] 
&=& 
\frac{z_{\textrm{\tiny inv}}(x+\textrm{i} t_{\textrm{\tiny E}}) - z_{\textrm{\tiny inv}}(x-\textrm{i} t_{\textrm{\tiny E}}) }{2} 
= t_{\textrm{\tiny inv}} (x, \textrm{i} t_{\textrm{\tiny E}})\,.
\eea

The inversion on the real line given in  (\ref{x-inv = z-inv t=0}) 
can be obtained by restricting (\ref{Haag-inversion-map-z-map}) to the real line 
or by evaluating (\ref{z-inv-1int}) at $\theta=\pi$, 
which is equivalent to (\ref{xi-def-gen}) evaluated at $\tau=\ri/2$,
namely
\be
\label{z-inv-1int-pi}
z_{\textrm{\tiny inv}}(x) = z(\pi, x) = \xi(\ri/2, x)
\;\;\qquad\;\;
x\in \RR \,.
\ee
This tells us that, in Fig.\,\ref{figure-inversion-Euclid-1int},
the endpoints of a black solid half circle in the upper half plane 
(or of the corresponding black dotted half circle in the lower half plane) belong to the real axis 
and are related through the inversion map (\ref{x-inv = z-inv t=0}).
Hence, any point in the interval $A$ (red segment) 
is uniquely mapped into a point in its complement on the line (union of the blue half lines), and viceversa.
From (\ref{z-inv-1int-pi}), we also observe that
(\ref{z-inv-1int}) e (\ref{Haag-inversion-map-z-map}) are related as follows
\be
\label{z-inv-1int-pi-complex}
z_{\textrm{\tiny inv}}(\zeta) = z(\pi, \zeta) 
\;\;\;\qquad\;\;
\zeta \in \mathbb{C} \,.
\ee
Finally, 
by  combining (\ref{z-inv-1int-pi}) with (\ref{x-t-inv-z-one-int}) and (\ref{z-inv-1int}) (or (\ref{xi-def-gen})),
one observes that 
the geometric action of the modular conjugation (\ref{Haag-inversion-map}) 
can be expressed through $w(x)$ in (\ref{wfund}).

%\newpage
%%%%%%%%%%%%%%%%%%%%%%%%%%%%%%%%%%%%%%%%%%%%%%%%
\section{Geodesic bit threads in Poincar\'e AdS$_3$}
\label{sec-AdS}
%%%%%%%%%%%%%%%%%%%%%%%%%%%%%%%%%%%%%%%%%%%%%%%%

In this section we relate the geodesic bit threads in Poincar\'e AdS$_3$ 
for the RT curve corresponding to an interval on the line
to the results discussed in Sec.\,\ref{sec-inversion-gs-lorentz} and Sec.\,\ref{sec-euclid-gs}.

%\noindent
%$\bullet$ {\bf Intro and RT curve}

\begin{figure}[t!]
\vspace{-.6cm}
\hspace{.3cm}
%\begin{center}
\includegraphics[width=.95\textwidth]{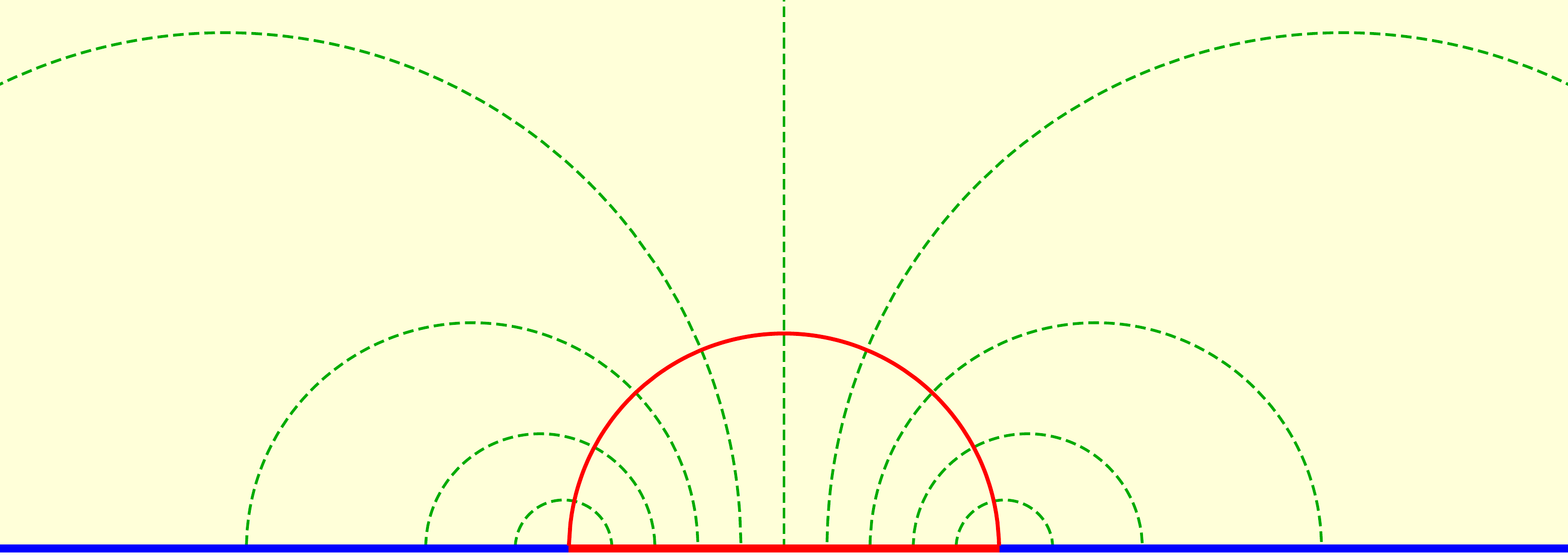}
% \end{center}
\vspace{.2cm}
\caption{
Geodesic bit threads (green dashed curves) in a fixed time slice of Poincar\'e AdS$_3$ (yellow region)
associated to an interval on the line (red segment), 
whose RT curve is the red solid half circle anchored to its endpoints. 
}
\label{figure-1int-AdS}
\end{figure}

The Poincar\'e  AdS$_3$ can be described by the coordinates $(t, x, \zeta)$, 
where $\zeta > 0$ parameterises the holographic direction, 
hence its boundary corresponds to $\zeta = 0$.
A $t =\textrm{const}$ slice of Poincar\'e AdS$_3$ is the upper half plane 
described by the coordinates $(x,\zeta)$ and equipped with the following metric
(yellow region in Fig.\,\ref{figure-1int-AdS})
\be
\label{UHP-metric}
ds^2 = \frac{dx^2 + d\zeta^2}{\zeta^2}
\;\;\;\;\qquad\;\;\;\;\;
\zeta > 0 \,.
\ee
In this Euclidean manifold, the set of geodesics having both the endpoints on the boundary
is the two-parameter family of half circles centered on the real axis, namely
\be
\label{geodesics AdS}
(x - x_s)^2 + \zeta^2 = R_s^2 \,.
\ee
The vertical lines orthogonal to the boundary provide the remaining set of geodesics.
According to the AdS/CFT correspondence, on the boundary of the Poincar\'e AdS$_3$
we have a dual 2D CFT on the real line and in its ground state.

In order to recover the same situation investigated in Sec.\,\ref{sec-line-single-interval},
the spatial direction on the boundary is bipartite by an interval of length $\ell \equiv 2b$ 
(red segment in Fig.\,\ref{figure-1int-AdS})
and its complement (union of the blue half lines in Fig.\,\ref{figure-1int-AdS}).
The interval can be placed symmetrically with respect to the origin (i.e. $ A=\{x\, ; |x| \leqslant b$\})
without loss of generality.
The RT formula \cite{Ryu:2006bv, Ryu:2006ef} provides
the holographic prescription to evaluate the holographic entanglement entropy through a bulk computation.
For static three dimensional gravitational spacetimes and a single interval $A$ in the dual CFT,
the RT prescription requires the minimal length curve anchored to the endpoints of $A$ 
(the RT curve, denoted by $\gamma_A$ in Sec.\,\ref{sec_intro}) 
because its regularised length multiplied by $1/(4G_{\textrm{\tiny N}})$
provides the holographic entanglement entropy.
In a fixed time slice of Poincar\'e AdS$_3$ and for the interval $A=(-b,b)$,
the RT curve is the half circle of radius $b$ centered in the origin (red solid curve in Fig.\,\ref{figure-1int-AdS}),
whose generic point $P_m$ 
(that must not be confused with the point indicated in the same way in Sec.\,\ref{sec-inversion-gs-lorentz})
have coordinates $(x_m , \zeta_m)$ satisfying 
\be
\label{RT-curve-AdS}
x_m^2 + \zeta_m^2 = b^2 \,.
\ee
It is worth reminding that also the curve $z(\theta=\pi/2, x)$ for $x\in A$ obtained from (\ref{z-inv-1int})
is the half circle centered in the mid-point of $A$ in the complex plane,
as already noticed in Sec.\,\ref{sec-euclid-gs}.

Among the holographic bit threads introduced in \cite{Freedman:2016zud},
let us consider the specific configuration given by the geodesic bit threads discussed in \cite{Agon:2018lwq}.
A geodesic bit thread for a RT curve is a geodesic such that
it connects a point $x\in A$ to a point that does not belong to $A$
and intersects the RT curve orthogonally.
At such intersection point, 
its unit norm tangent vector coincides with the unit norm outward-pointing vector normal to the RT curve.

For the RT curve (\ref{RT-curve-AdS}), 
the generic geodesic bit thread is (\ref{geodesics AdS}) with \cite{Agon:2018lwq}
\be
\label{geo-flow-AdS}
R_s = \frac{b\,\sqrt{b^2 - x_m^2}}{|x_m|}
\;\;\;\qquad\;\;\;
x_s = \frac{b^2}{x_m}
\ee
where  $(x_m , \zeta_m)$ are the coordinates of the 
intersection point between the geodesic bit thread and the RT curve, 
hence they are constrained by (\ref{RT-curve-AdS}).
Some geodesic bit threads
obtained by combining (\ref{geodesics AdS}),  (\ref{RT-curve-AdS}) and (\ref{geo-flow-AdS})
are shown in Fig.\,\ref{figure-1int-AdS} (green dashed half circles).

We find it more convenient to parameterise a geodesic bit thread in terms of the coordinate $x_A\in A$ of its endpoint in $A$.
The coordinate $x_m$ of the intersection point between the geodesic bit thread and the RT curve is related to $x_A$ 
as follows\footnote{The expression (\ref{xm-from-xA-AdS}) can be obtained by inverting the relation reported in Eq.\,(2.10) of \cite{Agon:2018lwq}.}
\be
\label{xm-from-xA-AdS}
x_m = \frac{2\, b^2 \,x_A}{b^2 + x_A^2}
\;\;\;\qquad\;\;\;
|x_A| \leqslant  b
\ee
which leads to write (\ref{geo-flow-AdS}) as
\be
\label{geo-flow-AdS-xA}
R_s = \frac{b^2 - x_A^2}{2\, |x_A|}
\;\;\;\;\qquad\;\;\;\;
x_s = \frac{b^2 + x_A^2}{2\, x_A}
\ee
implying that $(R_s , x_s) = (0, \pm b)$ when $x_A = \pm b$, as expected. 
From (\ref{geo-flow-AdS-xA}), one obtains
\be
\label{xs-pm-Rs-from-xA}
x_s + R_s = \left\{
\begin{array}{ll}
b^2/x_A  \hspace{.6cm}& x_A > 0
\\
\rule{0pt}{.5cm}
x_A & x_A < 0
\end{array}
\right.
\;\;\;\;\;\; \qquad \;\;\;\;\;\;
x_s - R_s = \left\{
\begin{array}{ll}
x_A  & x_A > 0
\\
\rule{0pt}{.5cm}
b^2/x_A \hspace{.6cm}& x_A < 0 \,.
\end{array}
\right.
\ee
Plugging (\ref{geo-flow-AdS-xA}) into (\ref{geodesics AdS}), we find 
that the geodesic bit thread anchored to $x_A \in A$ reads
\be
\label{bt-AdS-factor-form}
\zeta\,=\,\sqrt{(x - x_A)\left(\frac{b^2}{x_A} - x\right)}
\;\;\;\qquad\;\;\;
|x_A| \leqslant  b \,.
\ee
From (\ref{xs-pm-Rs-from-xA}) or (\ref{bt-AdS-factor-form}), 
we have that the geodesic bit thread intersects the AdS boundary 
in $A$ at $x = x_A$, as expected by construction, 
and in the complement of $A$ on the line at $x=x_B$, where
\be
\label{x_B-from-x_A-AdS}
x_B = \frac{b^2}{x_A} \,.
\ee
This relation between the endpoints of a geodesic bit thread
coincides with the inversion relation on the real axis for $t=0$ 
provided by the geometric action of the modular conjugation 
for the interval $A=(-b,b)$, which has been reported in (\ref{xt-inv-centered-t=0}).
Furthermore, let us observe that a geodesic bit thread anchored at $x_A \in A$ 
(see (\ref{bt-AdS-factor-form}) and the green dashed curves in Fig.\,\ref{figure-1int-AdS})
and the trajectory corresponding to the Euclidean modular evolution for $\theta \in (0, \pi)$ 
that starts at the same $x_A$ when $\theta =0$
%discussed in Sec.\,\ref{sec-euclid-gs}
(see (\ref{z-theta-circle}) and the black solid curves in Fig.\,\ref{figure-inversion-Euclid-1int})
are described by the same curve.
Indeed, they are both half circles whose endpoints are related in the same way. 
However, these two curves belong to different Euclidean manifolds.

Finally, comparing (\ref{geo-flow-AdS}) and (\ref{x_B-from-x_A-AdS}),
we find it worth remarking  that the same inversion relation occurs between $x_A$ and $x_B$
and between $x_m$ and $x_s$.

We remind that 
the contour function for the entanglement entropies \cite{ChenVidal2014, Coser:2017dtb}
can be associated holographically to the restriction on the boundary of 
the norm bounded and divergenceless vector field providing the holographic bit threads \cite{Tonni18:bariloche-talk}.
In particular, from the vector field for the geodesic bit threads in Poincar\'e AdS$_3$ found in \cite{Agon:2018lwq},
the function $\beta_0(x)$ for a CFT in the ground state (see (\ref{beta0-def-local}) and (\ref{wfund}))
is obtained \cite{Kudler-Flam:2019oru}.

A straightforward extension of the above analysis 
allows to address the case of a CFT in the ground state 
and in the higher dimensional Minkowski spacetime
bipartite by a sphere of radius $b$. 
For this bipartition it is convenient to adopt the polar
coordinates in the Minkowski spacetime
with the origin in the center of the sphere.
In this case, the geodesic bit threads are obtained from 
the expressions discussed above restricted to $x>0$,
which corresponds to the radial coordinate.

%\newpage
%%%%%%%%%%%%%%%%%%%%%%%%%%%%%%%%%%%%%%%%%%%%%%%%%%%%%%%%
\section{Single interval: Thermal state}
\label{sec-thermal}
%%%%%%%%%%%%%%%%%%%%%%%%%%%%%%%%%%%%%%%%%%%%%%%%%%%%%%%%

In this section we consider a CFT in a thermal state at temperature $1/\beta$
and on the line bipartite by an interval $A=(a,b)$ and its complement. 
This thermal state is the geometric one within the analysis performed in \cite{Camassa:2011te, Camassa:2011wk},
which satisfies the Stefan-Boltzmann law (\ref{T}).
In this case the function $w(u)$ to employ is reported in (\ref{wtherm}) \cite{Wong:2013gua, Cardy:2016fqc}.
The KMS condition is fulfilled, both in the modular time and in the physical time
(see (\ref{KMStau}) and (\ref{KMSt}) respectively).

The geometric action of the modular automorphism group of the diamond 
induced by the thermal state can be found by specialising (\ref{xi-def-gen})
to (\ref{wtherm}) and the result is \cite{Mintchev:2020uom}
\be
\label{xi-thermal-tau}
 \xi(\tau,u) \,=\,
\frac{\beta}{2\pi }\, \log\! \bigg(
 \frac{\e^{ \pi (b+a)/\beta} +  \e^{2\pi b/\beta}\, \e^{w(u)+2\pi \tau}
 }{ 
\e^{\pi (b-a)/\beta}  +  \e^{w(u)+2\pi \tau}}
\bigg)
\ee
where $u \in A$ and $\tau \in \RR$.

In the Euclidean spacetime, by specialising (\ref{z-euclid-def-gen}) to (\ref{wtherm}), one finds \cite{Cardy:2016fqc}
\bea
\label{z-inv-1int-beta}
z(\theta, x) 
& = &
w^{-1} \big( w(x) + \textrm{i} \theta \big)
%=
% \xi(\tfrac{\ri \theta}{2\pi},x)
%\\
%\rule{0pt}{.8cm}
%&=&
=
\frac{\beta}{2\pi }\, \log\! \bigg(
 \frac{\e^{ \pi (b+a)/\beta} +  \e^{2\pi b/\beta}\, \e^{w(x)+\ri \theta}
 }{ 
 \e^{\pi (b-a)/\beta}  +  \e^{w(x)+\ri \theta}}
\bigg)
\\
\label{z-inv-1int-beta-3}
\rule{0pt}{.9cm}
&=&
\frac{\beta}{2\pi }\, \log\! \Bigg(
 \frac{ \big( \e^{ 2 \pi b/\beta} - \e^{ 2 \pi x/\beta} \big) \e^{ 2 \pi a/\beta} +  \big( \e^{ 2 \pi x/\beta} - \e^{ 2 \pi a/\beta} \big) \, \e^{ 2 \pi b/\beta}  \, \e^{\ri \theta} 
 }{ 
 \e^{ 2 \pi b/\beta} - \e^{ 2 \pi x/\beta} +   \big( \e^{ 2 \pi x/\beta} - \e^{ 2 \pi a/\beta}\big) \, \e^{\ri \theta}  }
\Bigg)
\eea
where $x \in A$ and $\theta \in [0,2\pi)$.
In the zero temperature limit $\beta \to +\infty$,
the expressions (\ref{xi-thermal-tau}) and (\ref{z-inv-1int-beta}) 
become (\ref{xi-interval-line-gs}) and (\ref{z-inv-1int}) respectively, as expected.

In the limiting regime where $A$ becomes the half line, i.e. when $a =0$ and $b \to +\infty$, 
one finds that (\ref{xi-thermal-tau}) simplifies to
\cite{Borchers:1998ye}\footnote{See Eq.\,(4.14) in \cite{Borchers:1998ye}, with $ u_{\textrm{\tiny there}} = \tau_{\textrm{\tiny here}}$ and $f_{\textrm{\tiny there}}(x) = x$.}
\be
\label{xi-BY}
 \xi(\tau,x) \; \longrightarrow \; \frac{\beta}{2\pi} \, \log\! \big[  1+  \e^{2\pi \tau}  ( \e^{2\pi x/\beta} -1 ) \big] \,.
\ee
In the Euclidean spacetime, by taking the same limit in (\ref{z-inv-1int-beta}),
we  obtain (\ref{xi-BY}) with $2\pi \tau$ replaced by $\ri \theta$.

%\newpage
\subsection{Inversion on the real line}
\label{inversion-line-thermal}

By adapting the observation  (\ref{z-inv-1int-pi}), made for the ground state,
to  a CFT at finite temperature and on the bipartite line,
from (\ref{xi-thermal-tau}) or (\ref{z-inv-1int-beta}) 
we introduce the following inversion map 
\be
\label{x-inv-th-def}
z_{\textrm{\tiny inv}}(x) 
\,\equiv\,
 \xi(\tau =\textrm{i}/2\,,x) 
 \,=\,
 z(\theta=\pi , x) 
 \,=\,
 \frac{\beta}{2\pi }\, \log\! \bigg(
 \frac{\e^{ \pi (b+a)/\beta} - \e^{2\pi b/\beta}\, e^{w(x)}
 }{ 
 \e^{\pi (b-a)/\beta}  -  \e^{w(x)}}
\bigg)\,.
\ee

%\noindent
%$\bullet$ {\bf Thermal sub-interval}

In this expression the argument of the logarithm becomes negative for some values of $x\in A$.
The range of these values can be found by observing that
the denominator and the numerator of the ratio in the argument of the logarithm in (\ref{x-inv-th-def})
vanish when $x$ is equal respectively to
\bea
\label{a-beta-def}
a_\beta 
& \equiv &a + \frac{\ell - \ell_\beta}{2}
\,=\,
a + \frac{\ell}{2} \left( 1- \frac{\log [ \cosh( \pi \ell/\beta) ]}{\pi \ell/\beta} \right)
=\,
-\,\frac{\beta}{2\pi}\, \log\!\big[\big(\textrm{e}^{-2\pi a/\beta} + \textrm{e}^{-2\pi b/\beta} \big)/2\big]
\hspace{1cm}
\\
\label{b-beta-def}
\rule{0pt}{.7cm}
b_\beta 
& \equiv & 
b - \frac{\ell - \ell_\beta}{2}
\,=\,
b - \frac{\ell}{2} \left( 1- \frac{\log [ \cosh( \pi \ell/\beta) ]}{\pi \ell/\beta} \right)
=\,
\frac{\beta}{2\pi}\, \log\!\big[\big(\textrm{e}^{2\pi a/\beta} + \textrm{e}^{2\pi b/\beta} \big)/2\big]
\hspace{1cm}
\eea
where
\be
\label{thermal-length-def}
\ell_\beta 
\, \equiv \, b_\beta - a_\beta
\, = \, \frac{\beta}{\pi} \, \log\! \big[ \! \cosh( \pi \ell/\beta)\big]
\, = \, \ell + \frac{\beta}{\pi}\, \log\!\big[\big(1 + \textrm{e}^{-2\pi \ell/\beta} \big)/2\big]
\,<\, \ell \,.
\ee
The argument of the logarithm in (\ref{x-inv-th-def}) is negative when $x\in A_\beta$,
where $A_\beta \equiv ( a_\beta\, , b_\beta )$, which  is a proper subinterval of $A$
of length $\ell_\beta$.
From (\ref{a-beta-def}) and (\ref{b-beta-def}), one observes that 
$A_\beta$ and $A$ have the same mid-point.
Thus, when $a=-b$, also $a_\beta = - b_\beta$.

From (\ref{thermal-length-def}), we have that $\ell_\beta \to 0$ as $\beta \to +\infty$ 
and $\ell_\beta \to \ell$ as $\beta \to 0$;
namely $A_\beta$ disappears at zero temperature
while it coincides with $A$ in the large temperature regime.

%\textcolor{red}{[Is the following also in \cite{Borchers:1998ye}?]}
In the limit where $A$ becomes the half-line,
which is obtained by setting $b=a+\ell$ first and then taking $\ell \to +\infty$,
for the endpoints of $A_\beta$ in (\ref{a-beta-def}) and (\ref{b-beta-def})
we find that $a_\beta \to a+\tfrac{\beta}{2\pi} \log(2) $ and $b_\beta \to +\infty$ respectively.
%hence $(\ell - \ell_\beta)/2 \to \frac{\beta}{2\pi} \log(2)$.
%
Now, taking $a\to -\infty$ in this result, $A_\beta$ becomes the whole line.
This is found also by taking  $b\to +\infty$ and $a\to -\infty$ simultaneously in 
 (\ref{a-beta-def}) and (\ref{b-beta-def}).

Let us stress that $A_\beta$ has been introduced through the inversion map (\ref{x-inv-th-def}).
It would be insightful to find a footprint of $A_\beta$ without invoking  the modular conjugation.

%\noindent
%$\bullet$ {\bf Complete inversion map on the line.}

\begin{figure}[t!]
\vspace{-.6cm}
\hspace{.1cm}
%\begin{center}
\includegraphics[width=.98\textwidth]{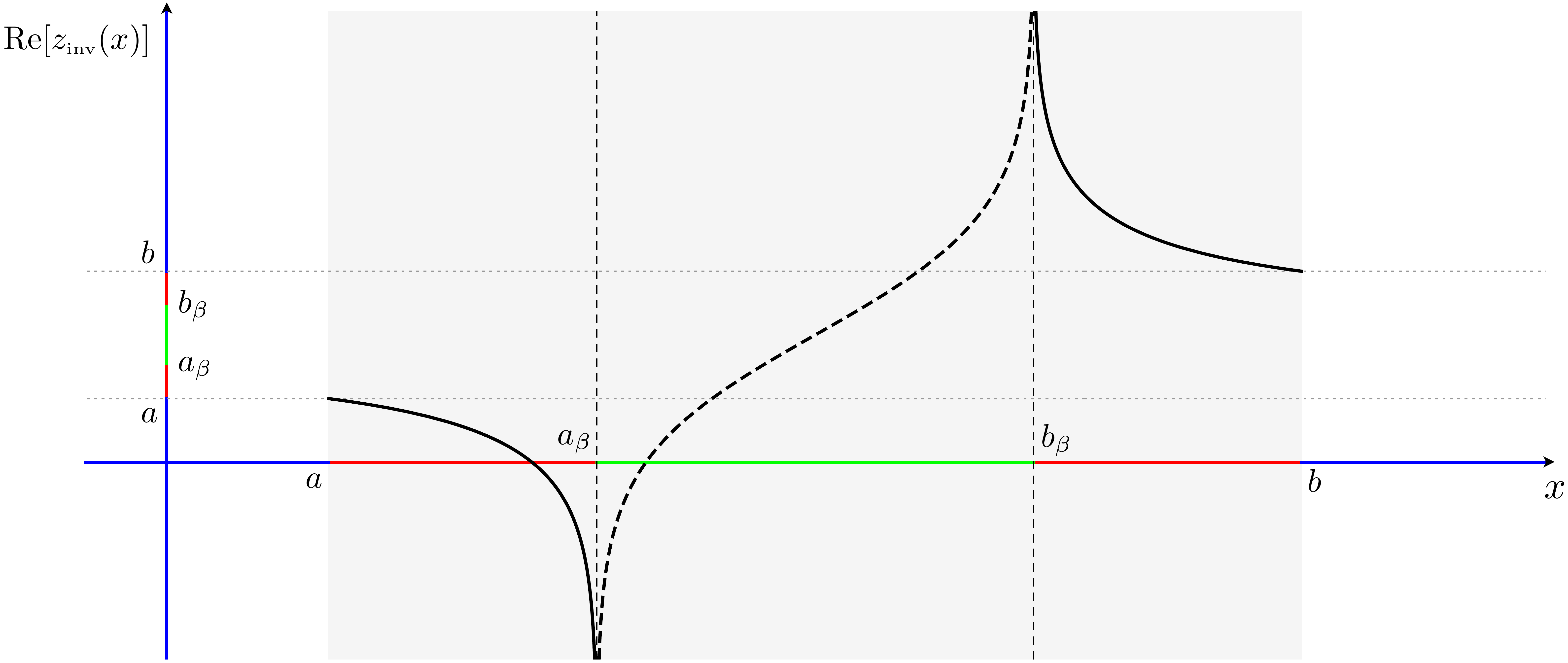}
% \end{center}
\vspace{.2cm}
\caption{The real part of the inversion map (\ref{x-inv-th-bis-split}).
The green segment corresponds to $A_\beta$, the red segments to $A \setminus A_\beta$
and the blue half lines to $B$.
}
\label{figure-xinv-thermal}
\end{figure}

Since the argument of the logarithm in (\ref{x-inv-th-def}) is negative when $x\in A_\beta$, 
the inversion (\ref{x-inv-th-def}) is a complex map that can be written as follows
\bea
\label{x-inv-th-bis}
z_{\textrm{\tiny inv}}(x) 
&=&
 \frac{\beta}{2\pi }\, \log\! 
 \bigg(\, \bigg|  \frac{ \e^{ \pi (b+a)/\beta} - \e^{2\pi b/\beta}\, \e^{w(x)} }{  \e^{\pi (b-a)/\beta}  -  \e^{w(x)}}  \bigg|\, \bigg)
\pm
\textrm{i}\, \frac{\beta}{2} \, \Theta_{A_\beta}(x)
\\
\label{x-inv-th-bis-split}
\rule{0pt}{1.6cm}
&=&
\left\{\begin{array}{ll}
\displaystyle
 \frac{\beta}{2\pi }\, \log\!  \bigg(   \frac{\e^{ \pi (b+a)/\beta} - \e^{2\pi b/\beta}\, \e^{w(x)} }{  \e^{\pi (b-a)/\beta}  -  \e^{w(x)}}   \bigg)
\hspace{2.4cm} &
x\in A\setminus A_\beta
\\
\rule{0pt}{1cm}
\displaystyle
 \frac{\beta}{2\pi }\, \log\!    \bigg(   \frac{ \e^{2\pi b/\beta}\, \e^{w(x)} - \e^{ \pi (b+a)/\beta}  }{  \e^{\pi (b-a)/\beta}  -  \e^{w(x)}}   \bigg)
 \pm
\textrm{i}\, \frac{\beta}{2} 
&
x\in  A_\beta
\end{array}\right.
\eea
where $\Theta_{A_\beta}(x)$ it the characteristic function\footnote{The characteristic function $\Theta_{R}(x)$ of a region $R$  is equal to one when $x\in R$ and zero otherwise.} 
of $A_\beta$.
%
%\textcolor{red}{ $\bullet$ [Which sign should we choose in (\ref{x-inv-th-bis})? Maybe plus.  It is used in (\ref{x-t-inv-z-one-int-thermal})]}
%
It is worth remarking that the real functions 
$z_{\textrm{\tiny inv}}(x)   : A\setminus A_\beta \to B$ and $\textrm{Re}[z_{\textrm{\tiny inv}}(x) ]  : A_\beta \to \mathbb{R}$ are both bijective.
They  are shown in Fig.\,\ref{figure-xinv-thermal} through the solid curves and the dashed curve respectively. 
The latter one tells us that $\textrm{Re}[z_{\textrm{\tiny inv}}(x) ]  : A_\beta \to \mathbb{R}$ is strictly increasing. 

Combining the two chiralities, in Sec.\,\ref{sec-thermal-inv-lorentz} 
we will see that the imaginary part occurring in the complex map $z_{\textrm{\tiny inv}}(x)   : A_\beta \to \mathbb{R}$
cancels and therefore naturally provides a second Minkowski space $\widetilde{\mathcal{M}}$,
which has been called second world in \cite{Borchers:1999short}
or shadow (virgin) world in \cite{Schroer:1998pj}.
We remark that this second world $\widetilde{\mathcal{M}}$, 
differently from the real one $\mathcal{M}$, is not bipartite.

In the special case of $A=(-b,b)$, the inversion map (\ref{x-inv-th-bis-split}) simplifies to
\be
\label{xinv-th-sym}
z_{\textrm{\tiny inv}}(x) 
\,=\,
\left\{\begin{array}{ll}
\displaystyle
 \frac{\beta}{2\pi }\, \log\!  \bigg(  \frac{\e^{2\pi x/\beta}  \cosh(2\pi b/\beta) - 1}{ \e^{2\pi x/\beta} - \cosh(2\pi b/\beta) } \bigg)
\hspace{2cm} &
x\in A\setminus A_\beta
\\
\rule{0pt}{1cm}
\displaystyle
 \frac{\beta}{2\pi }\, \log\!   \bigg(  \frac{\e^{2\pi x/\beta}  \cosh(2\pi b/\beta) - 1}{ \cosh(2\pi b/\beta) - \e^{2\pi x/\beta} }  \bigg)
 \pm
\textrm{i}\, \frac{\beta}{2} 
&
x\in  A_\beta \,.
\end{array}\right.
\ee
This result will be re-obtained holographically
also through the geodesic bit threads in the BTZ black brane background
in Sec.\,\ref{sec-BTZ}.

In the zero temperature limit $\beta \to +\infty$, we have that $A_\beta \to \emptyset$, as remarked above;
hence only the first expression in (\ref{x-inv-th-bis-split}) must be considered, 
which becomes (\ref{x-inv = z-inv t=0}) in this limit, as expected.

In the limiting regime $b \to +\infty$, the interval $A$ becomes the half-line $(a, +\infty)$ and,
by using also the above results $a_\beta \to a+\tfrac{\beta}{2\pi} \log(2) $ and $b_\beta \to +\infty$,
one finds that the map (\ref{x-inv-th-bis-split}) simplifies to 
\be
\label{x-split-half}
z_{\textrm{\tiny inv}}(x) 
\,=\,
\left\{\begin{array}{ll}
\displaystyle
a +   \frac{\beta}{2\pi }\, \log\!  \Big(   2 - \e^{2\pi(x-a)/\beta}  \Big)
\hspace{2.2cm} &
\displaystyle
a< x < a+ \frac{\beta}{2\pi} \log(2)
\\
\rule{0pt}{1cm}
\displaystyle
a +   \frac{\beta}{2\pi }\, \log\!  \Big(    \e^{2\pi(x-a)/\beta} - 2 \Big)
 \pm
\textrm{i}\, \frac{\beta}{2} 
&
\displaystyle
x > a +\frac{\beta}{2\pi} \log(2)
\end{array}\right.
\ee
which is real in $A\setminus A_\beta = (a, a+\tfrac{\beta}{2\pi} \log(2))$ 
and sends this limited interval onto the half line $B=(-\infty, a)$ complementary to $A$ in a bijective way. 
By employing the exponential map characterising the thermal state,
we can write (\ref{x-split-half}) in the following suggestive form
\be
\label{reflection-1-exp}
\e^{2\pi [z_{\textrm{\tiny inv}}(x) - a]/\beta} \,=\, 2 - \e^{2\pi (x-a)/\beta}
\ee
which becomes (\ref{Haag-inversion-map-upm-half-line}) in the zero temperature limit, as expected. 
In the non negative variable $y\equiv \e^{2\pi (x-a)/\beta}$, this map becomes to the reflection $y \to 2 - y$ with respect to $1$.
The reflection (\ref{reflection-1-exp}) suggests to continue this real positive variable $y$
by including also the semi-axis made by the negative real numbers $y < 0$, which corresponds to the second world
\cite{Schroer:1998pj, Borchers:1999short}.
Indeed, such reflection maps the points $y > 2$ into the negative real semi-axis in a bijective way.
%

%\noindent
%$\bullet$ {\bf Stefan law}

In order to explore the meaning of the sub-interval $A_\beta$,
we find it worth exploiting the contour function for the entanglement entropies \cite{ChenVidal2014, Coser:2017dtb}.
For a CFT at finite temperature on the line bipartite by an interval,
this function reads \cite{Coser:2017dtb}\footnote{See Eqs.\,(65) and (71) of \cite{Coser:2017dtb}.}
\be
\label{contour-func-def}
s^{(n)}_A(x) \,\equiv\, \frac{c}{12} \left( 1 + \frac{1}{n} \right) w'(x) + \frac{C}{\ell}
\ee
where $w(x)$ is given by (\ref{wtherm})
and $C$ is a non universal constant such that $s^{(n)}_A(x) > 0$ for every $x \in A$.
The main feature of the positive function (\ref{contour-func-def}) is that 
the entanglement entropies of $A$ found in \cite{Calabrese:2004eu} 
are obtained by integrating it on $A_\epsilon \equiv (a+\epsilon, b- \epsilon)$, 
where $\epsilon$ is the UV cutoff, i.e.
\be
\label{S_A-thermal-from-contour}
S^{(n)}_A 
= 
\int_{a+\epsilon}^{b-\epsilon} \!\! s^{(n)}_A(x) \, \rd x
\,=\,
\frac{c}{12} \left( 1 + \frac{1}{n} \right)
\log\!\left[\,\frac{\beta}{\pi \epsilon}\, \sinh (\pi \ell/ \beta)\,\right]
+O(1)
\ee
up to $O(1)$ non universal contributions as $\epsilon \to 0$.
In \cite{Calabrese:2004eu} it has been remarked that,
when $\ell \gg \beta$,
the entanglement entropy obtained from (\ref{S_A-thermal-from-contour}) 
becomes extensive and provides the Gibbs entropy $S_A \simeq \pi\,c \,\ell / (3\beta)$ 
of an isolated system of length $\ell$ 
coming from the Stefan-Boltzmann law for a 2D CFT (see (\ref{T}))
%the free energy $\beta F \sim - \,\pi\,c \,\ell / (6\beta)$ of a CFT
\cite{Bloete:1986qm, Affleck:1986bv, Cardy:2010fa}.

It is sometimes insightful to integrate the contour function for the entanglement entropies in (\ref{contour-func-def}) 
over a subsystem properly included in $A$ (see e.g. \cite{DiGiulio:2019lpb}).
In the case that we are exploring, 
the integration of (\ref{contour-func-def}) over $A_\beta$ gives
\be
\label{integral-contour-beta}
\int_{a_\beta}^{b_\beta}  \!\! s^{(n)}_A(x) \, \rd x
\,=\,
\frac{c}{12} \left( 1 + \frac{1}{n} \right) \big[ w(b_\beta) - w(a_\beta) \big] + \frac{C}{\ell}\, \ell_\beta
\,=\,
\frac{c}{6} \left( 1 + \frac{1}{n} \right) \frac{\pi\, \ell}{\beta} + \frac{\ell_\beta}{\ell}\, C \,.
\ee
where $\ell_\beta$ has been introduced in (\ref{thermal-length-def}).
When $n=1$ and $C=0$, the expression (\ref{integral-contour-beta}) provides the above mentioned Gibbs entropy $\pi\, c \,\ell / (3\beta)$ 
for any value of $\ell$ and $\beta$.

Considering the sub-interval $A_\beta$, 
from (\ref{wtherm}) and its endpoints (\ref{a-beta-def}) and (\ref{b-beta-def}),
one finds
\be
\label{Delta-w-A-beta}
w(b_\beta) - w(a_\beta) = \frac{2\pi \,(b-a)}{\beta} \equiv \tau_\beta 
\ee
which can be interpreted through the modular parameter $\tau$.
Indeed, by considering (\ref{xi-def-gen}) for $x = a_\beta$, 
the expression (\ref{Delta-w-A-beta}) corresponds to 
the value $\tau_\beta$ of the  modular parameter such that $\xi(\tau_\beta , a_\beta) = b_\beta$.
As a consequence, also (\ref{integral-contour-beta}) can be interpreted through the modular parameter $\tau$
because it is proportional to (\ref{Delta-w-A-beta}) when $C=0$.

%\newpage
\subsection{Geometric action in Euclidean spacetime}
\label{sec-thermal-inv-euclid}

\begin{figure}[t!]
\vspace{-.6cm}
\hspace{.3cm}
%\begin{center}
\includegraphics[width=.95\textwidth]{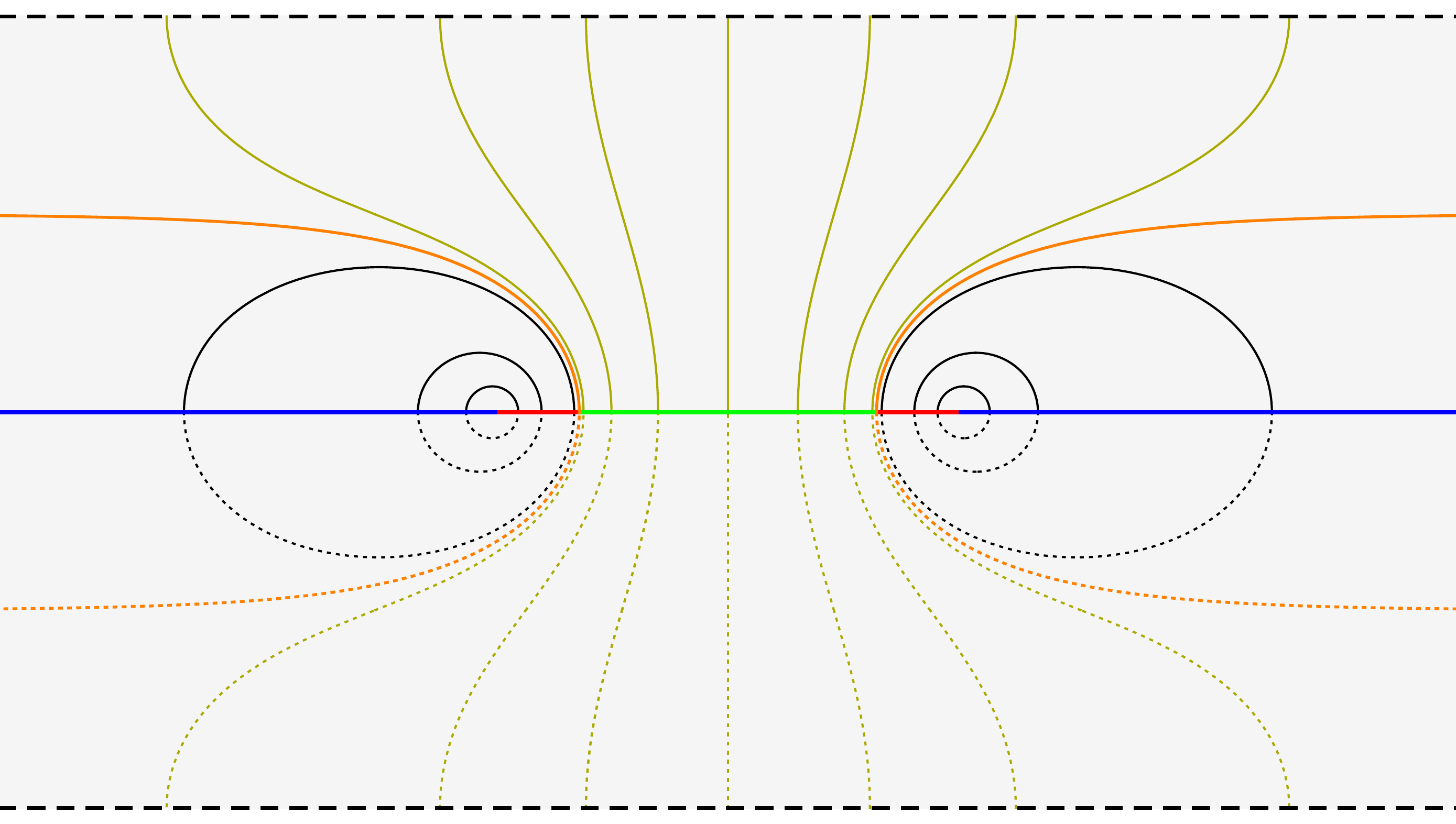}
% \end{center}
\vspace{.2cm}
\caption{Euclidean modular evolution for a CFT  at finite temperature $1/\beta$ on the bipartite line,
obtained from (\ref{z-inv-1int-beta}),
with $\theta \in (0,\pi)$ (solid arcs) or $\theta \in (\pi,2\pi)$ (dotted arcs),
for some $x\in A$.
The green segment corresponds to $A_\beta$.
The solid black curves (or the dotted black ones equivalently) 
map $A\setminus A_\beta$ into $B$ in a bijective way,
while the solid yellow curves (or the dotted yellow ones equivalently) 
map $A_\beta$ into the horizontal dashed black line at $\textrm{Im}(z) = + \beta/2$
(which is identified with the horizontal dashed black line at $\textrm{Im}(z) = - \beta/2$)
 in a bijective way.
}
\label{figure-inversion-Euclid-1int-thermal}
\end{figure}

In the finite temperature case that we are investigating, 
the geometric action of the modular conjugation can be studied by adapting the observations made 
in Sec.\,\ref{sec-inversion-gs-lorentz} and Sec.\,\ref{sec-euclid-gs} for the ground state. 
It is more instructive to discuss first the Euclidean spacetime.

For a CFT at finite temperature $1/\beta$ on the line,
the Euclidean spacetime to consider is 
the infinite cylinder whose section along the compactified direction is a circle having length equal to $\beta$.
It can be equivalently represented by the horizontal strip 
$\{z\in \mathbb{C} ; |\textrm{Im}(z)| \leqslant \beta/2 \}$ 
in the complex plane (parameterised by $z$)
with vertical width equal to $\beta$ (grey region in Fig.\,\ref{figure-inversion-Euclid-1int-thermal}),
where the horizontal lines at $\textrm{Im}(z) = \pm \beta/2$ delimiting the strip
(horizontal black dashed lines in Fig.\,\ref{figure-inversion-Euclid-1int-thermal})
are identified.
The space where the CFT is defined corresponds to the real axis $\textrm{Im}(z) = 0$,
which is bipartite by the interval $A$ and its complement.
This bipartition is characterised by the entangling points $x=a$ and $x=b$ on the real axis, 
which correspond to  the points where a red segment and a blue segment join in Fig.\,\ref{figure-inversion-Euclid-1int-thermal};
hence $A$ is the union of the two red segments and the green segment,
while $B$ corresponds to the union of the blue half lines.

The curves shown in Fig.\,\ref{figure-inversion-Euclid-1int-thermal}
are obtained from (\ref{z-inv-1int-beta}) for some values of $x\in A$:
the solid ones have $\theta \in (0, \pi)$,
while $\theta \in (\pi, 2\pi)$ for the dotted ones.
Any black solid arc (or equivalently the dotted one having the same endpoints)
associates a point in the red segments to a point in $B$ in a unique way.
Instead, any yellow arc (or equivalently the dotted one having the same endpoints)
associates a point in the green segment to a point on the horizontal black dashed lines
at $\textrm{Im}(z) = \pm \beta/2$ (which are identified).
This map is (\ref{x-inv-th-bis-split})
and the green segment in Fig.\,\ref{figure-inversion-Euclid-1int-thermal} denotes the sub-interval $A_\beta$.
The two orange curves in Fig.\,\ref{figure-inversion-Euclid-1int-thermal} 
intersect the real axis at $x=a_\beta$ or $x=b_\beta$
and connect these points to $-\infty$ or $+\infty$ respectively;
hence they provide the extremal lines 
separating the set of the black curves and the set of the yellow curves,
whose behaviours are qualitatively different.
In this Euclidean setup, the horizontal black dashed lines at $\textrm{Im}(z) = \pm \beta/2$ 
corresponds to the space of the second world \cite{Borchers:1999short, Schroer:1998pj}
introduced in Sec.\,\ref{inversion-line-thermal}.

In this Euclidean strip,
it is useful to consider also the curves obtained from (\ref{z-inv-1int-beta}) at fixed $\theta \in (0,2\pi )$, 
parameterised by $x\in A$.
For these curves, we find that $\partial_x z(\theta, x)  \to  \e^{\ri \theta}$ as $x\to a$
and  $\partial_x z(\theta, x)  \to  \e^{-\ri \theta}$ as $x\to b$;
hence the curve corresponding to $\theta = \pi/2$ intersects orthogonally the real axis at $x=a$ and $x=b$.
This result is independent of $\beta$ 
and for the ground state it has been already remarked in Sec.\,\ref{sec-euclid-gs}.
\\

%\newpage
\subsection{Geometric action in Minkowski spacetime}
\label{sec-thermal-inv-lorentz}

\begin{figure}[t!]
\vspace{-.4cm}
%\hspace{3.6cm}
%\includegraphics[width=.55\textwidth]{fig-th-lor-out}
%\\
%\rule{0pt}{9.7cm}
\hspace{-1.1cm}
\includegraphics[width=.55\textwidth]{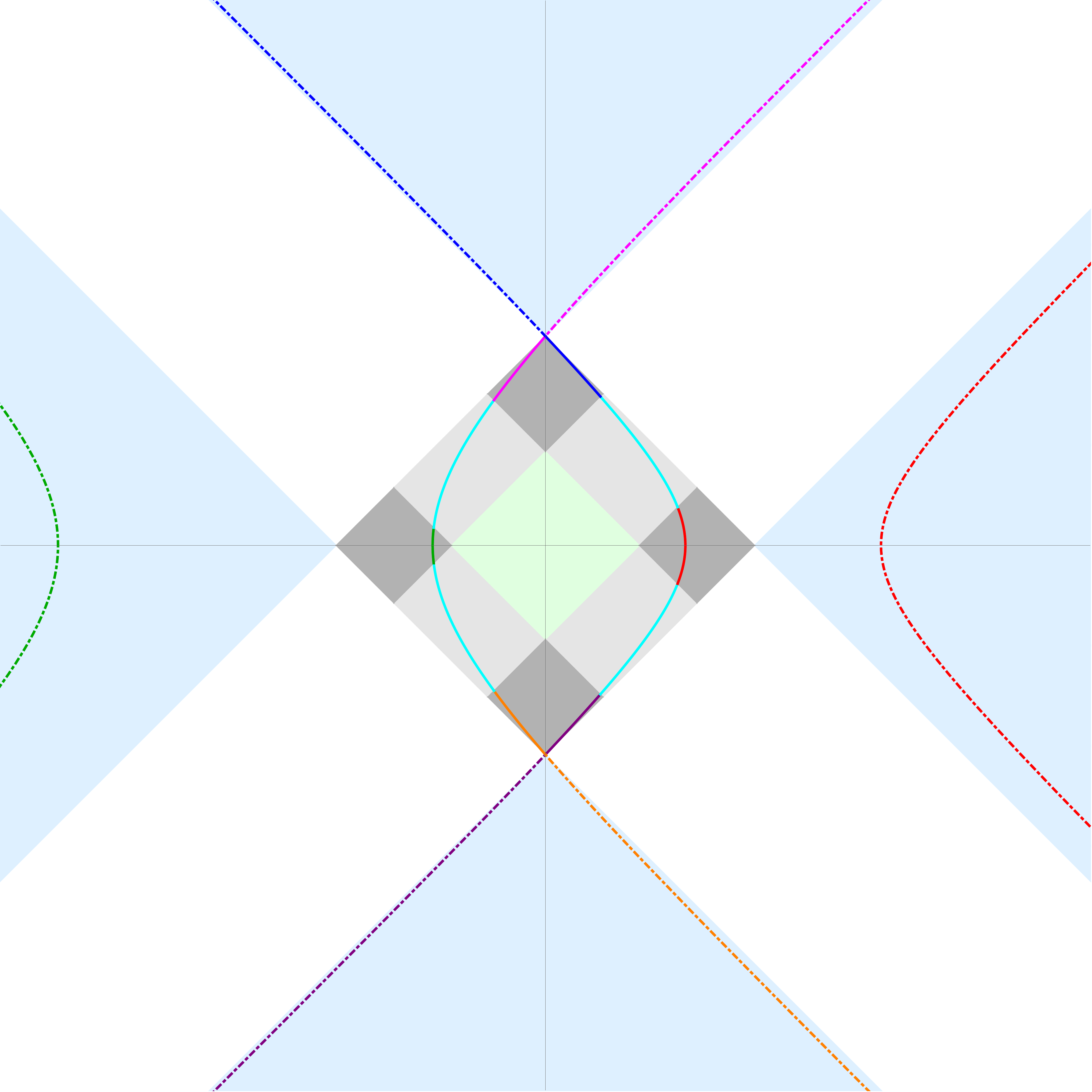}
\hspace{.7cm}
\includegraphics[width=.55\textwidth]{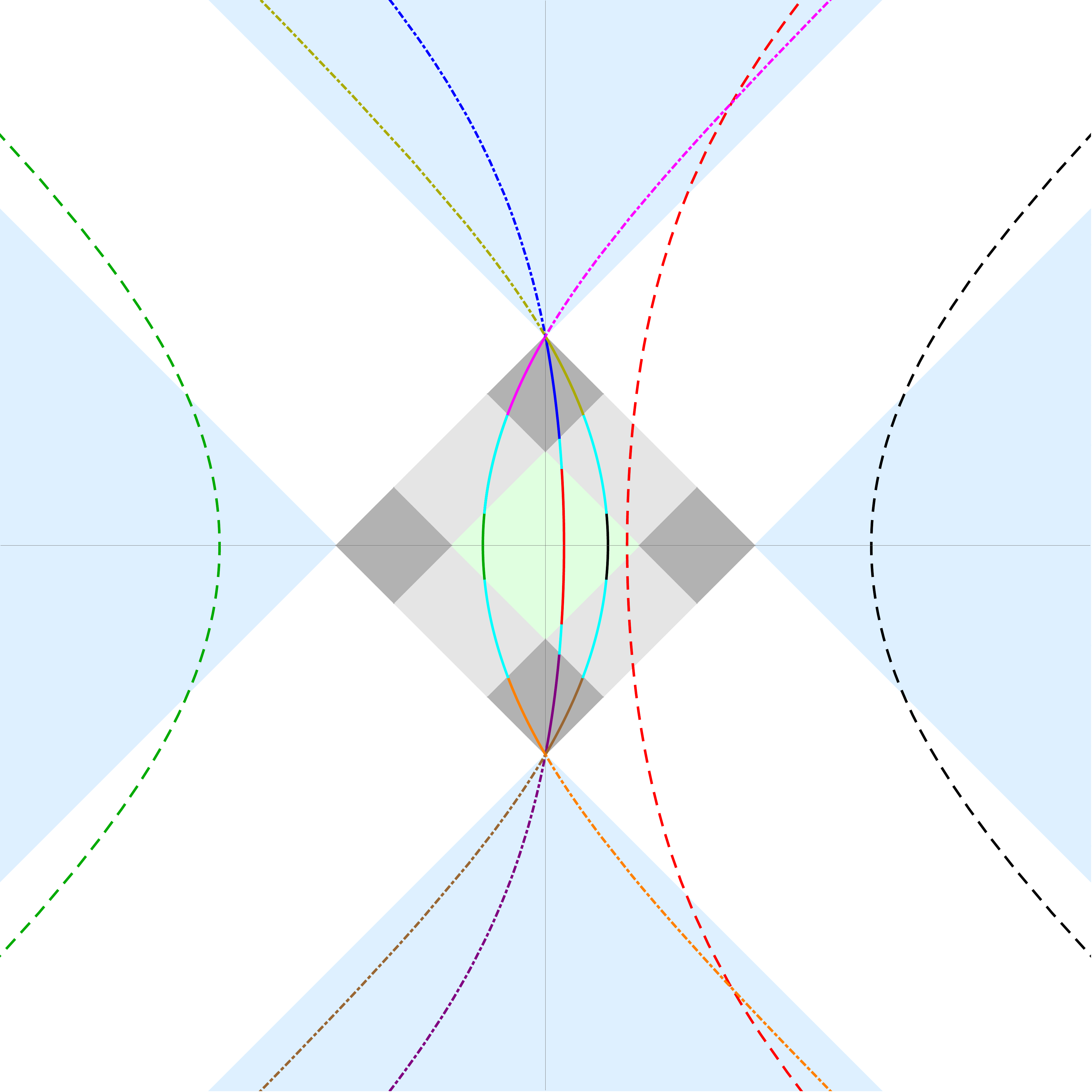}
\vspace{-.2cm}
\caption{Modular trajectories (solid lines) in $\mathcal{D}_A$ passing at some $x\in A$ when $t=0$ 
and their images (dot-dashed and dashed lines) under (\ref{x-t-inv-th-final}).
Left: two values of $x\in A\setminus A_\beta$.
Right: three values of $x\in A_\beta$.
The arcs denoted through different kind of lines but having the same colour are related through (\ref{x-t-inv-th-final}).
While the dot-dashed lines span $\mathcal{W}_A$ (light blue region, see (\ref{WA-dec-gs})) 
as $x\in A\setminus A_\beta$,
the dashed lines span the entire Minkowski spacetime (the second world $\widetilde{\mathcal{M}}$)
as $x\in A_\beta$.
}
\label{figure-inv-thermal-2-world}
\end{figure}

In the Minkowski spacetime $\mathcal{M}$ parameterised by the coordinates $(x,t)$,
let us consider the diamond $\mathcal{D}_A$ (see Sec.\,\ref{sec-mod-evo-gs}).
The occurrence of the sub-interval $A_\beta$ introduced in Sec.\,\ref{inversion-line-thermal},
whose endpoints on the spatial line are (\ref{a-beta-def}) and (\ref{b-beta-def}),
naturally leads to the following partition 
\be
\label{DA-partition-th}
\mathcal{D}_A = \mathcal{D}_{A_\beta} \cup \widetilde{\mathcal{D}}_A \cup \mathcal{R}_A 
\ee
identified by the null rays departing from the points $(a_\beta, 0)$ and $(b_\beta, 0)$
(see Fig.\,\ref{figure-inv-thermal-2-world}).
In (\ref{DA-partition-th}), $\mathcal{D}_{A_\beta}$ is the domain of dependence (diamond) of $A_\beta$
(green region in Fig.\,\ref{figure-inv-thermal-2-world}), 
while $ \widetilde{\mathcal{D}}_A$ and $\mathcal{R}_A $ are defined as follows
\be
\label{CA-RA-dec-thermal}
\widetilde{\mathcal{D}}_A \,\equiv \,
\widetilde{\mathcal{D}}_{\textrm{\tiny R}} \cup \widetilde{\mathcal{D}}_{\textrm{\tiny L}} \cup 
\widetilde{\mathcal{D}}_{\textrm{\tiny F}} \cup \widetilde{\mathcal{D}}_{\textrm{\tiny P}} 
\;\;\;\;\qquad\;\;\;\;
\mathcal{R}_A \,\equiv \,
\mathcal{R}_{++} \cup \mathcal{R}_{-+} \cup 
\mathcal{R}_{--} \cup \mathcal{R}_{+-}  \;.
\ee
The domains $\widetilde{\mathcal{D}}_k$ with $k \in \{ \textrm{R} , \textrm{L} , \textrm{F} , \textrm{P}\}$
are the dark grey diamonds in Fig.\,\ref{figure-inv-thermal-2-world}
and the letters for the subindex,
which stand for right, left, future and past respectively (see also (\ref{DA-dec-gs})),
indicate their position with respect to $\mathcal{D}_{A_\beta}$.
Each of these diamonds shares a vertex with $\mathcal{D}_{A_\beta}$.
The domains $\mathcal{R}_{rs}$ with $r,s \in \{ +, -\}$ 
are the light grey rectangles in Fig.\,\ref{figure-inv-thermal-2-world}
and the subindices indicate their position with respect to $\mathcal{D}_{A_\beta}$
(e.g. $++$ and $+-$ correspond to the top right and top left light grey rectangle respectively).

In the limit where $A$ becomes the half line, 
$\mathcal{D}_A $ and $\mathcal{D}_{A_\beta}$ become the right wedges
with vertex in $(a,0)$ and $(a+\tfrac{\beta}{2\pi} \log(2) , 0)$ respectively. 
Furthermore, only $\widetilde{\mathcal{D}}_{\textrm{\tiny L}}$ remains a finite diamond in this limit, 
while the other dark grey regions $\widetilde{\mathcal{D}}_{k}$ with $k \neq \textrm{L}$
disappear at infinity.

In the zero temperature limit $\beta \to +\infty$,
we have that $\mathcal{D}_{A_\beta}$ and $\mathcal{R}_A$ 
become zero measure regions in Minkowski space
and $\widetilde{\mathcal{D}}_A  \to \mathcal{D}_A $.
Moreover, considering the partitions of $ \mathcal{D}_A$ introduced in (\ref{DA-dec-gs}) and (\ref{CA-RA-dec-thermal}),
in this limit we have that $\widetilde{\mathcal{D}}_k \to \mathcal{D}_k$ for each $k \in \{ \textrm{R} , \textrm{L} , \textrm{F} , \textrm{P}\}$,
up to zero measure domains.

A modular trajectory in Minkowski space is obtained by combining (\ref{xi-thermal-tau}) and (\ref{mod-trajec-1int}).
It depends on the assigned value $x\in A$ at $\tau =0$ and it entirely belongs to $\mathcal{D}_A$.
The partition (\ref{DA-partition-th}) of $\mathcal{D}_A$ induces a partition of a modular trajectory into (at most) five arcs
corresponding to its intersections with $\mathcal{D}_{A_\beta}$, $\widetilde{\mathcal{D}}_A$ and $\mathcal{R}_A$
(arcs denoted with different colours that partition each solid curve in Fig.\,\ref{figure-inv-thermal-2-world}).
The only exception is the modular trajectory corresponding to $x=\tfrac{a+b}{2}$,
which is the vertical segment connecting the top and bottom vertices of $\mathcal{D}_A$;
indeed, its partition is made only by three vertical segments because it does not intersect $\mathcal{R}_A$.
The five arcs partitioning the modular trajectory can be found
also through the corresponding values of the modular parameter.
The ranges of $\tau$ that provide the different arcs are identified 
by $\pm |\tau_{a_\beta}|$ and $\pm |\tau_{b_\beta}|$, where
\be
\label{tau-ab-beta-x}
\tau_{a_\beta} \equiv \frac{w(a_\beta) - w(x)}{2\pi}
\;\;\qquad\;\;
\tau_{b_\beta} \equiv \frac{w(b_\beta) - w(x)}{2\pi} \,.
\ee

A modular trajectory is characterised by its point $(x,0)$ with $x\in A$ at $\tau=0$.
At finite temperature, we have either $x\in A\setminus A_\beta$ (left panel in Fig.\,\ref{figure-inv-thermal-2-world})
or $x\in A_\beta$ (right panel in Fig.\,\ref{figure-inv-thermal-2-world}).
Considering the partition (\ref{DA-partition-th}), 
notice that, while all the modular trajectories have non vanishing intersections with $\widetilde{\mathcal{D}}_A$ ad $\mathcal{R}_A$,
only those ones corresponding to $x\in A_\beta$ intersect also $\mathcal{D}_{A_\beta} $.

In order to adapt to the thermal case
the results discussed  in Sec.\,\ref{sec-inversion-gs-lorentz} for the ground state
(see (\ref{Haag-inversion-map}) and (\ref{x-t-inv-z-one-int})), 
we introduce the geometric action of the modular conjugation as follows
\be
\label{x-t-inv-z-one-int-thermal}
x_{\textrm{\tiny inv}} (x, t)
=
\frac{z_{\textrm{\tiny inv}}(x+t) + z_{\textrm{\tiny inv}}(x-t) }{2} 
\;\;\qquad\;\;
t_{\textrm{\tiny inv}} (x, t)
=
\frac{z_{\textrm{\tiny inv}}(x+t) - z_{\textrm{\tiny inv}}(x-t) }{2} 
\ee
where the function $z_{\textrm{\tiny inv}}(x) $ is given by (\ref{x-inv-th-bis}).
In terms of the light ray coordinates (\ref{lc1}), it is convenient to introduce
$u_{\pm,\textrm{\tiny inv}} \equiv x_{\textrm{\tiny inv}}  \pm t_{\textrm{\tiny inv}} $, finding that
\be
u_{\pm,\textrm{\tiny inv}}
=
z_{\textrm{\tiny inv}}(u_\pm) \,. 
\ee
From (\ref{x-t-inv-z-one-int-thermal}) we obtain 
\be
\label{x-t-inv-th-final}
\left\{
\begin{array}{l}
\displaystyle
x_{\textrm{\tiny inv}} (x, t)
=\,
 \frac{\beta}{4\pi }\, \log\!  \Bigg[
  \frac{ \big( \e^{ \pi (b+a)/\beta} - \e^{2\pi b/\beta}\, \e^{w(x+t)} \big) \big(\e^{ \pi (b+a)/\beta} - \e^{2\pi b/\beta}\, \e^{w(x-t)}  \big)
 }{ 
 \big( \e^{\pi (b-a)/\beta}  -  \e^{w(x+t)} \big) \big( \e^{\pi (b-a)/\beta}  -  \e^{w(x-t)}\big)} 
 \Bigg]
\\
\rule{0pt}{1.1cm}
\displaystyle
t_{\textrm{\tiny inv}} (x, t)
\,=\,
 \frac{\beta}{4\pi }\, \log\!  \Bigg[
  \frac{ \big( \e^{ \pi (b+a)/\beta} - \e^{2\pi b/\beta}\, \e^{w(x+t)} \big) \big( \e^{\pi (b-a)/\beta}  -  \e^{w(x-t)} \big)
 }{ 
 \big( \e^{\pi (b-a)/\beta}  -  \e^{w(x+t)} \big) \big(  \e^{ \pi (b+a)/\beta} - \e^{2\pi b/\beta}\, \e^{w(x-t)} \big)} 
 \Bigg] \,.
 \end{array}
 \right.
\ee
This map satisfies some consistency relations. 
For $t=0$ we have $t_{\textrm{\tiny inv}} (x, t=0) = 0$, as expected,
and, interestingly, $x_{\textrm{\tiny inv}} (x, t=0) = \textrm{Re}[z_{\textrm{\tiny inv}} (x)]$, 
which is shown in Fig.\,\ref{figure-xinv-thermal}.
In the zero temperature limit $\beta \to +\infty$,
the inversion map (\ref{Haag-inversion-map}) is recovered, as expected. 
We remark that (\ref{x-t-inv-th-final}) has been obtained by merging the two logarithmic functions
coming from $z_{\textrm{\tiny inv}}(x+t)$ and $z_{\textrm{\tiny inv}}(x-t)$ into a single logarithmic function
and that such seemingly innocent simplification leads to interesting consequences, as discussed below.

Consider the partitions in  (\ref{DA-partition-th}) and (\ref{CA-RA-dec-thermal}),
together with the partition (\ref{WA-dec-gs}) for the light blue region in both panels of Fig.\,\ref{figure-inv-thermal-2-world}.
When $(x,t) \in \widetilde{\mathcal{D}}_A$,
the inversion map (\ref{x-t-inv-z-one-int-thermal}) is equivalent to (\ref{x-t-inv-th-final})
and it defines the four bijective maps 
$\widetilde{\mathcal{D}}_{\textrm{\tiny R}}  \to \mathcal{W}_{\textrm{\tiny R}}$,
$\widetilde{\mathcal{D}}_{\textrm{\tiny L}}  \to \mathcal{W}_{\textrm{\tiny L}}$,
$\widetilde{\mathcal{D}}_{\textrm{\tiny F}}  \to \mathcal{V}_{\textrm{\tiny F}}$
and $\widetilde{\mathcal{D}}_{\textrm{\tiny P}}  \to \mathcal{V}_{\textrm{\tiny P}}$.

The action of the map (\ref{x-t-inv-th-final})
 in $\mathcal{D}_{A_\beta} $ and $\mathcal{R}_{A}$
 is very interesting. 
When $(x,t) \in \mathcal{D}_{A_\beta} $,
the arguments of the logarithm in $z_{\textrm{\tiny inv}}(x+t)$ and $z_{\textrm{\tiny inv}}(x-t)$ are both negative;
hence (\ref{x-t-inv-z-one-int-thermal}) gives $\textrm{Im}(t_{\textrm{\tiny inv}} ) = 0$  and $| \textrm{Im}(x_{\textrm{\tiny inv}} ) | = \beta/2$.
Instead, in (\ref{x-t-inv-th-final}) the arguments of the logarithms remains positive;
hence (\ref{x-t-inv-th-final}) is a real map from $\mathcal{D}_{A_\beta} $ to 
the entire Minkowski spacetime $\widetilde{\mathcal{M}}$,
which is identified with the second world.
This is an interesting consequence of the combination of the two different chiralities;
indeed, they separately provide results with non vanishing imaginary parts. 
Notice that, differently from $\mathcal{M}$, 
the space direction of $\widetilde{\mathcal{M}}$ is not bipartite.
%
%Furthermore, this map assigns to any point of $\mathcal{D}_{A_\beta} $ a point in $\RR^2$ and viceversa. 
%This is highlighted by 
In the right panel of Fig.\,\ref{figure-inv-thermal-2-world},
a dashed curve of certain colour is the image under (\ref{x-t-inv-th-final}) of the solid arc in $\mathcal{D}_{A_\beta}$ having the same colour.
As $x\in A_\beta$, these dashed curves span the entire Minkowski spacetime $\widetilde{\mathcal{M}}$ (the second world).
Hence, 
{\it the structure of the "real world" is transported with the help of the conjugation $J$ to the "second world"},
as Borchers observed in \cite{Borchers:1999short}.

The above discussion tells us that the full modular Hamiltonian in $\mathcal{M} \cup \widetilde{\mathcal{M}}$ reads
\be
\label{K-full-ojima}
K
=
K_A \otimes \boldsymbol{1}_{B \cup \widetilde{\mathcal{M}}}
-
\boldsymbol{1}_{A} \otimes J K_A   J \,.
\ee
In the zero temperature limit $\beta \to +\infty$, 
we have that the second world $\widetilde{\mathcal{M}}$ disappears
and (\ref{K-full-ojima}) tends to $K = K_A \otimes \boldsymbol{1}_B - \boldsymbol{1}_A \otimes K_B$,
where $K_B = J K_A  J$, 
being $J$ the modular conjugation whose geometric action has been described in Sec.\,\ref{sec-line-single-interval}.

Instead, the image of a point $(x,t) \in \mathcal{R}_{A} $ under (\ref{x-t-inv-z-one-int-thermal}) or (\ref{x-t-inv-th-final}) 
has  $| \textrm{Im}(t_{\textrm{\tiny inv}} ) | =  | \textrm{Im}(x_{\textrm{\tiny inv}} ) | = \beta/4$;
hence the reality condition suggest to consider this result unphysical.
In Fig.\,\ref{figure-inv-thermal-2-world}, this implies that the cyan arcs do not have a dot-dashed or dashed counterpart.

When $A$ becomes the half line,
i.e. in the limit $b \to +\infty$, 
the inversion (\ref{x-t-inv-th-final}) simplifies to 
\be
\label{x-t-inv-th-final-half-line}
\left\{
\begin{array}{l}
\displaystyle
x_{\textrm{\tiny inv}} (x, t)
=\,
a + 
 \frac{\beta}{4\pi } \log\! \Big[ \big( \e^{2\pi(x-a+t)/\beta} - 2 \big)  \big( \e^{2\pi(x-a-t)/\beta} - 2 \big)  \Big]
\\
\rule{0pt}{1.cm}
\displaystyle
t_{\textrm{\tiny inv}} (x, t)
\,=\,
 \frac{\beta}{4\pi } \, \log\!  \bigg( \frac{\e^{2\pi(x-a+t)/\beta} - 2 }{ \e^{2\pi(x-a-t)/\beta} - 2  }  \bigg)
 \end{array}
 \right.
\ee
whose zero temperature limit gives (\ref{mod-conj-wedge}), as expected. 

In terms of the light ray coordinates $u_\pm$ and of $u_{\pm, \textrm{\tiny inv}} \equiv x_{\textrm{\tiny inv}}  \pm t_{\textrm{\tiny inv}} $, 
the map (\ref{x-t-inv-th-final-half-line}) can be written in the form (\ref{x-split-half}) or (\ref{reflection-1-exp}),
as expected.

\newpage
%%%%%%%%%%%%%%%%%%%%%%%%%%%%%%%%%%%%%%%%%%%%%%%%
\section{Geodesic bit threads in BTZ black brane}
\label{sec-BTZ}
%%%%%%%%%%%%%%%%%%%%%%%%%%%%%%%%%%%%%%%%%%%%%%%%

In this section we show that the inversion map on the real line discussed in Sec.\,\ref{inversion-line-thermal}
can be found through the geodesic bit threads in the BTZ black brane geometry described in \cite{Agon:2018lwq}.

\begin{figure}[t!]
\vspace{-.6cm}
\hspace{.3cm}
%\begin{center}
\includegraphics[width=.95\textwidth]{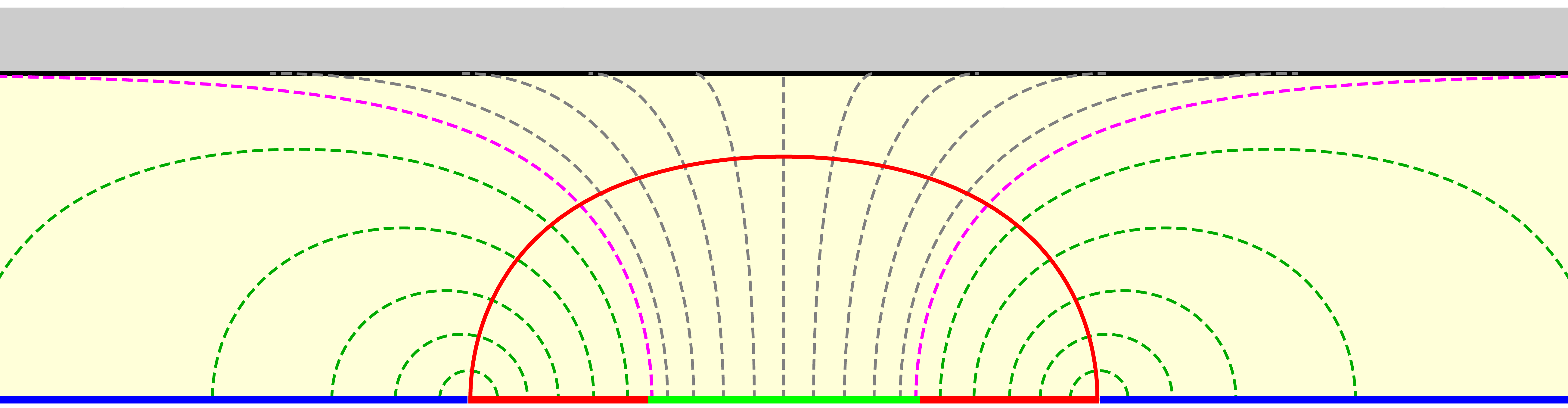}
% \end{center}
\vspace{.2cm}
\caption{
Geodesic bit threads in BTZ black brane geometry for an interval on the line.
The red solid curve is the RT curve (\ref{RT-curve-BTZ}).
The green segment corresponds to $A_\beta$ (see Sec.\,\ref{inversion-line-thermal}).
}
\label{figure-1int-BTZ}
\end{figure}

The BTZ black brane is a three dimensional gravitational background solving
the Einstein equations with negative cosmological constant and without matter term.
A horizon occurs and it is a two dimensional plane. 
This geometry can be described in terms of the coordinates $(t, x, \zeta)$,
where $t$ is the time coordinate and $\zeta >0$ is the holographic coordinate.
The boundary at $\zeta = 0$ is a $1+1$ dimensional Minkowski spacetime $\mathcal{M}$ in the coordinates $(x,t)$.
A constant time slice of the BTZ black brane geometry is equipped with the following induced metric
\be
\label{BTZ-metric}
ds^2 = \frac{1}{\zeta^2} \left( dx^2 + \frac{d\zeta^2}{1 -(\zeta/\zeta_h)^2} \right)
\;\;\;\;\qquad\;\;\;\;\;
\zeta >0 
\ee
where $\zeta_h$ denotes the position of the horizon.
This two dimensional geometry is considered in Fig.\,\ref{figure-1int-BTZ},
where the yellow region corresponds to $0 < \zeta < \zeta_h$
and the grey region to $\zeta > \zeta_h$.
The AdS/CFT correspondence relates $\zeta_h$
to the temperature $1/\beta$ of the dual 2D CFT 
in the Minkowski spacetime at the boundary of the BTZ black brane 
as follows
\be
\zeta_h = \frac{\beta}{2\pi} \,.
\ee

%\noindent
%$\bullet$ {\bf RT curve.}

Consider the bipartition of the spatial line of the dual CFT given by an interval,
which can be set to $A=(-b,b) \subset \mathbb{R}$ without loss of generality
(see the union of the red segments and of the green segment in Fig.\,\ref{figure-1int-BTZ},
while the union of the blue half lines is the complement of $A$ on the line), 
as already done in Sec.\,\ref{sec-AdS}.
In the geometry (\ref{BTZ-metric}),
the minimal length curve anchored to the endpoints of $A$ (the RT curve)
is made by the points $P_m$ 
(that must not be confused with the point indicated in the same way in Sec.\,\ref{sec-inversion-gs-lorentz})
whose coordinates $(x_m , \zeta_m)$ satisfy the following constraint
\be
\label{RT-curve-BTZ}
\zeta_m 
\,=\,
\sqrt{ \frac{\zeta_h^2 + \zeta_\ast^2}{2} - \frac{\zeta_h^2 - \zeta_\ast^2}{2}  \, \cosh(2x_m/\zeta_h)}
\,=\,
\zeta_h\,
\frac{\sqrt{\cosh(2b/\zeta_h) - \cosh(2x_m/\zeta_h)}}{\sqrt{2}\; \cosh(b/\zeta_h)}
\ee
where $\zeta_\ast$ denotes the maximum depth of this geodesic,
which can be expressed in terms of $\zeta_h$ and $b$ as follows
\be
\label{z-star-def}
\zeta_\ast = \zeta_h  \tanh(b/\zeta_h) \,.
\ee
In Fig.\,\ref{figure-1int-BTZ}, the RT curve (\ref{RT-curve-BTZ}) corresponds 
to the red solid curve  in the yellow region. 
This RT curve does not coincide with the one 
discussed in the last paragraph of Sec.\,\ref{sec-thermal-inv-euclid}.
%corresponding to $\theta=\pi/2$.
%
These two curves overlap in the zero temperature limit, as remarked in Sec.\,\ref{sec-AdS}.

%\noindent
%$\bullet$ {\bf Geodesic bit threads.}

The geodesic bit threads for the BTZ black brane geometry (\ref{BTZ-metric}) 
and the RT curve (\ref{RT-curve-BTZ}) have been found in \cite{Agon:2018lwq}.
In the following we report some of their results,
trying to keep their notation wherever it is possible.  

Some geodesic bit threads (dashed lines) are shown in Fig.\,\ref{figure-1int-BTZ} 
(see also the left panel of Fig.\,4 in \cite{Agon:2018lwq}).
A crucial feature of these geodesic bit threads is that
those ones
whose intersection with the RT curve (\ref{RT-curve-BTZ})
has $x_m$ such that $x_{m,\textrm{\tiny max}}   < |x_m| < b$ 
with
\be
\label{x-max-BTZ}
x_{m,\textrm{\tiny max}} \equiv \zeta_h \, \textrm{arccoth}\big(\zeta_h^2/\zeta_\ast^2\big)
\ee
have both the endpoints on the boundary at $\zeta = 0$;
hence they connect a point $A$ to a point in its complement $B$ on the line
(green dashed curves in Fig.\,\ref{figure-1int-BTZ}).
Instead, the geodesic bit threads intersecting the RT curve (\ref{RT-curve-BTZ}) in a point whose 
coordinate $x_m$ satisfies $|x_m| \leqslant x_{m,\textrm{\tiny max}}$ reach the horizon;
hence they connect a point in $A$ to a point on the horizon
(grey dashed curves in Fig.\,\ref{figure-1int-BTZ}).

A convenient parameterisation for the geodesic bit threads connecting $A$ to its complement on the real axis is
\be
\label{geo-bt-BTZ-para-1}
\zeta\,=\,\sqrt{C_1 + C_2 \, \cosh\!\big[2(x-x_m)/\zeta_h\big] + C_3 \, \sinh\!\big[2(x-x_m)/\zeta_h\big] }
\ee
where the constants $C_1$, $C_2$ and $C_3$ are defined respectively as 
\be
\label{C-123-def}
C_1 \equiv \frac{(\zeta_h^2 + \zeta_m^2)(\zeta_h^2 - \zeta_m^2) +  \zeta_h^2 \, \zeta_m^2\, s_m^2}{2\, (\zeta_h^2 - \zeta_m^2)}
\;\qquad\;
C_2 \equiv - \frac{(\zeta_h^2 - \zeta_m^2)^2 + \zeta_h^2 \, \zeta_m^2\, s_m^2}{2\, (\zeta_h^2 - \zeta_m^2)}
\;\qquad\;
C_3 \equiv \zeta_h \, \zeta_m\, s_m
\ee
being $s_m$ the slope of the outward-pointing normal vector at a given point $(x_m, \zeta_m)$ of the RT curve (\ref{RT-curve-BTZ}),
that reads
\be
\label{s_m-def}
s_m 
\,\equiv\, 
\frac{\zeta_m}{\zeta_h}\, \coth(x_m / \zeta_h)
\,=\, 
\pm \,\frac{\zeta_m}{\zeta_h} \, \sqrt{\frac{\zeta_h^2 - \zeta_m^2}{\zeta_\ast^2 -\zeta_m^2}} \,.
\ee

\noindent
As for the geodesic bit threads connecting a point in $A$ to a point of the horizon,
they can be written also as 
\be
\label{geo-bit-BTZ-par2}
x\,=\, x_m + \sigma\, \zeta_m\, \log\left(
\frac{\sqrt{\zeta_h^2 - \zeta^2} + \sqrt{D_1^2 +\zeta_m^2 - \zeta^2 } }{ \sqrt{\zeta_h^2 - \zeta_m^2} -\sigma\, D_1 }
\,\right)
\ee
being the constants $D_1 $ and $\sigma$ defined as follows
\be
D_1  \equiv \frac{ \zeta_h \, \zeta_m\, s_m}{\sqrt{\zeta_h^2 - \zeta_m^2}}
\;\;\;\;\qquad\;\;\;\;
\sigma\in \{+1, -1 \}
\ee
where $\sigma$ and $s_m$ must have opposite signs. 
Both the parameterisations 
(\ref{geo-bt-BTZ-para-1}) and (\ref{geo-bit-BTZ-par2}) can be adopted
to describe each one of the two different types of geodesic bit threads,
by employing two branches if necessary.

%\noindent
%$\bullet$ {\bf Comments to Fig.\,\ref{figure-1int-BTZ}.}

In Fig.\,\ref{figure-1int-BTZ} we show some geodesic bit threads
(dashed lines in the yellow region):
the green dashed lines connect a point in $A$ to a point in $B$, 
hence both these points are on the boundary;
instead, the grey ones connect a point in $A$ to a point on the horizon. 
The magenta dashed curves denote the geodesic bit threads 
corresponding to the transition between these two different behaviours,
hence they intersect the horizon (or the boundary) at infinity.

The thermal sub-interval $A_\beta \subset A$ introduced in Sec.\,\ref{inversion-line-thermal},
whose endpoints are (\ref{a-beta-def}) and  (\ref{b-beta-def}),
can be found also through the geodesic bit threads  in the BTZ black brane geometry 
described above.
Consider the transition geodesic bit thread 
given by (\ref{geo-bit-BTZ-par2}) with $\sigma = -1$ (hence $s_m >0$) 
and having $x_m = x_{m,\textrm{\tiny max}} $  in (\ref{x-max-BTZ}),
which corresponds to the magenta dashed curve on the right in Fig.\,\ref{figure-1int-BTZ}.
Setting $\zeta = 0$  in the resulting expression, 
we recover $b_\beta $ in (\ref{b-beta-def}) in the special case of $a= -\,b$.
Because of the symmetry of $A$,
the magenta dashed curve on the left in Fig.\,\ref{figure-1int-BTZ}
provides $a_\beta $ in (\ref{a-beta-def}) when $a= -\,b$.
Thus, the green segment on the boundary in Fig.\,\ref{figure-1int-BTZ}, 
obtained holographically through the geodesic bit threads corresponding to the magenta dashed curves,
can be identified  with $A_\beta$, 
which is also the green segment on the real line in Fig.\,\ref{figure-inversion-Euclid-1int-thermal}
and the segment on the real axis providing the green diamond in Fig.\ref{figure-inv-thermal-2-world}.

%\noindent
%$\Longrightarrow$ {\bf [I-B]. A new relation to interpret}

Plugging the expression for $\zeta_\ast$ in (\ref{z-star-def}) into (\ref{x-max-BTZ}) 
and comparing the result with $b_\beta$ in (\ref{b-beta-def}) specialised to the case $a=-b$,
we observe that
\be
x_{m,\textrm{\tiny max}}  \,=\,  b_{\frac{\beta}{2}} \,.
\ee
This interesting holographic relation deserves some interpretation.

%\noindent
%$\bullet$ {\bf Our results: [II]}

Also the inversion map (\ref{xinv-th-sym}) can be interpret through the geodesic bit threads.

\begin{figure}[t!]
\vspace{-1.5cm}
\hspace{.3cm}
\begin{center}
\includegraphics[width=.7\textwidth]{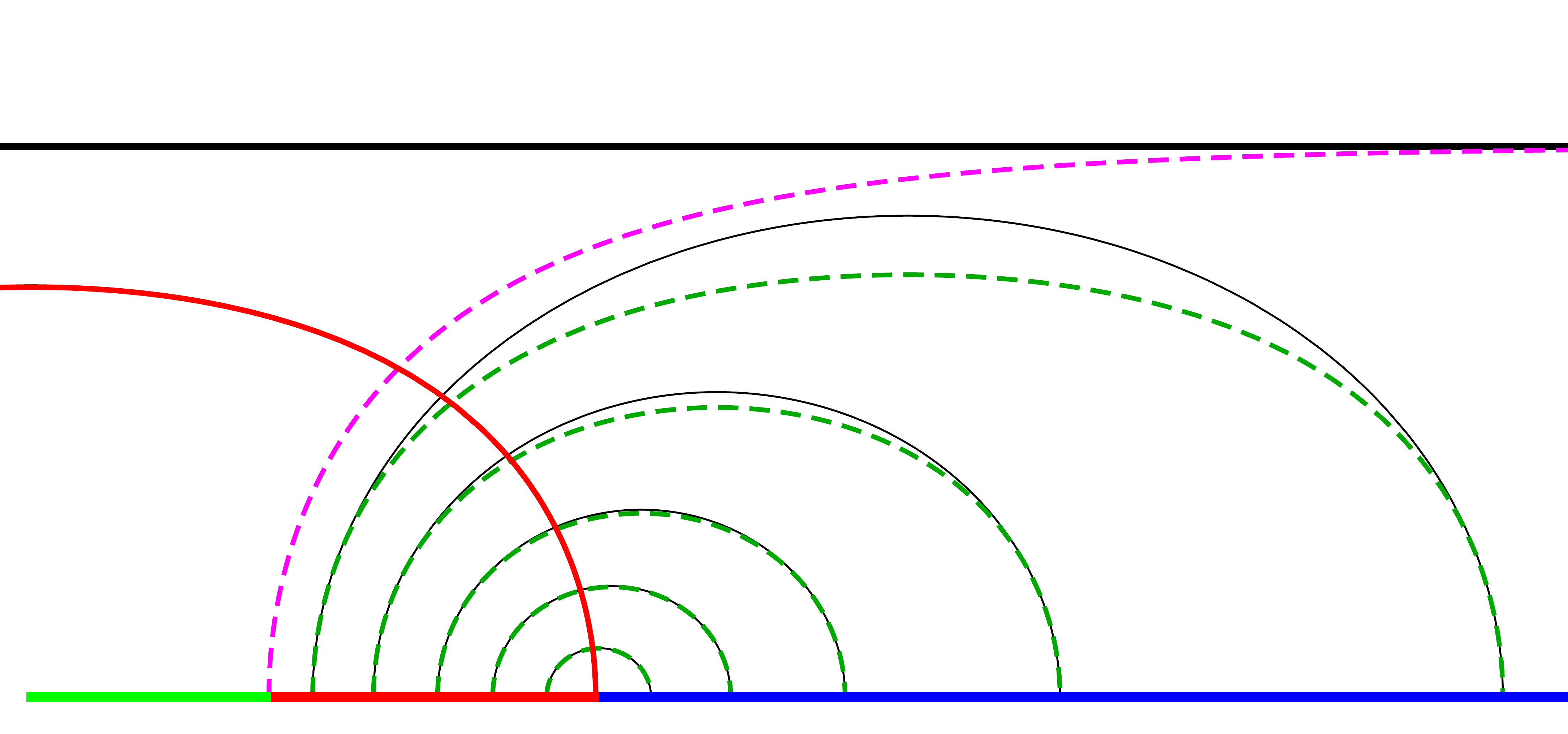}
 \end{center}
\vspace{-.2cm}
\caption{Comparison between
geodesic bit threads (green dashed curves) 
connecting a point $x_A \in A\setminus A_\beta$ to a point $x_B \in B$
(see also Fig.\,\ref{figure-1int-BTZ})
and the corresponding lines (starting at the same $x_A$'s)
for the modular evolution in the Euclidean spacetime described in Sec.\,\ref{sec-thermal-inv-euclid}
(black solid curves, see also Fig.\,\ref{figure-inversion-Euclid-1int-thermal}).
}
\label{figure-1int-BTZ-half-out}
\end{figure}

Consider first a point on the boundary at $x_A \in A\setminus A_\beta$,
which is therefore connected through a geodesic bit thread described by (\ref{geo-bt-BTZ-para-1})
to another point on the boundary at some $x_B \in B$.
In order to find $x_B$ in terms of $x_A$, 
notice that (\ref{geo-bt-BTZ-para-1}) vanishes for $x=x_\pm$ given by
\be
\label{x-pm-def-BTZ}
x_\pm
\,=\,
x_m
+ 
\frac{\zeta_h}{2}\,
\log\!\bigg(
\frac{C_1 \pm \sqrt{C_1^2 - C_2^2 + C_3^2} }{-\, C_2 - C_3}
\,\bigg)
\ee
where, from (\ref{C-123-def}), we have
\be
\label{x-pm-C123-combs}
- C_2 - C_3 = \frac{\big(\zeta_h^2 -\zeta_h\, \zeta_m\, s_m -\zeta_m^2\big)^2}{2\,(\zeta_h^2 - \zeta_m^2)}
\;\;\qquad\;\;
C_1^2 - C_2^2 + C_3^2 
=  \frac{\zeta_h^2\,\zeta_m^2\big[(1+s_m^2)\,\zeta_h^2 -\zeta_m^2\big]}{\zeta_h^2 - \zeta_m^2} \,.
%= \zeta_h^2\, \zeta_{\textrm{\tiny max}}^2
\ee
%where $\zeta_{\textrm{\tiny max}}$ has been introduced in  (\ref{zeta-max-def}).
Since $\zeta_h > \zeta_m$, 
the constant $C_1$ and both the expressions in (\ref{x-pm-C123-combs}) are positive quantities;
hence one concludes that $x_+ > x_-$, from (\ref{x-pm-def-BTZ}).
The symmetric choice of $A$ with respect to the origin allows us 
to assume $x_A > b_\beta >0$ without loss of generality;
hence $x_A = x_-$ and $x_B = x_+$ (see Fig.\,\ref{figure-1int-BTZ-half-out}).

To express $x_m$ in terms of $x_A = x_-$, 
%\footnote{The expression for $x_+$ in (\ref{x-pm-def-BTZ}) leads to the same result.} 
first employ (\ref{s_m-def})  and then (\ref{RT-curve-BTZ}) into $x_-$ given by (\ref{x-pm-def-BTZ});
and finally invert the resulting expression.
This gives different solutions, 
but the one providing (\ref{xm-from-xA-AdS}) when $\zeta_h \to +\infty$ reads
\be
\label{xm-from-xA-BTZ-out}
x_m
=
\frac{\zeta_h}{2}\,
\log\!\bigg(
\frac{
\e^{2x_A / \zeta_h} \big(\e^{4b / \zeta_h}  + 1\big) + 2\, \e^{2b / \zeta_h} - 2\,  \e^{(b+x_A) / \zeta_h} \big(\e^{2b / \zeta_h}  + 1\big)  
}{
2\, \e^{2(x_A + b) / \zeta_h} + \e^{4b / \zeta_h}  + 1 - 2\,  \e^{(b+x_A) / \zeta_h} \big(\e^{2b / \zeta_h}  + 1\big)  
}
\bigg) \,.
\ee
Now $x_B$ can be written in terms of $x_A$ by first using (\ref{s_m-def})  and (\ref{RT-curve-BTZ})
 into $x_+=x_B$ given by (\ref{x-pm-def-BTZ}),
 and then employing (\ref{xm-from-xA-BTZ-out}) in the resulting formula. 
After some algebra, we obtain 
\be
\label{x_B-from-x_A-BTZ-out}
\e^{2x_B / \zeta_h} \,
=
\bigg(\frac{\e^{x_A/\zeta_h} \cosh(b/\zeta_h) - 1}{\e^{x_A/\zeta_h} - \cosh(b/\zeta_h)} \bigg)^2
\ee
where the ratio within the round brackets
in the r.h.s. is positive when $x_A\in A\setminus A_\beta$.
Thus,  the CFT result (\ref{xinv-th-sym}) in $A\setminus A_\beta$ has been obtained 
holographically through the geodesic bit threads. 
As consistency check, notice that
the limit  $\zeta_h \to +\infty$ 
of the expression for $x_B$ obtained from (\ref{x_B-from-x_A-BTZ-out}) when $x_A\in A\setminus A_\beta$
leads to (\ref{x_B-from-x_A-AdS}), as expected. 

\begin{figure}[t!]
\vspace{-1cm}
\hspace{.3cm}
\begin{center}
\includegraphics[width=.7\textwidth]{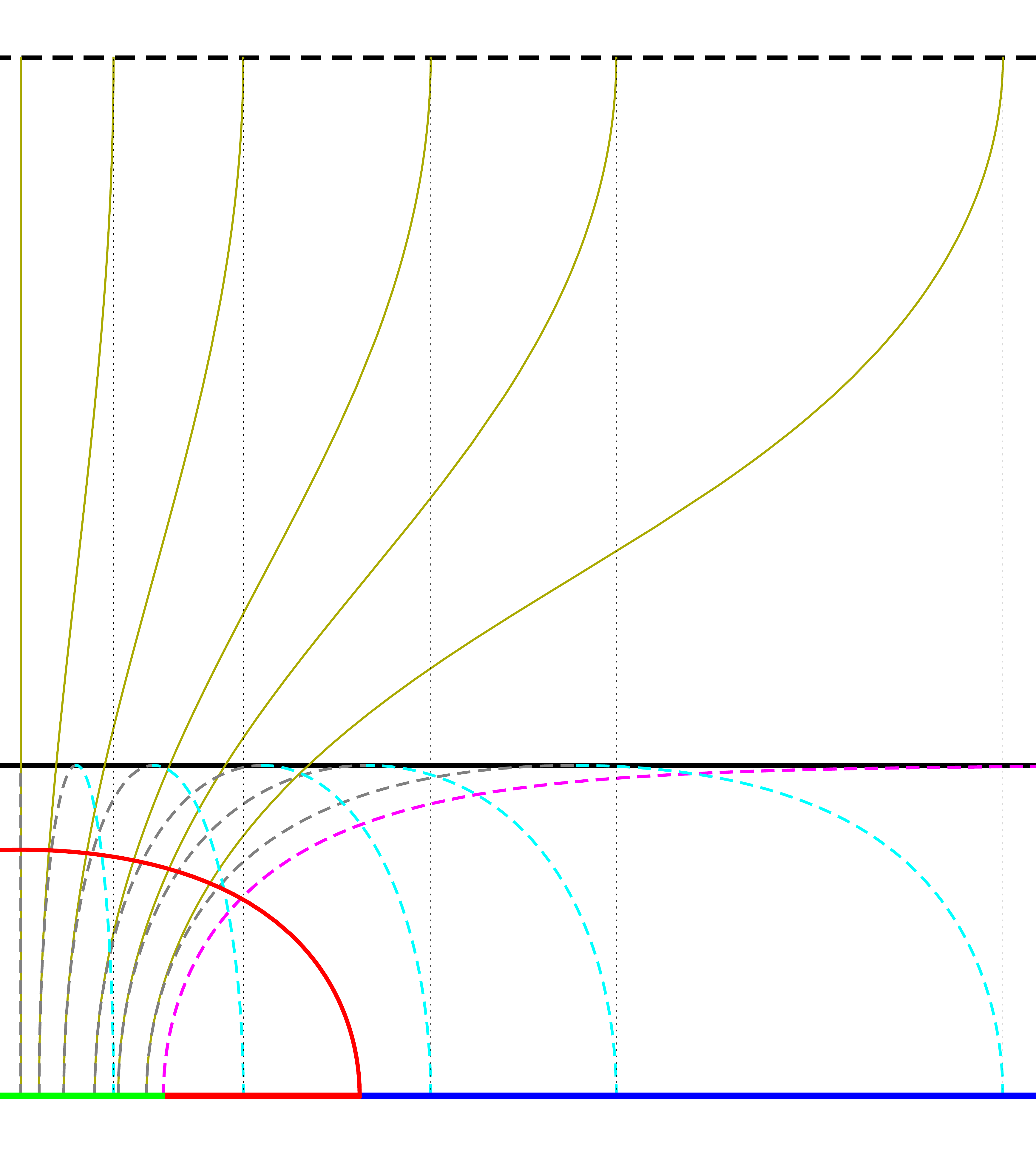}
 \end{center}
\vspace{-.2cm}
\caption{
Comparison between 
geodesic bit threads (grey dashed curves) 
connecting a point $x_A \in A_\beta$ to a point on the horizon
(see also Fig.\,\ref{figure-1int-BTZ})
and the corresponding lines (starting at the same $x_A$'s)
of the modular evolution in the Euclidean spacetime 
(yellow solid curves, see also Fig.\,\ref{figure-inversion-Euclid-1int-thermal}).
}
\label{figure-1int-BTZ-half}
\end{figure}

In Fig.\,\ref{figure-1int-BTZ-half-out},
considering only the positive half real axis without loss of generality, 
we show some geodesic bit threads having both the endpoints on the boundary (green dashed curves).
%as done also in Fig.\,\ref{figure-1int-BTZ}.
These bit threads are compared with
the corresponding curves (black solid lines)
starting at the same values of $x_A \in A\setminus A_\beta$
obtained from the CFT expression (\ref{z-inv-1int-beta}) for the same value of $\beta$,
which have been shown also in Fig.\,\ref{figure-inversion-Euclid-1int-thermal}.
We remark that these two sets of curves belong to two different Euclidean manifolds. 
These geodesic bit threads share the same endpoints with the corresponding CFT lines,
as expected from the analytic result discussed above,
but they do not coincide for the other points in general. 
There is no reason to expect such overlap.
However, the overlap approximatively occurs for the geodesics bit threads close to the boundary,
which connect points close to the entangling point $b$. 
Indeed, since these geodesics bit threads 
are weakly influenced by the occurrence of the horizon,
they become approximatively half circles, i.e. geodesics in Poincar\'e AdS$_3$.

The inversion map (\ref{xinv-th-sym}) obtained from CFT
can be recovered holographically through the geodesic bit threads also when $x_A \in A_\beta$.
The geodesic bit thread starting at this point on the boundary 
reaches the horizon at some point $(x_h , \zeta_h)$
(grey dashed curves in Fig.\,\ref{figure-1int-BTZ} and Fig.\,\ref{figure-1int-BTZ-half}).
We observe that this bit thread is tangent to the horizon;
indeed, from (\ref{geo-bit-BTZ-par2}) one obtains 
$ \partial_\zeta x = -\,\sigma\, \zeta_h \zeta  \,\big[\big(\zeta_h^2 -\zeta^2\big) \big(D_1^2 + \zeta_m^2 -\zeta^2\big)\big]^{-1/2} $,
which diverges as $\zeta \to \zeta_h$.
Such observation suggests us to merge this geodesic bit thread 
with the geodesic starting at $(x_h , \zeta_h)$ tangentially to the horizon and ending at some point $(\tilde{x}, 0)$ on the boundary
(cyan dashed curves in Fig.\,\ref{figure-1int-BTZ-half}).
The latter geodesic can be obtained by reflecting the geodesic bit thread starting at $(x_A, 0)$
with respect to the vertical axis at $x=x_h$.
This construction implies that $\tilde{x} = x_A +2(x_h-x_A)$.

The same procedure described above to obtain $x_B$ from $x_A \in A \setminus A_\beta$
can be employed to find $\tilde{x}$ in terms of $x_A \in A_\beta$
(in particular, (\ref{xm-from-xA-BTZ-out}) holds also in this case),
obtaining (\ref{x_B-from-x_A-BTZ-out}) with $x_B$ replaced by $\tilde{x}$.
The crucial difference is that the ratio in the r.h.s. of (\ref{x_B-from-x_A-BTZ-out}) is negative when $x_A \in A_\beta$;
hence the proper sign must be introduced taking the square root of (\ref{x_B-from-x_A-BTZ-out}).
This leads to $\tilde{x} = \textrm{Re}[z_{\textrm{\tiny inv}}(x_A)]$,
where $z_{\textrm{\tiny inv}}(x_A)$ is given in (\ref{xinv-th-sym}) with $x_A \in A_\beta$.
We remind that $\tilde{x}(x_A)$ is a bijective function from $A_\beta$ onto $\mathbb{R}$.

In Fig.\,\ref{figure-1int-BTZ-half},
considering only  the positive half real axis without loss of generality, 
we display some geodesic bit threads that start at various  $x_A \in A_\beta$,
and therefore reach the horizon (grey dashed curves).
%as done also in Fig.\,\ref{figure-inversion-Euclid-1int-thermal}.
%
Each geodesic bit thread of this kind 
is merged smoothly at the horizon with the corresponding geodesic (cyan dashed curves) 
that connects the horizon to the point $(\tilde{x},0)$ on the boundary, 
where $\tilde{x} = \textrm{Re}[z_{\textrm{\tiny inv}}(x_A)]$,
as discussed above. 
The curves obtained from these geodesics are compared with the corresponding 
CFT curves described in Sec.\,\ref{sec-thermal-inv-euclid}
(yellow solid curves, shown also in Fig.\,\ref{figure-inversion-Euclid-1int-thermal}),
that start at the same values of $x_A \in A_\beta$,
grow in the strip $\{ z\in \mathbb{C} \,; |\textrm{Im}(z)| \leqslant \beta/2 \}$
according to (\ref{z-inv-1int-beta}) with $\theta \in (0,\pi)$
and end at $ \textrm{Re}[z_{\textrm{\tiny inv}}(x_A)] +\ri \frac{\beta}{2}$ when $\theta = \pi$.
The relation $\tilde{x} = \textrm{Re}[z_{\textrm{\tiny inv}}(x_A)]$ 
is highlighted by the vertical dotted lines in Fig.\,\ref{figure-1int-BTZ-half}.

%\newpage
%%%%%%%%%%%%%%%%%%%%%%%%%%%%%%%%%%%%%%%%%%%%%%%%
\section{Two disjoint intervals: Massless Dirac field, ground state}
\label{sec-line-two-interval}
%%%%%%%%%%%%%%%%%%%%%%%%%%%%%%%%%%%%%%%%%%%%%%%%

In this section we consider the massless Dirac fermion in its ground state
on the line bipartite by the union $A =A_1 \cup A_2$
of two disjoint intervals $A_1 = (a_1, b_1)$ and $A_2=(a_2, b_2)$.

\subsection{Internal modular evolution}
\label{subsec-2int-internal}

The modular Hamiltonian of two disjoint intervals
for a CFT in its ground state and on the line depends on the model.
For the massless Dirac field, 
this modular Hamiltonian and the corresponding modular flow for the Dirac field
have been described in \cite{Casini:2009vk,Longo:2009mn}.
Casini and Huerta \cite{Casini:2009vk} 
found that this operator can be written as $K_A  = K_A^{\textrm{\tiny loc}} + K_A^{\textrm{\tiny bi-loc}} $,
where 
\be
\label{K_A-2int-terms-local}
K_A^{\textrm{\tiny loc}} 
=
2\pi 
\int_A 
\beta_{\textrm{\tiny loc}}(x) \, T_{tt}(t,x)\, \rd x 
\ee
is a local operator determined by the energy density  $T_{tt}(t,x)$.
For the massless Dirac field, the energy density reads\footnote{See the decomposition in Eqs.\,(2.22)-(2.23) of \cite{Mintchev:2020uom}.}
\be
\label{T00-lambda-def}
T_{tt}(t,x) 
\,\equiv\,
\,\frac{\textrm{i}}{2}
:\! \!
\Big[ \Big (
\psi^\ast_1\, (\partial_x \psi_1) 
-
(\partial_x \psi^\ast_1)\, \psi_1 
\Big )(x+t)
- 
\Big(
\psi^\ast_2\, (\partial_x \psi_2)
-
(\partial_x \psi^\ast_2)\,  \psi_2 
\Big) (x-t)
\Big]\!\! : 
\ee
in terms of the chiral components $\psi_1$ and $\psi_2$ of the massless Dirac field.
The bi-local term $K_A^{\textrm{\tiny bi-loc}}$ is 
\be
\label{K_A-2int-terms-bi-local}
K_A^{\textrm{\tiny bi-loc}} 
=
2\pi 
\int_A 
\beta_{\textrm{\tiny bi-loc}}(x) \, T_{\textrm{\tiny bi-loc}}(t,x, x_{\textrm{\tiny c}} ) \, \rd x 
\ee
where the bi-local operator $T_{\textrm{\tiny bi-loc}}(t,x, y) $ is 
the following quadratic operator
\bea
\label{T-bilocal-2int}
T_{\textrm{\tiny bi-loc}}(t,x, y) 
 &\equiv &
\frac{\textrm{i}}{2}\;
\bigg\{ \,
\!:\!\!\Big[\, \psi^\ast_1(x+t) \,  \psi_1(y+t) - \psi^\ast_1(y+t) \,  \psi_1(x+t) \, \Big]\!\!: 
\nonumber
\\
& & \hspace{.8cm}
+ \,
\!:\!\! \Big[\,  \psi^\ast_2(x-t) \,  \psi_2(y-t) - \psi^\ast_2(y-t) \,  \psi_2(x-t) \, \Big] \!\!: \!
\bigg\} \,.
\eea

The weight functions in (\ref{K_A-2int-terms-local}) and (\ref{K_A-2int-terms-bi-local}) can be written as 
\be
\label{betas-2int-gen}
\beta_{\textrm{\tiny loc}}(x)  = \frac{1}{w'(x)} 
\;\;\qquad\;\;
\beta_{\textrm{\tiny bi-loc}}(x)  = \frac{\beta_{\textrm{\tiny loc}}(x_{\textrm{\tiny c}}(x))}{x - x_{\textrm{\tiny c}}(x)} 
\ee
where 
\be
\label{w-function-2int}
w(x) \,=\,  
\log \!\left [ - \frac{(x-a_1)(x-a_2)}{(x-b_1)(x- b_2)} \right ] 
\ee
which is the crucial function of our analysis throughout this section.
This function naturally leads to introduce the point $x_{\textrm{\tiny c}}(x)$ conjugate to $x\in A$ 
through the condition $w(x_{\textrm{\tiny c}}(x))=w(x)$.
The explicit expression for $x_{\textrm{\tiny c}}(x)$ is \cite{Casini:2009vk}
\be
\label{x-conjugate-2int}
x_{\textrm{\tiny c}}(x)
\, \equiv \,
\frac{(b_1 b_2 - a_1 a_2)\, x +(b_1 + b_2)\, a_1 a_2 - (a_1 + a_2)\, b_1 b_2}{(b_1 - a_1 + b_2 - a_2)\, x + a_1 a_2 - b_1 b_2 }
\,=\,
x_0 - \frac{R_0^2}{x - x_0}
\ee
where 
%\cite{Eisler:2022rnp}
\be
x_0 \equiv  \frac{b_1 b_2 - a_1 a_2}{b_1 - a_1 + b_2 - a_2}
\;\;\;\qquad\;\;\;
R_0^2 \equiv 
\frac{(b_1 - a_1)(b_2 - a_2)(b_2 - a_1)(a_2 - b_1)}{(b_1 - a_1 + b_2 - a_2)^2} \,.
\ee
Notice that a point $x$ and its conjugate point $x_{\textrm{\tiny c}}(x) $ belong to different intervals in $A$.
Furthermore, $x_{\textrm{\tiny c}}(x)$ in (\ref{x-conjugate-2int}) is invariant under a cyclic permutation of the sequence of endpoints  $(a_1, b_1, a_2, b_2)$.
It is sometimes convenient to write these expressions in terms of 
the lengths of the intervals $\ell_j \equiv b_j - a_j$ with $j \in \{1,2\}$
and of the distance $d \equiv a_2 - b_1$ separating them.

The function providing the geometric action of the modular group of automorphisms 
induced by the vacuum reads  \cite{Casini:2009vk,Longo:2009mn, Mintchev:2020uom}
\be
\label{xi-2int-gen-def}
\xi(\tau, x) \,\equiv\, \Theta_{A_1}(x) \, \xi_{-} (\tau, x) + \Theta_{A_2}(x) \, \xi_{+} (\tau, x)
\ee
where $\Theta_{A_j}(x)$ is the characteristic function of $A_j$ with $j \in \{1,2\}$, and  
\be
\label{xi-pm-2int}
\xi_\pm (\tau, x) \,\equiv\, w^{-1}_\pm  \big( w(x) + 2\pi \tau \big)
\ee
satisfying $\xi(0, x) = x$ for any $x\in A$.
The functions in (\ref{xi-pm-2int}) are
written in terms of the inverse functions $w^{-1}_\pm$ of (\ref{w-function-2int}), which are
\be
\label{inverse-w-pm-2int}
w_\pm^{-1} (x) 
\,=\,
\frac{a_1 + a_2 + (b_1 + b_2)\, e^{x}}{2(1+e^{x})}
\pm\,
\frac{\sqrt{ \big(a_1 + a_2 + (b_1 + b_2)\, e^{x} \big)^2 - 4 (1+e^{x})\,(a_1  a_2+b_1 b_2\,e^{x} ) }}{2 (1+e^{x} )}
\ee
where the expression under the square root is a quadratic polynomial in the variable $e^x$, 
whose discriminant is $-16(b_1 - a_1)(b_2 - a_2)(b_2 - a_1)(a_2 - b_1)$;
hence it is positive for any $x\in \RR$ when $a_1 <b_1 <a_2<b_2$.

The map sending  a generic configuration $A$ of two intervals on the line
into the symmetric configuration $A_{\textrm{\tiny sym}} \equiv (-b,-a) \cup (a,b)$, with $0<a<b$, 
reads
\be
\label{f-map-def}
f(x) \,\equiv\, 
\frac{a \big[ a_1 a_2 -2 a_1 b_2 +a_2 b_2 +(a_1-2a_2+b_2)\, x \big]  + b \,(b_2 -a_1)(x-a_2) }{
\, b \big[ a_1 a_2 -2 a_1 b_2 +a_2 b_2 +(a_1-2a_2+b_2)\, x \big] + a\,(b_2 -a_1)(x-a_2)}\;b
\ee
where 
\be
\frac{b}{a} =   
\frac{a_1 b_1 + a_2 b_2 + b_1 b_2 + a_1 a_2 - 2(a_1 b_2 +a_2 b_1)  + 2\, \sqrt{(b_1-a_1) (b_2-a_2) (a_2-a_1) (b_2-b_1) }}{(b_2-a_1)(a_2 - b_1)} \,.
\ee
Considering a point $x\in A$ and its conjugate $x_{\textrm{\tiny c}}(x)$ given by (\ref{x-conjugate-2int}), 
we observe that (\ref{f-map-def}) satisfies
\be
\label{f-f-x_c}
f(x) \, f(x_{\textrm{\tiny c}}(x)) \,=\, -\, a\, b \,.
\ee

Let us introduce the points $c_j \in A_j$ with $j \in \{1,2\}$ such that
\be
\label{f-conj}
f ( x_{\textrm{\tiny c}}(c_j) ) \,=\, -\, f(c_j) \,.
\ee 
Combining this definition with (\ref{f-f-x_c}), one finds that $f(c_j)^2 = a \,b $ for these two points. 
From (\ref{f-conj}), their explicit expressions are
\bea
\label{c1-point}
c_1 
&\equiv &
\frac{a_2 b_2 - a_1 b_1}{a_2 + b_2 - a_1 - b_1} - \frac{\sqrt{(a_2 - a_1)(a_2 - b_1)(b_2 - a_1)(b_2 - b_1)}}{a_2 + b_2 - a_1 - b_1}
\\
\label{c1-point-d-ell}
\rule{0pt}{.8cm}
& = &
a_1  + \frac{\sqrt{(\ell_1 + d)(\ell_1 + \ell_2 + d)}}{\ell_1 + \ell_2 + 2d}\,  \Big( \sqrt{(\ell_1+d)(\ell_1 + \ell_2 + d)} - \sqrt{(\ell_2 + d) d}\, \Big)
\eea
and
\bea
\label{c2-point}
c_2
&\equiv &
\frac{a_2 b_2 - a_1 b_1}{a_2 + b_2 - a_1 - b_1} + \frac{\sqrt{(a_2 - a_1)(a_2 - b_1)(b_2 - a_1)(b_2 - b_1)}}{a_2 + b_2 - a_1 - b_1}
\\
\label{c2-point-d-ell}
\rule{0pt}{.8cm}
& = &
a_2  + \frac{\sqrt{(\ell_1 + d) \,d}}{\ell_1 + \ell_2 + 2d}\,  \Big( \sqrt{(\ell_2+d)(\ell_1 + \ell_2 + d)} - \sqrt{(\ell_1 + d) d}\, \Big) \,.
\eea
We also notice that $x_{\textrm{\tiny c}} (c_i) = c_j$ with $i \neq j $ and that
\be
\label{w-c12}
w(c_1) = w(c_2) =\log\! \bigg( \frac{a_2- a_1}{b_2 - b_1} \bigg)\,.
\ee

In the case of the symmetric configuration $A_{\textrm{\tiny sym}}$, 
the function $f$ in (\ref{f-map-def}) is the identity map
and the expression (\ref{x-conjugate-2int}) for the conjugate point simplifies to $x_{\textrm{\tiny c}}(x) =-ab/x$.
In this case the condition (\ref{f-conj}) becomes $c_j^2 = ab$,  
which is the square of the geometric mean of $a$ and $b$;
hence $c_2 = \sqrt{ab} \in (a,b)$ and $c_1 = - \sqrt{ab} \in (-b,-a)$.

In the limit of large separation, i.e. when $d \to +\infty$,
from (\ref{c1-point-d-ell}) and (\ref{c2-point-d-ell}) we have that $c_j = a_j + \ell_j / 2 +O(1/d)$, with $j \in \{1,2\}$.
In the opposite limit of adjacent intervals, i.e. when $d \to 0$ and the two intervals become the single interval $(a_1, b_2)$,
from (\ref{c1-point-d-ell}) and (\ref{c2-point-d-ell}) one finds that both $c_1$ and $c_2$ converge to the point $a_2 = b_1$ inside the final single interval. 

\begin{figure}[t!]
\vspace{-.6cm}
\hspace{-.4cm}
\vspace{1.cm}
%\begin{center}
\includegraphics[width=1.05\textwidth]{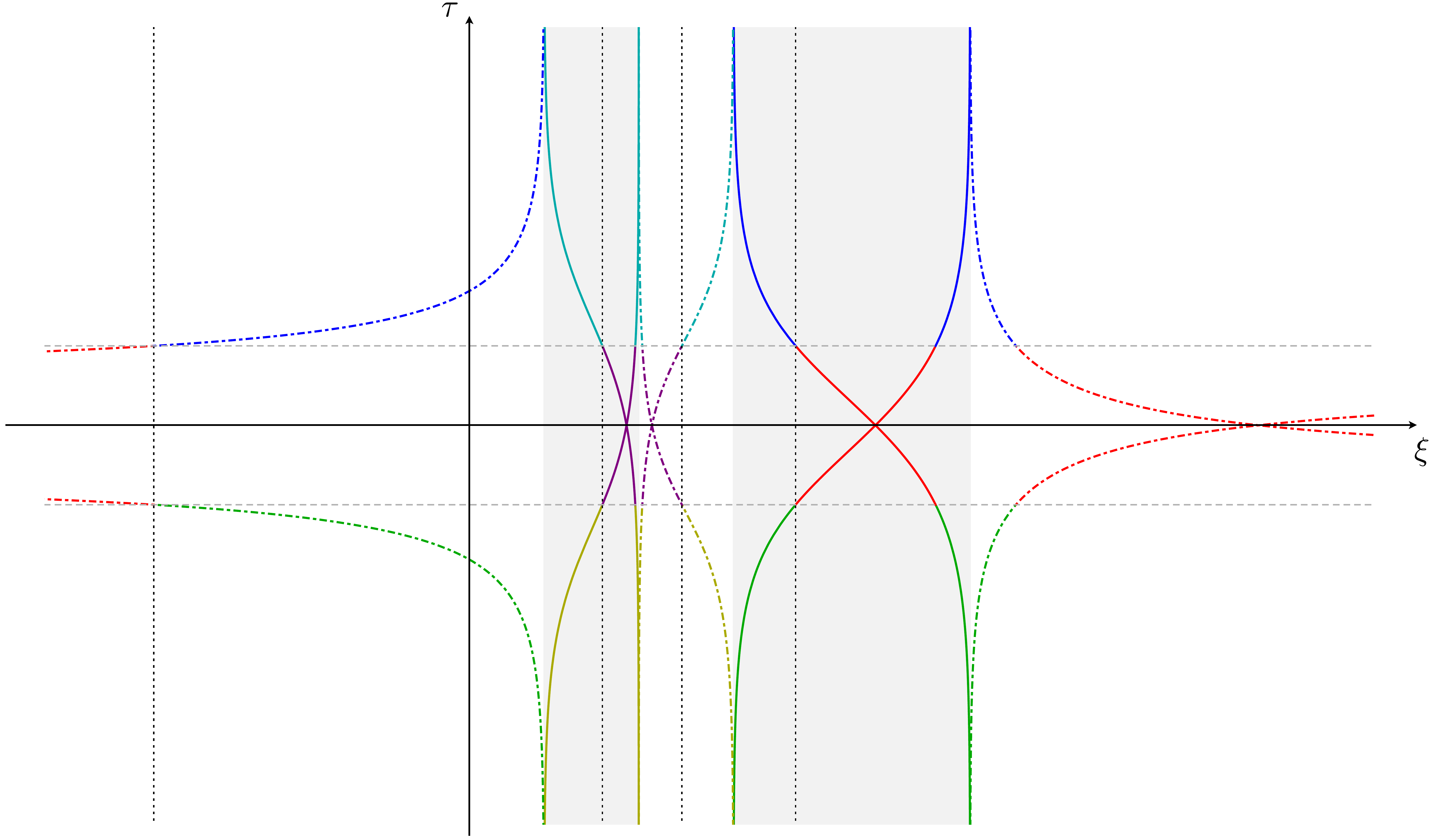}
\\
\vspace{-.4cm}
\hspace{-.5cm}
\includegraphics[width=1.05\textwidth]{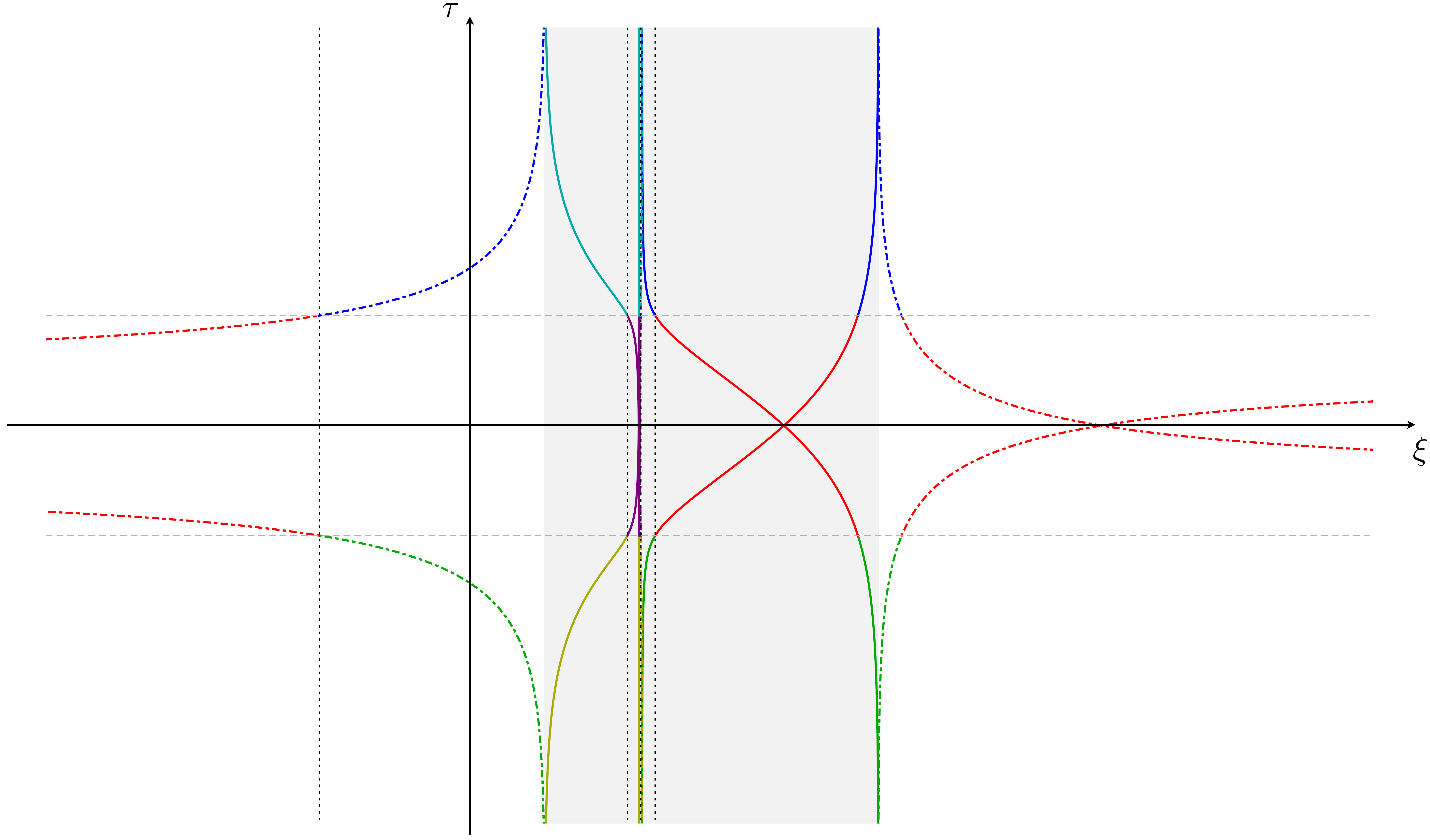}
% \end{center}
\vspace{.0cm}
\caption{The curves $\xi(\tau, x)$ and $\xi(-\tau, x)$ (solid lines) and their images under the inversion map (dot-dashed  lines)
in the plane $(\xi, \tau)$, obtained from (\ref{xi-2int-gen-def}) and (\ref{inv-map-2int-0}).
The vertical dotted lines in the grey strips correspond to (\ref{c1-point}) and (\ref{c2-point}).
%The horizontal dashed lines are obtained from (\ref{tau_c-2int}).
The bottom panel illustrates the limiting regime of adjacent intervals.
}
\label{figure-xi-2int}
\end{figure}

In the top panel of Fig.\,\ref{figure-xi-2int} we consider the plane 
$\RR^2$ parameterised by the coordinates $(\xi,\tau)$.
The subsystem $A= A_1 \cup A_2$ provides the two grey vertical strips where $\xi \in A$,
which contain the solid curves corresponding to $\xi(\tau,x)$ and $\xi(-\tau,x)$,  
obtained from (\ref{xi-2int-gen-def}) for a generic point $x\in A$ 
and its conjugate $x_{\textrm{\tiny c}}(x)$ at $\tau=0$.
The points (\ref{c1-point}) and (\ref{c2-point}) give the dotted vertical lines in these two grey strips.
For any choice of $x\in A$ at $\tau=0$, the intersection of these vertical lines with the solid curves
is characterised by the condition $\xi(\tau_c , x) = c_j$ with $j\in \{1,2\}$,
which identifies the following value $\tau_c$ of the modular parameter
\be
\label{tau_c-2int}
\tau_c(x) \equiv \frac{w(c_j) - w(x)}{2\pi}
\;\;\qquad\;\;
j\in \{1,2\}
\ee
where $w(c_j)$ is given by (\ref{w-c12}).
From (\ref{w-c12}) and (\ref{tau_c-2int}),  
we have that both $c_1$ and $c_2$ lead to the same value for $\tau_c(x)$ for any $x\in A$.
This value naturally  induces a partition of the solid curves obtained from (\ref{xi-2int-gen-def})
into three arcs corresponding to the partition 
$\RR = (-\infty, -|\tau_c| ) \cup (-|\tau_c|, |\tau_c|) \cup (|\tau_c|, + \infty)$
for the domain of $\tau$,
which are indicated through different colours in Fig.\,\ref{figure-xi-2int}.
The bottom panel of Fig.\,\ref{figure-xi-2int} 
displays the limit of adjacent intervals,
identifying the arcs that remain non trivial 
and the ones that collapse either to vertical segments 
or to vertical half lines 
in this limit.

\begin{figure}[t!]
\vspace{.5cm}
\hspace{-.6cm}
%\begin{center}
\includegraphics[width=1.1\textwidth]{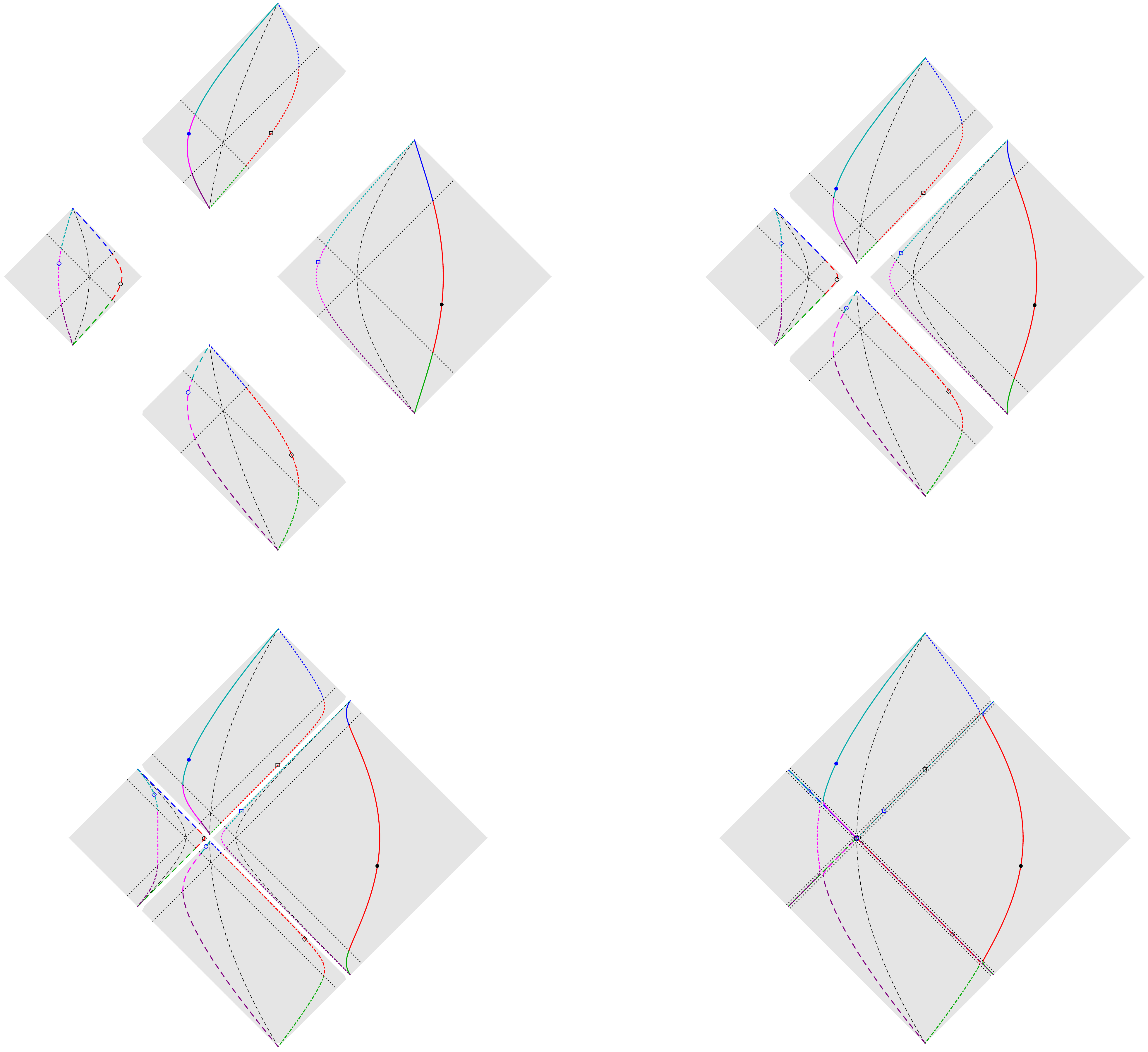}
% \end{center}
\vspace{.5cm}
\caption{ 
%\textcolor{red}{\bf [scrivere una volta definito il main text. Toppe ai rettangoli]}
Spacetime domain $\mathcal{D}_{A}$
for the union of two disjoint intervals $A_1$ and $A_2$ on the line,
whose lengths are $\ell_1=\ell$ and $\ell_2=2\ell$ respectively.
The distance $d$ between the intervals changes in the different panels, while $\ell$ is kept fixed:
$d/\ell = 1$ (top left),  $d/\ell = 1/5$ (top right),  
$d/\ell = 1/20$ (bottom left) and  $d/\ell = 1/1000$ (bottom right).
The modular trajectories corresponding to following two choices of the initial point $P$ are displayed: 
the black filled circle in $\mathcal{D}_{2}$ and the blue filled circle in $\mathcal{D}_{12}^+$.
The empty symbols having the same colour denote the conjugate point $P_{\textrm{\tiny c}}$ (empty circle) and the auxiliary points
$P_{\textrm{\tiny c},\pm}$ (empty square and empty rhombus) obtained from $P$.
}
\label{figure-diamond-2int}
\end{figure}

In the Minkowski spacetime parameterised by the coordinates $(x,t)$,
we have to consider the bipartition given by the union of two disjoint intervals $A_1$ and $A_2$
and its complement along each light ray direction (\ref{lc1}).
This leads to the spacetime domain
$\mathcal{D}_A \equiv  \mathcal{D}_1 \cup \mathcal{D}_2 \cup \mathcal{D}_{12}^{+} \cup \mathcal{D}_{12}^{-} $,
made by the union of the four grey regions 
shown in each the panels of Fig.\,\ref{figure-diamond-2int}:
$\mathcal{D}_j$ are the diamonds corresponding to $A_j$, with $j \in \{1,2\}$,
and the two regions $\mathcal{D}_{12}^{\pm}$, which are symmetric with respect to the $x$-axis,
are made by points of the spacetime where a light ray from $A_1$ and a light ray from $A_2$ intersect,
travelling either in the future for $\mathcal{D}_{12}^{+} $ or in the past for $\mathcal{D}_{12}^{-} $.
%
%\textcolor{red}{$\bullet$ \bf [stress that we have to consider this $\mathcal{D}_A$ because we are putting together two different chiralities]}

A modular trajectory in $\mathcal{D}_A$
can be obtained by using (\ref{xi-2int-gen-def}) for each light ray coordinate.
This leads to the following spacetime coordinates for 
the generic point of a modular trajectory
\be
\label{x-t-trajectory-2int}
x(\tau) =
%\sqrt{2}\; 
\frac{\xi(\tau , p_{0,+}) + \xi(-\tau , p_{0,-}) }{2}
\;\;\qquad\;\;
t(\tau) =
%\sqrt{2}\; 
\frac{\xi(\tau , p_{0,+}) - \xi(-\tau , p_{0,-}) }{2}
\;\;\qquad\;\;
\tau \in \RR
\ee
where $(u_+, u_-)=(p_{0,+}, p_{0,-})$ are the light ray coordinates of the initial point at $\tau=0$.

Denote by $P=(u_{+,0} , u_{-,0})$ the light ray coordinates of the initial point of the modular evolution at $\tau=0$,
which must be both different from  (\ref{c1-point}) and (\ref{c2-point}) to explore the most generic case.
The light ray coordinates of the point $P_{\textrm{\tiny c}}$ conjugate to $P$
are $( x_{\textrm{\tiny c}} (u_{+,0}) , x_{\textrm{\tiny c}} (u_{-,0}) )$.
In Fig.\,\ref{figure-diamond-2int}, two examples for $P$ are shown:
$P\in \mathcal{D}_{2}$ (black filled circle) and  $P\in \mathcal{D}_{12}^{+}$ (blue filled circle).
The corresponding $P_{\textrm{\tiny c}}$'s are indicated with empty circles of the same colour. 
In order to obtain the expected modular trajectories 
for the single interval in the limit of adjacent intervals, 
beside the initial point $P$ and its conjugate point $P_{\textrm{\tiny c}}$, 
we introduce also the two auxiliary points 
$P_{\textrm{\tiny c},-} \equiv ( u_{+,0} , x_{\textrm{\tiny c}} (u_{-,0}))$ 
and $P_{\textrm{\tiny c},+} \equiv ( x_{\textrm{\tiny c}} (u_{+,0}) , u_{-,0} )$,
 denoted by empty squares and empty rhombi in Fig.\,\ref{figure-diamond-2int}.

The occurrence of the four points,
which are the initial point $P$ and its associated points $P_{\textrm{\tiny c}}$, $P_{\textrm{\tiny c},+}$ and $P_{\textrm{\tiny c},-}$,
comes from the combination of the two different chiralities. 
Each region among
$\mathcal{D}_1$, $\mathcal{D}_2$, $\mathcal{D}_{12}^{+}$ and $\mathcal{D}_{12}^{-}$ 
contains one and only one of these points. 
Choosing one of these four points at $\tau=0$ 
and plugging its coordinates into (\ref{x-t-trajectory-2int}),
a modular trajectory is obtained in the corresponding region as $\tau \in \RR$.
Thus, we can associate four disjoint  modular trajectories in $\mathcal{D}_A$ to any point $P \in \mathcal{D}_A$, 
which are in one-to-one correspondence with the four regions $\mathcal{D}_1$, $\mathcal{D}_2$, $\mathcal{D}_{12}^{+}$ and $\mathcal{D}_{12}^{-}$.

In Fig.\,\ref{figure-diamond-2int} we show $\mathcal{D}_A$ for two disjoint intervals having assigned lengths,
while the various panels correspond to different values of the distance separating them. 
The two choices for $P$ mentioned above are performed and 
each panel displays the four curves corresponding to each $P$.
For a given $P$, the solid curve, the dashed curve, 
the dotted and the dot-dashed  curves
pass through $P$,  $P_{\textrm{\tiny c}}$ and $P_{\textrm{\tiny c},\pm}$
respectively.

We find it instructive to consider also the four points 
$C_1$, $C_2$, $C_{12}^{+}$ and $C_{12}^{-}$
whose light ray coordinates  are obtained by combining (\ref{c1-point}) and (\ref{c2-point}) in all the four possible ways.
These points correspond to the intersections of the black dotted segments 
in each panel of Fig.\,\ref{figure-diamond-2int}
and are in one-to-one correspondence with the regions 
$\mathcal{D}_1$, $\mathcal{D}_2$, $\mathcal{D}_{12}^{+}$ and $\mathcal{D}_{12}^{-}$;
hence they are denoted in the same way. 
The null rays departing from each one of these four points 
%$C_1$, $C_2$, $C_{12}^{+}$ and $C_{12}^{-}$
(black dotted segments in Fig.\,\ref{figure-diamond-2int})
partition the corresponding grey region  into four subregions. 
As a consequence, each one of the four points $C_1$, $C_2$, $C_{12}^{+}$ and $C_{12}^{-}$
induces a natural partition of any modular trajectory belonging to the same region,
which can be found by plugging its coordinates into (\ref{tau_c-2int}) first and then
partitioning the range $\RR$ of the modular parameter accordingly. 
In Fig.\,\ref{figure-diamond-2int}, the components of the partition of each curve are 
indicated through different colours.
The dashed black curves in Fig.\,\ref{figure-diamond-2int}
correspond to the modular trajectories passing through the points $C_1$, $C_2$, $C_{12}^{+}$ and $C_{12}^{-}$.

In Fig.\,\ref{figure-diamond-2int},  the sequence of panels shows the domain
$\mathcal{D}_{A}$ of two disjoint intervals $A_1$ and $A_2$ with given lengths $\ell_1$ and $\ell_2$ 
for four decreasing values of the distance $d$ separating them.
This sequence illustrates how the four modular trajectories corresponding to a generic initial point $P$
are deformed by the limiting procedure
and finally provide the single modular trajectory in the final diamond
of a single interval of length $\ell_1 + \ell_2$, 
obtained when $A_1$  and $A_2$ are adjacent and therefore share one endpoint.
Various arcs of the different modular trajectories do not contribute to the final 
modular trajectory in the diamond of the adjacent intervals because they collapse to segments along null rays. 
In particular,  taking $d \to 0$ in (\ref{c1-point}) and (\ref{c2-point}), 
one finds that all the points $C_1$, $C_2$, $C_{12}^{+}$ and $C_{12}^{-}$ collapse to the shared endpoint.
This implies that, considering the partition of each one of the four regions 
$\mathcal{D}_1$, $\mathcal{D}_2$, $\mathcal{D}_{12}^{+}$ and $\mathcal{D}_{12}^{-}$ identified by these points, 
only one component of the partition of each one of these four regions 
is non vanishing after the adjacent intervals limit.
As for the four modular trajectories contained in $\mathcal{D}_{A}$,
%$D_1$, $\mathcal{D}_2$, $\mathcal{D}_{12}^{+}$ and $\mathcal{D}_{12}^{-}$,
since each of them is partitioned accordingly to the partition 
induced by either $C_1$ or $C_2$ or $C_{12}^{+}$ or $C_{12}^{-}$ of the region it belongs to,
at most one component of their partition contributes 
to the final modular trajectory in the diamond of the adjacent intervals. 
The other arcs collapse on the light rays of the final shared endpoint,
as shown in the bottom panels of Fig.\,\ref{figure-diamond-2int}.
Notice that, beside the modular trajectory passing through the initial point $P$, 
only two other modular trajectories among the ones passing through  $P_{\textrm{\tiny c}}$ or $P_{\textrm{\tiny c},\pm}$
contribute to the final modular trajectory after the limit $d\to 0$.
%which passes through $P$ in the diamond of the final interval obtained when the intervals $A_1$ and $A_2$ are adjacent
It is instructive to compare the bottom right panel of Fig.\,\ref{figure-diamond-2int} 
with the right panel of Fig.\,\ref{figure-xi-double-cone}.
%

%\newpage
\subsection{Geometric action in Euclidean spacetime}
\label{sec-2int-inv-Eucid}

In order to explore the geometric action of the modular conjugation,
we adapt to this case the analyses presented in Sec.\,\ref{sec-line-single-interval} and Sec.\,\ref{sec-thermal}.
This leads us first to write an inversion map $z_{\textrm{\tiny inv}}(x)$ on the real axis at $t=0$ 
for $A= A_1\cup A_2$ and its complement $B$ on the line,
and then to its extension to an inversion map on $\mathcal{D}_A$.

\begin{figure}[t!]
\vspace{-1cm}
\hspace{-.9cm}
%\begin{center}
\includegraphics[width=1.1\textwidth]{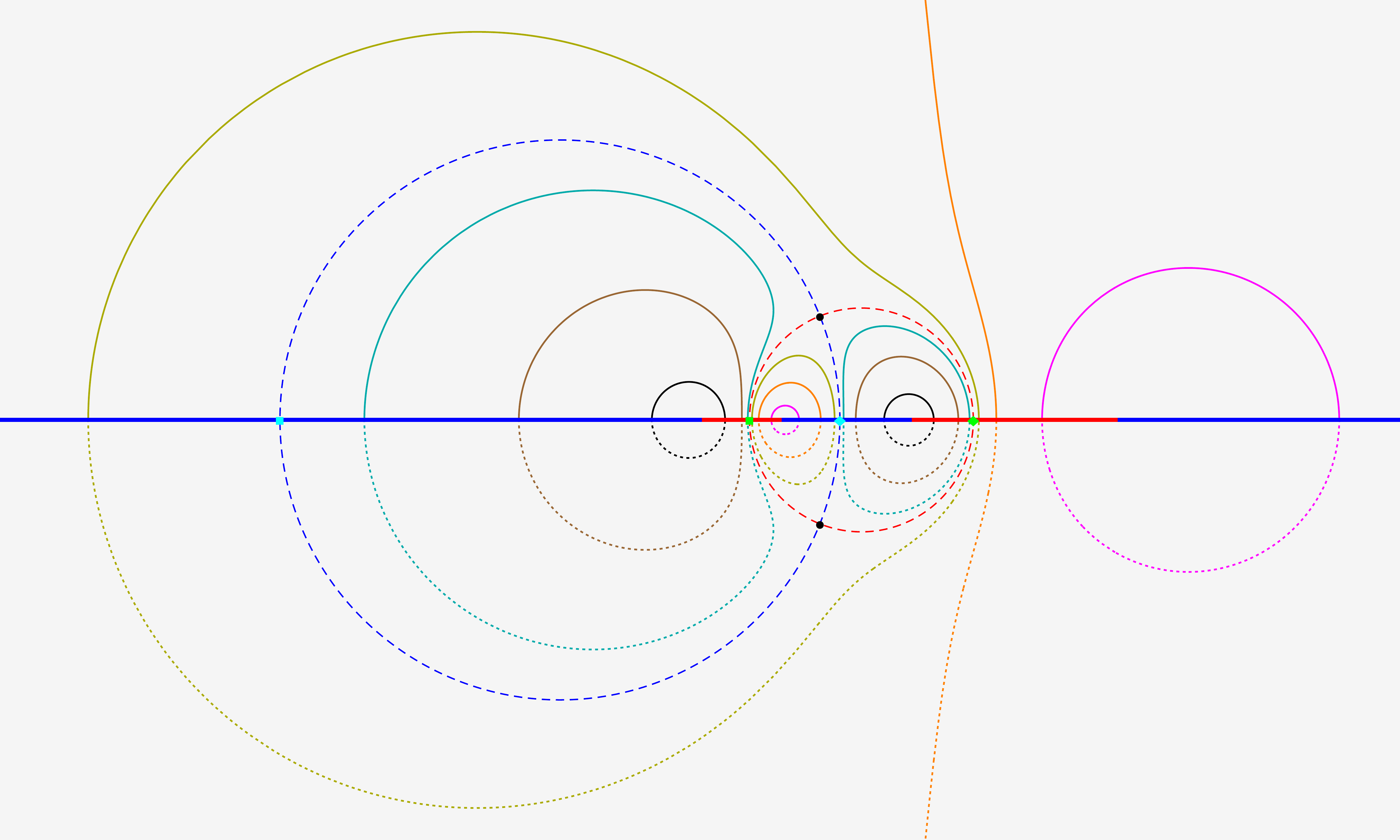}
% \end{center}
\vspace{-.4cm}
\caption{
Euclidean modular evolution for the union $A=A_1 \cup A_2$ of two disjoint intervals (red segments),
from (\ref{z-pm-2int}) with either $\theta \in (0,\pi)$ (solid lines) or  $\theta \in (\pi,2\pi)$ (dot-dashed  lines),
which lead to the inversion map (\ref{z-inv-full-lin-2int}) on the line.
Curves having the same colour correspond to conjugates points in $A$.
}
\label{figure-inversion-Euclid-2int}
\end{figure}

An inversion map for $A= A_1\cup A_2$ and its complement $B$ on the line 
can be constructed by exploiting the analysis in the Euclidean spacetime carried out in \cite{Cardy:2016fqc}.
In the complex plane parameterised by the complex variable $z$,
the union of two disjoint intervals is identified through the endpoints of $A_1$ and $A_2$ 
(the entangling points) on the real axis.
In Fig.\,\ref{figure-inversion-Euclid-2int}, the subsystem $A$ corresponds to 
the union of the two red segments 
and its complement $B$ to the union of the blue ones;
hence the four entangling points separate the red and blue segments, partitioning the real line. 
From the function $w(x)$ in (\ref{w-function-2int}) and its inverse functions (\ref{inverse-w-pm-2int}),
one constructs the following geometric action in the imaginary time \cite{Cardy:2016fqc}
\be
\label{z-pm-2int}
z_\pm(\theta, x) \equiv w^{-1}_\pm \big( w(x) + \textrm{i} \theta \big)
\;\;\qquad\;\;
\theta \in [0,2\pi)
\;\;\qquad\;\;
x\in A
\ee
which can be also obtained by replacing $2\pi \tau$ with $\textrm{i} \theta $ in (\ref{xi-pm-2int}).

For any $x\in A$, the complex maps (\ref{z-pm-2int}) allow to find a closed curve in the complex plane 
which intersects only once the real axis in $B$, when $\theta = \pi$ in (\ref{z-pm-2int}).
Similarly to the cases studied in Sec.\,\ref{sec-line-single-interval} and Sec.\,\ref{sec-thermal},
this observation provides a natural way to construct an inversion map $z_{\textrm{\tiny inv}, A} (x) $ 
sending $A$ to its complement $B$ on the real axis at $t=0$.
We require that a point close to an entangling point in $A$ is mapped by this inversion 
to a point close to the same entangling point in $B$, and viceversa. 
In Fig.\,\ref{figure-inversion-Euclid-2int} we show the curves obtained from (\ref{z-pm-2int})
for some $x\in A$ and for their conjugate points $x_{\textrm{\tiny c}}(x)$ given by (\ref{x-conjugate-2int})
(the curves with the same colour intersect $A$ in $x$ and $x_{\textrm{\tiny c}}(x)$),
as done in  Fig.\,9 of \cite{Cardy:2016fqc} for the special case of $A_{\textrm{\tiny sym}}$.

In order to write $z_{\textrm{\tiny inv},A} (x) $, 
let us consider the points (\ref{c1-point}) and (\ref{c2-point})
(green points in Fig.\,\ref{figure-inversion-Euclid-2int}).
As observed in \cite{Cardy:2016fqc},
these points can be found from the points in the complex plane where $w'(z)=0$,
which are
\be
\label{z0-pm-def}
z_{0, \pm}
\equiv
c_0 \pm \,\textrm{i}\,\frac{\sqrt{(b_1 - a_1)(a_2 - b_1)(b_2 - a_1)(b_2 - a_2)}}{b_1 + b_2 - a_1 - a_2} 
\;\;\qquad\;\;
c_0 \equiv \frac{b_1 b_2 -a_1 a_2}{b_1 + b_2 - a_1 - a_2} \,.
\ee
The points $z_{0, \pm}$ (black dots in Fig.\,\ref{figure-inversion-Euclid-2int})
are the intersections of the circles $C_a$ and $C_b$ 
whose centers are on the real axis,  in $(a_1+a_2)/2$ and $(b_1+b_2)/2$ respectively.
%They are denoted by the black dots in Fig.\,\ref{figure-inversion-Euclid-2int}.
%%
In $z_{0, \pm}$, the straight lines tangent to $C_a$ and $C_b$ provide two supplementary angles
whose bisectors allow to identify in a unique way two circles among the ones passing through $z_{0, \pm}$.
One of these two circles (red dashed circle in Fig.\,\ref{figure-inversion-Euclid-2int})
intersects the real axis at $c_1$ and $c_2$ given by (\ref{c1-point}) and (\ref{c2-point})
(green points),
while the other one (blue dashed circle)
intersects the real axis at $x_{\textrm{\tiny c}}(c_1)$ and $x_{\textrm{\tiny c}}(c_2)$
(cyan points).

In \cite{Cardy:2016fqc} it has been argued  that
the points (\ref{z0-pm-def}) induce the occurrence of two critical values for $\theta\in (0,2\pi)$ whose sum is $2\pi$.
One of these angles is $ \theta_0 \equiv \big| \textrm{Im} [ w(z_{0, \pm})] \big| $; hence
\be
\big| \tan \theta_0\, \big| 
\,=\,
\bigg| \frac{\textrm{Im}\big[e^{w(z_{0,\pm})} \big] }{ \textrm{Re}\big[e^{w(z_{0,\pm})} \big]} \bigg|
\,=\,
 \frac{2 \sqrt{(b_1 - a_1)(b_2 - a_2)(a_2 - b_1)(b_2 - a_1)}}{\big| 2(a_1 a_2 + b_1 b_2) - (a_1 + a_2)(b_1 + b_2)\big|} \,.
\ee

From (\ref{z-pm-2int}),  (\ref{c1-point}) and (\ref{c2-point})
we construct the following inversion map on the line bipartite by 
the union of two disjoint intervals
\be
\label{inv-map-2int-0}
\begin{array}{lll}
%x\in A_1 
%& \hspace{1.5cm}&
\displaystyle
z_{\textrm{\tiny inv},A} (x) 
% = x_{\textrm{\tiny inv}} (x, t=0) 
\equiv
\left\{\begin{array}{ll}
z_-(\pi,x) \hspace{1cm}& x < c_1 
\\
\rule{0pt}{.5cm}
z_+(\pi,x) & x > c_1
\end{array}\right.
& \hspace{1.5cm}&
x\in A_1 
\\
\rule{0pt}{1.1cm}
\displaystyle
z_{\textrm{\tiny inv},A} (x) 
% = x_{\textrm{\tiny inv}} (x, t=0) 
\equiv
\left\{\begin{array}{ll}
z_+(\pi,x) \hspace{1cm}& x < c_2 
\\
\rule{0pt}{.5cm}
z_-(\pi,x) & x > c_2
\end{array}\right.
& \hspace{1.5cm}&
x\in A_2
\end{array}
\ee
which is one-to-one from $A$ to $B$ and 
maps points close to an entangling point into points close to the same entangling point.

At this stage let us introduce
\be
\label{xi-2int-gen-invert}
\xi(\tau, x) \,\equiv\, \Theta_{B_1}(x) \, \xi_{-} (-\tau, x) + \Theta_{B_2}(x) \, \xi_{+} (-\tau, x)
\;\;\qquad\;\;
x\in B_1 \cup B_2
\ee
which has been employed to draw the dot-dashed  curves in Fig.\,\ref{figure-xi-2int},
obtained  from (\ref{xi-2int-gen-invert}) with $x$ given by the image 
under (\ref{inv-map-2int-0}) of the initial point in $A$ 
providing the solid curves in the grey strips through (\ref{xi-2int-gen-def}).

In order to construct an inversion map $z_{\textrm{\tiny inv}} (x) $ on the entire line, 
we have to describe also the one-to-one map inversion map 
$ z_{\textrm{\tiny inv},B} (x):B \to A$ as the inverse of (\ref{inv-map-2int-0}).
This can be done by employing again the curves defined through (\ref{z-pm-2int}), 
but now in the range  $\theta \in (\pi, 2\pi)$.
This leads to the curves in the lower half plane
shown in Fig.\,\ref{figure-inversion-Euclid-2int}  (dot-dashed  lines).
Since the complement $B=B_1 \cup B_2 \in \RR$ of $A$ on the line 
is the union of  $B_1 \equiv (b_1, a_2)$ and $B_2
\equiv (b_2, +\infty) \cup (-\infty, a_1)$,
we find that $z_{\textrm{\tiny inv},B}(x)$ reads
\be
\label{inv-map-2int-0-BA}
\begin{array}{lll}
%x\in A_1 
%& \hspace{1.5cm}&
\displaystyle
z_{\textrm{\tiny inv},B} (x) 
% = x_{\textrm{\tiny inv}} (x, t=0) 
\equiv
\left\{\begin{array}{ll}
z_-(\pi,x) \hspace{1cm}& x < d_1 
\\
\rule{0pt}{.5cm}
z_+(\pi,x) & x > d_1
\end{array}\right.
& \hspace{1.5cm}&
x\in B_1 
\\
\rule{0pt}{1.1cm}
\displaystyle
z_{\textrm{\tiny inv},B} (x) 
% = x_{\textrm{\tiny inv}} (x, t=0) 
\equiv
\left\{\begin{array}{ll}
z_-(\pi,x) \hspace{1cm}& x < d_2 
\\
\rule{0pt}{.5cm}
z_+(\pi,x) & x > d_2
\end{array}\right.
& \hspace{1.5cm}&
x\in B_2
\end{array}
\ee
where $d_j \in B_j$ with $j \in \{1,2\}$ are the points on the line
(cyan points in Fig.\,\ref{figure-inversion-Euclid-2int})
obtained from (\ref{c1-point}) and (\ref{c2-point})
modified through the cyclic shift of the ordered sequence of the endpoints 
$(a_1, b_1, a_2, b_2) \to (b_1, a_2, b_2,a_1)$.

%\noindent
%$\bullet$ {\bf Inversion on the full line}

Combining (\ref{inv-map-2int-0}) and (\ref{inv-map-2int-0-BA}),
we obtain the following inversion map on the line
\be
\label{z-inv-full-lin-2int}
z_{\textrm{\tiny inv}} (x) : \RR \to \RR
\hspace{2cm}
z_{\textrm{\tiny inv}} (x) 
% = x_{\textrm{\tiny inv}} (x, t=0) 
\equiv
\left\{\begin{array}{ll}
z_{\textrm{\tiny inv},A} (x)  \hspace{1cm}& x \in A 
\\
\rule{0pt}{.5cm}
z_{\textrm{\tiny inv},B} (x)  \hspace{1cm}& x \in B
\end{array}\right.
\ee
which is idempotent, as expected.

\newpage
\subsection{Geometric action in Minkowski spacetime}
\label{sec-2int-inv-lorentz}

The idempotent map (\ref{z-inv-full-lin-2int})
allows to introduce the following inversion map for  $\mathcal{D}_{A}$ 
\be
\label{x-t-inv-z-2int}
x_{\textrm{\tiny inv}} (x, t)
=
\frac{z_{\textrm{\tiny inv}}(x+t) + z_{\textrm{\tiny inv}}(x-t) }{2} 
\;\;\qquad\;\;
t_{\textrm{\tiny inv}} (x, t)
=
\frac{z_{\textrm{\tiny inv}}(x+t) - z_{\textrm{\tiny inv}}(x-t) }{2} 
\ee
which has the same structure of (\ref{x-t-inv-z-one-int}) and (\ref{x-t-inv-z-one-int-thermal}),
explored in the previous analyses.

%\noindent
%$\bullet$ {\bf Comments Fig.\,\ref{figure-scacchiera-2int}.}

\begin{figure}[t!]
\vspace{-.8cm}
\hspace{1.12cm}
\vspace{1.cm}
%\begin{center}
\includegraphics[width=.81\textwidth]{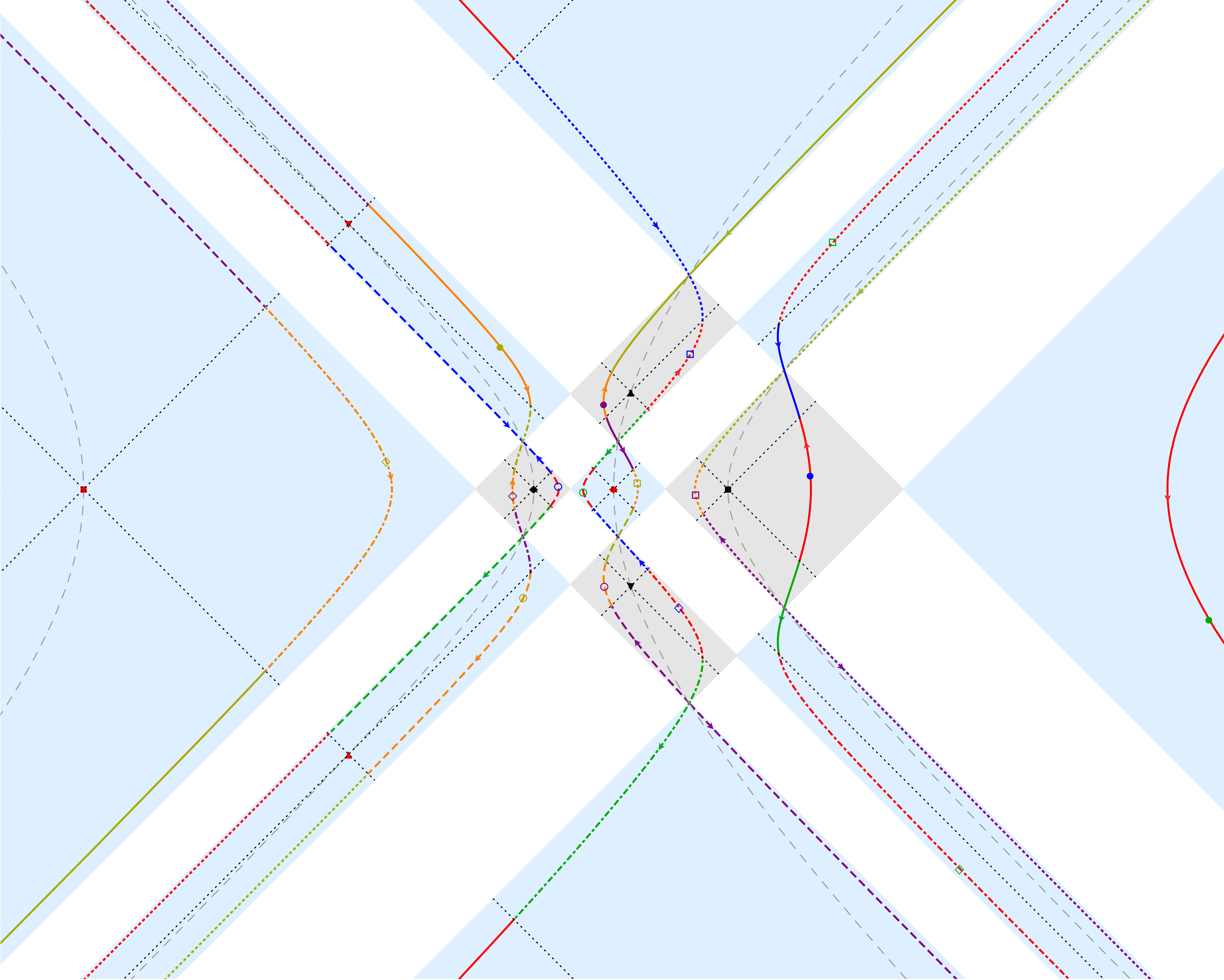}
\\
\vspace{.1cm}
\hspace{1.cm}
\includegraphics[width=.81\textwidth]{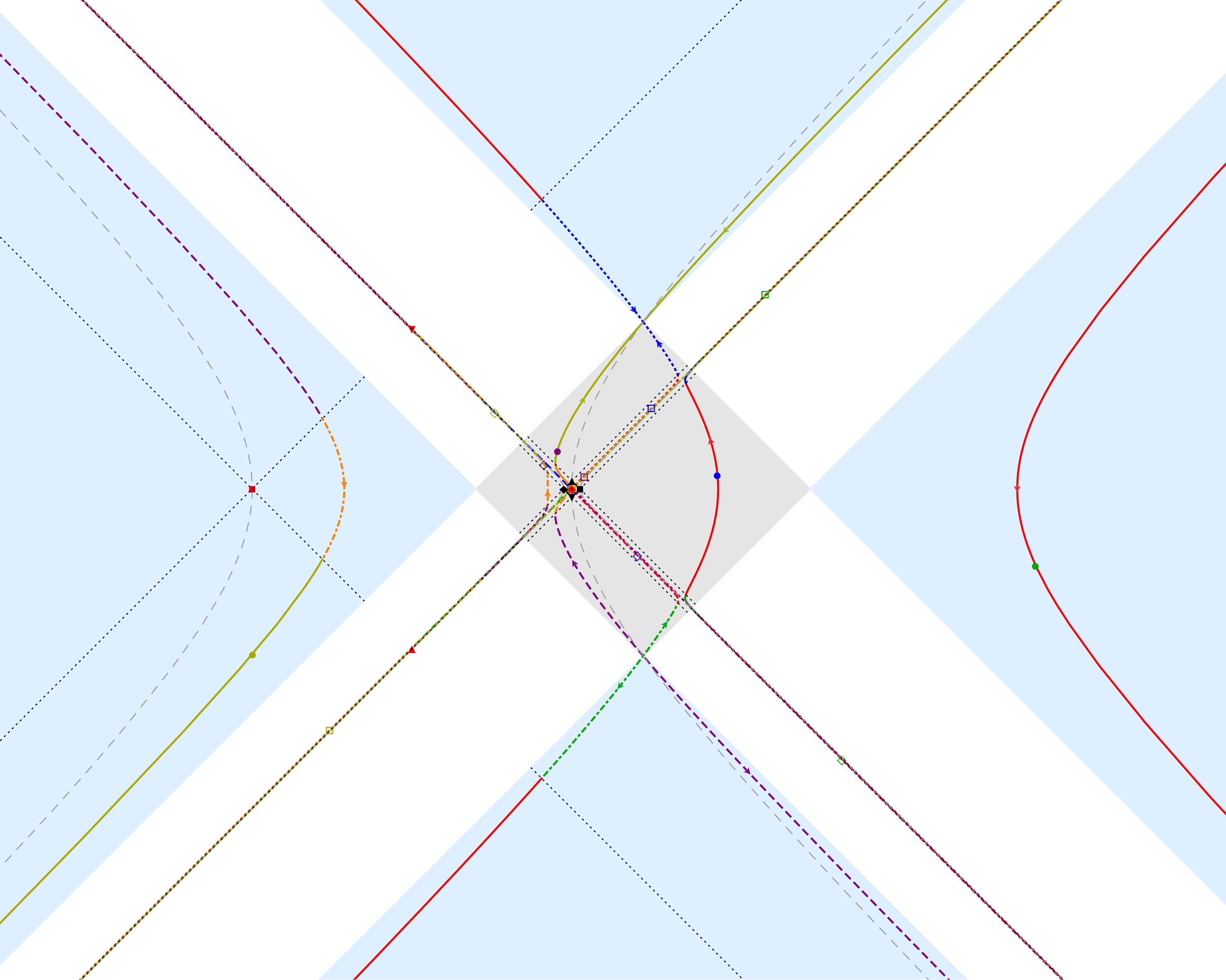}
% \end{center}
\vspace{.3cm}
\caption{
Modular trajectories in $\mathcal{D}_{A_1\cup A_2}$ 
and their images under the inversion map (\ref{x-t-inv-z-2int}).
The bottom panel shows the limiting regime of adjacent intervals
(see the right panel of Fig.\,\ref{figure-xi-double-cone}).
}
\label{figure-scacchiera-2int}
\end{figure}

In Fig.\,\ref{figure-scacchiera-2int},
the map (\ref{x-t-inv-z-2int}) sends $\mathcal{D}_{A} = \mathcal{D}_1 \cup \mathcal{D}_2 \cup \mathcal{D}_{12}^{+} \cup \mathcal{D}_{12}^{-}  $  
(grey region, like in Fig.\,\ref{figure-diamond-2int})
into $\mathcal{W}_{A}$ (light blue region), and viceversa.
Moreover, considering the vertices shared by $\mathcal{D}_{A}$ and $\mathcal{W}_{A}$,
this map sends a point close to a vertex in the grey region into a point close to the same vertex in the blue region,
and viceversa. 

The four points labelled by black filled markers in $\mathcal{D}_{A} $
are obtained by combining  (\ref{c1-point}) and (\ref{c2-point}) in all the four possible ways. 
Each one of these black points, 
which provides a natural partition of the rhombus it belongs to in four subregions
(as already discussed in Sec.\,\ref{subsec-2int-internal} for Fig.\,\ref{figure-diamond-2int})
 is mapped through the inversion (\ref{x-t-inv-z-2int})
 to the point in the light blue region labelled by the same marker coloured in red.
 Such red point induces a partition of the connected light blue region it belongs to
 in  four subregions through the null rays departing from it. 
Anyone of the four subregions partitioning one of the four grey regions 
$\mathcal{D}_1$, $\mathcal{D}_2$, $\mathcal{D}_{12}^{+}$ and $\mathcal{D}_{12}^{-}$
is mapped by (\ref{x-t-inv-z-2int}) into the subregion in the light blue domain that
shares a vertex with it.

In Fig.\,\ref{figure-scacchiera-2int} we display the modular trajectories corresponding to 
two initial points in $\mathcal{D}_{A}$ labelled by the blue and purple filled circles,
as also done in Fig.\,\ref{figure-diamond-2int} and by adopting the same conventions.
In this figure we also show the images of these modular trajectories under the inversion map (\ref{x-t-inv-z-2int}), 
that belong to $\mathcal{W}_{A}$.
The images of the blue (purple) points are the green (yellow) points, labelled by the same kind of marker. 
The points labelled by empty markers are obtained 
from the one corresponding to the filled circle having the same colour,
as discussed above for the points $P_{\textrm{\tiny c}}$ and $P_{\textrm{\tiny c},\pm}$ of Fig.\,\ref{figure-diamond-2int}.
In Fig.\,\ref{figure-scacchiera-2int} 
the image of each arc in $\mathcal{D}_{A}$ under (\ref{x-t-inv-z-2int})
is the arc in $\mathcal{W}_{A}$ having the same colour and the same kind of line. 
The modular parameter $\tau$ grows along the curves as indicated by the arrows;
hence the upper and lower vertices in 
$\mathcal{D}_1$, $\mathcal{D}_2$, $\mathcal{D}_{12}^{+}$ and $\mathcal{D}_{12}^{-}$
correspond to $\tau \to +\infty$ and $\tau \to -\infty$ respectively. 

In Fig.\,\ref{figure-scacchiera-2int} 
the black markers in $\mathcal{D}_A$ correspond to 
the four points $C_1$, $C_2$, $C_{12}^{+}$ and $C_{12}^{-}$
introduced in Sec.\,\ref{subsec-2int-internal}
and the red markers in $\mathcal{W}_A$
to their images under the inversion (\ref{x-t-inv-z-2int}).
The modular trajectories in $\mathcal{D}_A$
and their images under (\ref{x-t-inv-z-2int}) in $\mathcal{W}_A$
are denoted by the grey dashed curves in the figure
and we find it worth remarking that they do not display oscillations.

The bottom panel of Fig.\,\ref{figure-scacchiera-2int} considers the curves 
displayed in the top panel of the same figure in the limit of adjacent intervals,
showing that the right panel of Fig.\,\ref{figure-xi-double-cone} is recovered
in this limiting regime. 

We stress that the results described in this section hold
for the massless Dirac field.

\newpage
%%%%%%%%%%%%%%%%%%%%%%%%%%%%%%%%%%%%%%%%%%%
\section{Conclusions}
\label{sec_conclusions}

%\noindent
%$\bullet$ {\bf Results.}

In this manuscript we have studied a holographic relation 
between the geometric action of some modular conjugations in 2D CFT
and the geodesic bit threads considered in \cite{Freedman:2016zud, Agon:2018lwq}
in a constant time slice of the corresponding dual gravitational backgrounds. 
This comparison has been performed for three cases and 
in all of them the bipartition of the space is given by a single interval:
for a CFT in the ground state, either on the line (Sec.\,\ref{sec-line-single-interval} and Sec.\,\ref{sec-AdS})
or on the circle (Appendices\;\ref{sec_app_circle} and \ref{sec_app_global_AdS}),
and in the thermal state on the line (Sec.\,\ref{sec-thermal} and Sec.\,\ref{sec-BTZ}).

When the CFT is in the ground state and on the line bipartite by an interval, 
the geometric actions of the modular operator and of the modular conjugation
are described by the theorem of Hislop and Longo \cite{Haag:1992hx, Hislop:1981uh, Brunetti:1992zf}
and are given by (\ref{xi-interval-line-gs}) and (\ref{Haag-inversion-map}) respectively. 
Considering for simplicity and without loss of generality the interval $A=(-b,b)$ on the line,
the geometric action for the modular conjugation at $t=0$ simplifies to (\ref{xt-inv-centered-t=0}).
In the context of the gauge/gravity correspondence, 
this inversion relation is obtained between the endpoints of a geodesic bit thread
on a constant time slice of Poincar\'e AdS$_3$ (see Fig.\,\ref{figure-1int-AdS} and (\ref{x_B-from-x_A-AdS})).
The case of a CFT in its ground state and on the circle bipartite by an interval
is a straightforward generalisation whose geometric actions for the 
modular operator and the modular conjugation
are (\ref{xi-circle-gs-tau}) and (\ref{z-inv-circle-t=0}) respectively. 
The corresponding holographic dual description 
of the geometric action of the modular conjugation
is obtained through the geodesic bit threads in a constant time slice of global AdS$_3$
(see Fig.\,\ref{figure-1int-AdS-global}).

For a CFT in a thermal state at temperature $1/\beta$ and on the line bipartite by an interval, 
the geometric action of the modular conjugation is more insightful.
In each chiral sector, the inversion map (\ref{x-inv-th-def}) on the bipartite line (see also (\ref{x-inv-th-bis-split}))
obtained from the modular evolution in the Euclidean spacetime \cite{Cardy:2016fqc}
naturally leads to introduce the subinterval $A_\beta \subset A$,
whose endpoints are (\ref{a-beta-def}) and (\ref{b-beta-def}).
Interestingly, denoting by $\ell$ the size of the interval $A$,
we find that the integration over $A_\beta$ of the corresponding contour function 
for the entanglement entropy \cite{Coser:2017dtb} (see (\ref{integral-contour-beta}))
provides the Gibbs entropy $\pi \, c\, \ell/(3\beta)$ of an isolated system of length $\ell$
coming from the Stefan-Boltzmann law for a 2D CFT 
\cite{Bloete:1986qm, Affleck:1986bv, Cardy:2010fa}
for any value of $\ell$ and $\beta$.
The inversion map (\ref{x-inv-th-def}) for a given chirality is a complex function 
which acquires a constant non vanishing imaginary part on $A_\beta$
and whose real part maps $A_\beta$ into the second world (see Fig.\,\ref{figure-xinv-thermal}).
The occurrence of such spacetime is induced by the non triviality of the commutant for a thermal state,
as discussed in \cite{Borchers:1999short, Schroer:1998pj, Longo:2000ui, Camassa:2011te, Camassa:2011wk}.
The combination of the two different chiralities allows 
to extend this picture to the Minkowski spacetime $\mathcal{M}$
(see Fig.\,\ref{figure-inv-thermal-2-world}),
where the geometric action of the modular conjugation is (\ref{x-t-inv-th-final}).
Interestingly, this map is real also on the domain of dependence of $A_\beta$
(green region in Fig.\,\ref{figure-inv-thermal-2-world}) and sends this diamond 
onto the entire Minkowski spacetime $\widetilde{\mathcal{M}}$ 
(which is interpreted as the second world)
in a bijective way. 

In the gauge/gravity correspondence,
the restriction to the real line at $t=0$  
of this inversion map obtained for a 2D CFT in the (geometric) thermal state 
coincides with the relation between the endpoints of the geodesic bit threads 
on a constant time slice of the BTZ black brane geometry 
(see Fig.\,\ref{figure-1int-BTZ-half-out} and Fig.\,\ref{figure-1int-BTZ-half}  
for the geodesic bit threads anchored in $A\setminus A_\beta$ and in $A_\beta$ respectively). 
The geodesic bit threads anchored in $A_\beta$ reach the horizon.

It is instructive to compare the modular conjugation for the thermal case derived in Sec.\,\ref{sec-thermal}
with the thermo field dynamic (TFD) setting of \cite{Takahashi:1996zn}.
Both constructions are based on the Gibbs state generated by the Hamiltonian $H$ of the system 
and adopt \cite{Ojima:1981ma} the Tomita-Takesaki modular theory. 
In the TFD case one applies this theory to the whole system and the modular Hamiltonian is $H - JHJ$. 
Instead, in our case the modular Hamiltonian (\ref{K-full-ojima}) is induced by the bipartition of the system in two parts $A$ and $B$. 
For this reason the modular evolution and the modular conjugation in the two settings are different. 
Heuristically, our construction approaches the TFD setting when the interval $A$ becomes the whole system.

%\noindent
%$\bullet$ {\bf Two intervals.}

Finally, for the massless Dirac fermion in the ground state 
and on the line bipartite by the union of two disjoint intervals, 
in Sec.\,\ref{sec-2int-inv-Eucid}
we have proposed the inversion map (\ref{inv-map-2int-0}),
obtained by following \cite{Cardy:2016fqc} (see Fig.\,\ref{figure-inversion-Euclid-2int})
and adapting the procedure established in the previous cases.
This inversion map can be extended to the Minkowski spacetime through (\ref{x-t-inv-z-2int})
(see Fig.\,\ref{figure-scacchiera-2int}).
It is not difficult to realise that a holographic description of this inversion map on the real line 
cannot be found only through the geodesic bit threads in Poincar\'e AdS$_3$ 
(see the corresponding discussion in \cite{Agon:2018lwq}).
However, it would be interesting to reproduce it holographically
by employing other holographic bit threads in Poincar\'e AdS$_3$.

%\noindent
%$\bullet$ {\bf Open directions.}

The analyses discussed in this manuscript can be extended in various directions.
In $1+1$ spacetime dimensions,  
more complicated configurations can be investigated by considering e.g.
a CFT at finite temperature and finite volume when the subsystem is still an interval
\cite{Headrick:2007km, Hubeny:2013gta, Hollands:2019hje, Blanco:2019xwi, Fries:2019ozf}
or a CFT in the ground state on the line when the bipartition 
is given by the union of an arbitrary number of disjoint intervals
\cite{Casini:2009vk, Longo:2009mn, Headrick:2010zt, Tonni:2010pv, Coser:2013qda}.
In higher dimensional CFT, 
it would be interesting to explore 
the action of the modular operators and of the modular conjugations
for subregions having different shapes,
comparing the results with the corresponding configurations of geodesic bit threads
(see e.g. \cite{Klebanov:2012yf, Allais:2014ata, Fonda:2014cca, Fonda:2015nma, Seminara:2018pmr, Cavini:2019wyb, Bueno:2021fxb} 
for the shape dependence of the holographic entanglement entropy).
Finally, we find it worth mentioning that
various modular Hamiltonians in free models 
have been studied by taking the continuum limit of the corresponding operators 
(often called entanglement Hamiltonians) 
on the lattice
\cite{Peschel:2003rdm, Casini:2009sr, EislerPeschel:2009review, Banchi:2015aaa, Arias:2016nip, Eisler:2017cqi, Eisler:2018ugn, 
Eisler:2019rnr, DiGiulio:2019cxv, Javerzat:2021hxt, Eisler:2020lyn, Rottoli:2022ego, Eisler:2022rnp}.
It would be insightful to study also other operators 
occurring in the Tomita-Takesaki modular theory in quantum many-body systems.

%\newpage
%%%%%%%%%%%%%%%%%%%%%%%%

\vskip 20pt 
\centerline{\bf Acknowledgments} 
\vskip 5pt 

We are grateful to
Cesar Ag\'on, Matthew Headrick,  Giuseppe Mussardo, Yoh Tanimoto
and in particular to Roberto Longo and Diego Pontello for useful discussions and correspondence. 
ET acknowledges Galileo Galilei Institute for warm hospitality and financial support during part of this work
through the program {\it Reconstructing the Gravitational Hologram with Quantum Information}.
ET’s research has been conducted within the framework of the Trieste Institute for Theoretical Quantum Technologies (TQT).

\vskip 30pt

\appendix

\newpage
%%%%%%%%%%%%%%%%%%%%%%%%%%%%%%%%%%%%%%%%%%%%%%%%%%%%%%%%
\section{Single interval on the circle: Ground state}
\label{sec_app_circle}
%%%%%%%%%%%%%%%%%%%%%%%%%%%%%%%%%%%%%%%%%%%%%%%%%%%%%%%%

In this Appendix we consider a CFT in its ground state
and  on a circle of length $L$
bipartite by an interval $A=(a,b)$ and its complement.
The circle can be represented by the interval $[-L/2\, , L/2] $
whose endpoints $\pm L/2$ are identified.

In this case, the function $w(u)$ in the light ray coordinate $u$
to employ in Sec.\,\ref{sec_mod_flow} 
reads \cite{Wong:2013gua, Cardy:2016fqc}
\be
\label{w-function-circle}
 w(u) =  \log\!\bigg( \frac{\sin[\pi (u-a)/L]}{\sin[\pi (b-u)/L]} \bigg) 
\ee
hence the weight function (\ref{beta0-def-local}) for the entanglement Hamiltonian (\ref{K-pm-A-def}) is
\be
 \beta_0(u) \equiv  \frac{L}{\pi}\; \frac{\sin[\pi (b-u)/L]\, \sin[\pi(u-a)/L]}{\sin[\pi(b-a)/L]} \,.
\ee

Specialising  (\ref{xi-def-gen}) to (\ref{w-function-circle}),
one finds the geometric action of the modular automorphism group of the diamond 
induced by the ground state \cite{Mintchev:2020uom}
\be
\label{xi-circle-gs-tau}
 \xi(\tau,u)
% \equiv 
%w^{-1} \big( w(x) + 2\pi \tau \big)
  \,=\,
\frac{L}{2\pi\,  \textrm{i}}\, \log\! \bigg(
 \frac{e^{\textrm{i} \pi (b+a)/L} +  e^{\textrm{i} 2\pi b/L}\, e^{w(u)+2\pi \tau}
 }{ 
 e^{\textrm{i} \pi (b-a)/L}  +  e^{w(u)+2\pi \tau}}
\bigg)
\ee
where $x \in A$ and $\tau \in \RR$.
In the Euclidean spacetime, 
the modular evolution (\ref{z-euclid-def-gen}) for this case becomes \cite{Cardy:2016fqc}
\be
\label{z-inv-1int-circle-gs}
z(\theta,x) 
%\equiv 
%w^{-1} \big( w(x) + \textrm{i} \theta \big)
\,=\,
\frac{L}{2\pi\,  \textrm{i}}\, \log\! \bigg(
 \frac{e^{\textrm{i} \pi (b+a)/L} +  e^{\textrm{i} 2\pi b/L}\, e^{w(x)+\textrm{i}\theta}
 }{ 
 e^{\textrm{i} \pi (b-a)/L}  +  e^{w(x)+\textrm{i}\theta}}
\bigg)
\ee
where $x \in A$ and $\theta \in [0,2\pi)$.
In the limit $L \to +\infty$,
the expressions (\ref{xi-circle-gs-tau}) and (\ref{z-inv-1int-circle-gs}) 
become (\ref{xi-interval-line-gs}) and (\ref{z-inv-1int}) respectively, as expected.

\begin{figure}[t!]
\vspace{-.6cm}
\hspace{1.85cm}
%\begin{center}
\includegraphics[width=.75\textwidth]{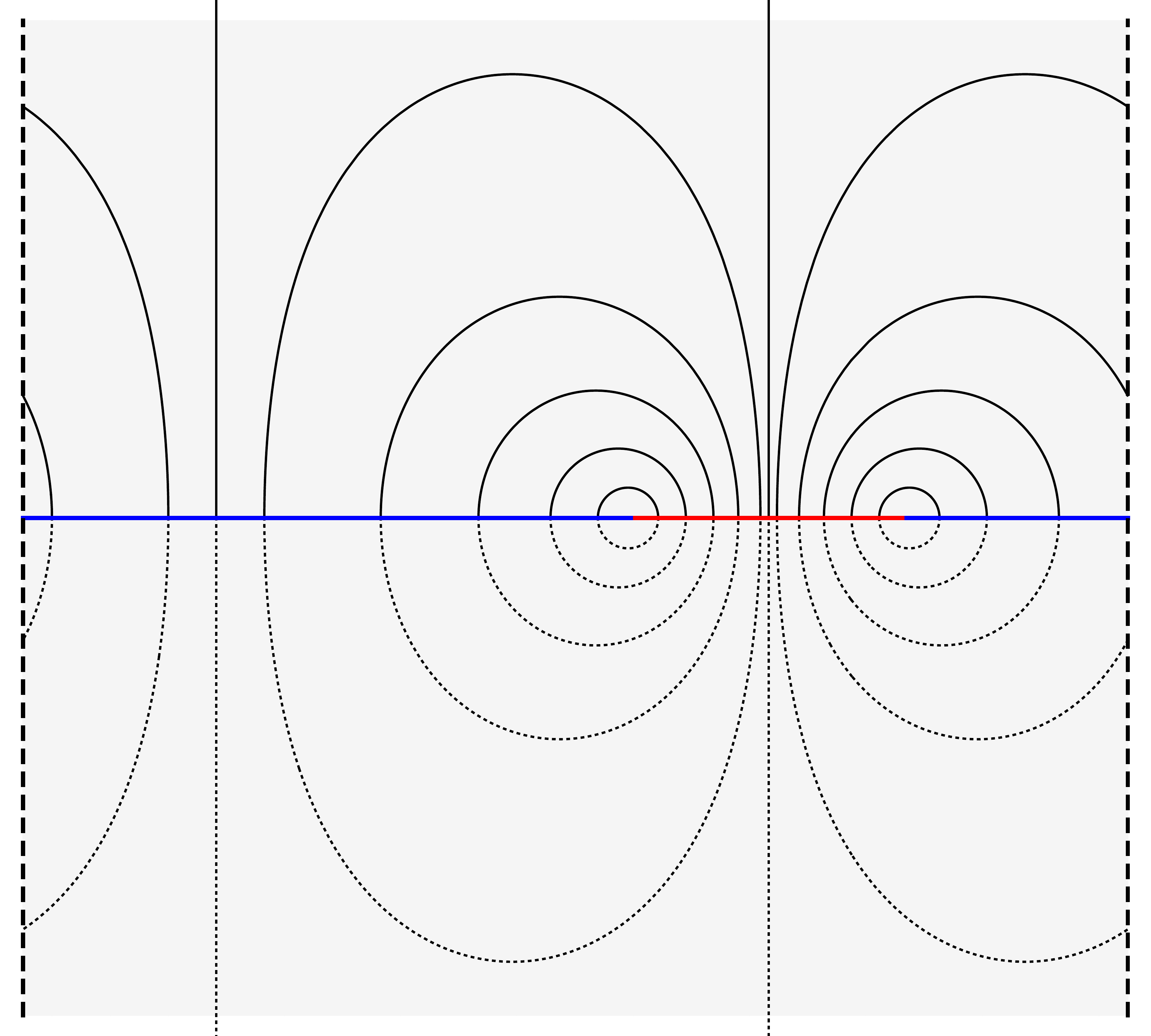}
% \end{center}
\vspace{.2cm}
\caption{Euclidean modular evolution for a CFT in the ground state 
and on the circle bipartite by an interval (the dashed vertical line are identified).
The curves are obtained from (\ref{z-inv-1int-circle-gs}) for some $x\in A$,
with either $\theta \in (0,\pi)$ (solid arcs) or $\theta \in (\pi,2\pi)$ (dotted arcs).
They map a point in $A$ to a point in its complement on the circle (and viceversa), 
related by (\ref{z-inv-circle-t=0}).
}
\label{figure-inversion-Euclid-1int-circle}
\end{figure}

By adapting (\ref{z-inv-1int-pi}) and (\ref{x-inv-th-def})
(that hold for the CFT on the line at zero and finite temperature respectively)
to this case,
from (\ref{xi-circle-gs-tau}) and (\ref{z-inv-1int-circle-gs}) 
we introduce the inversion map for the CFT in the ground state 
and on the bipartite circle as follows
\bea
\label{z-inv-circle-t=0}
z_{\textrm{\tiny inv}}(x) 
\,\equiv\,
 \xi(\tau=\textrm{i}/2\,,x) 
 =
  z(\theta=\pi \,,x) 
&=&
\frac{L}{2\pi\,  \textrm{i}}\, \log\! \bigg(
 \frac{e^{\textrm{i} \pi (b+a)/L} -  e^{\textrm{i} 2\pi b/L}\, e^{w(x)}
 }{ 
 e^{\textrm{i} \pi (b-a)/L}  -  e^{w(x)}}
\bigg)
\\
\label{xinv-cft-circle}
\rule{0pt}{.8cm}
&=&
b
+
\frac{L}{2\pi\,  \textrm{i}}\, \log\! \bigg(
 \frac{
 e^{w(x)} -  e^{-\textrm{i} \pi (b-a)/L} 
 }{ 
e^{w(x)} - e^{\textrm{i} \pi (b-a)/L}  
 }
\bigg) \,.
\eea
This function is real 
(see (\ref{xinv-cft-circle}), where the argument of the logarithm is purely imaginary)
and maps the interval $A$ into its complement $B\equiv [-L/2\, , L/2] \setminus A$.

In Fig.\,\ref{figure-inversion-Euclid-1int-circle} the Euclidean spacetime 
$\{ z\in \mathbb{C} ; | \textrm{Re}(z)| \leqslant L/2 \}$ is considered (grey region),
where the two black vertical dashed lines at $| \textrm{Re}(z)| = L/2$ are identified. 
The red segment corresponds to the interval $A$ and the blue segments to its complement $B$.
The curves have been obtained from (\ref{z-inv-1int-circle-gs}) for some $x \in A$,
with either $\theta \in (0, \pi)$ (black solid curves) or $\theta \in (\pi, 2\pi)$ (black dotted curves).
Their endpoints are related through the inversion map (\ref{z-inv-circle-t=0}).

Let us discuss a consistency condition between 
the inversion map on the line (\ref{x-inv = z-inv t=0}) 
and the one (\ref{z-inv-circle-t=0}) for the circle. 
The upper half plane in the complex coordinate $\mu$ 
and the unit disk in the complex coordinate $\nu$ 
are related by the Cayley map
\be
\label{cayley-map}
\nu(\mu) \,=\, -\,\frac{ \mu - \ri }{ \mu + \ri }
\;\;\;\;\qquad\;\;\;\;
\mu(\nu) \,=\, \ri\,\frac{1-\nu}{1+\nu} \,.
\ee
This map relates also the boundaries of these two domains in the complex plane,
i.e. the real line $\textrm{Re}(\mu)$ and the unit circle 
$\big\{ \nu= \e^{\ri \theta} ; \theta\in (-\pi, \pi]\big\}$ 
(notice that $\mu=0$ is sent to $\theta = 0$ and $\mu \to \pm \infty$ to $\theta \to  \pm \pi$).
More precisely, a point $p \in \RR$ is sent to $\e^{\ri \theta_p}$ on the circle such that 
\be
\label{theta-p-from-cayley}
\cos \theta_p = \frac{1-p^2}{1+p^2}
\;\;\;\;\qquad\;\;\;\;
\sin \theta_p = \frac{2\, p}{1+p^2} \,.
\ee

Then, the map $\nu \to \frac{L}{2\pi  \ri}  \log(\nu)$ can be used to send
the unit circle onto the segment $[-L/2, L/2]$ whose endpoints are identified.
Composing this transformation with the Cayley map, 
one finds a function that sends a point $p \in \RR $ to a point $s \in [-L/2, L/2]$ in the above segment 
as follows
\be
s(p) = \frac{L}{2\pi  \ri}  \log[\nu(p)] = \frac{L \,\theta_p}{2\pi} \,.
\ee

Considering the inversion map $z_{\textrm{\tiny inv}}(x)$ on the line 
(see (\ref{Haag-inversion-map-upm})  and (\ref{x-inv = z-inv t=0})) for $x\in (a,b) \subset \RR$ 
and denoting (\ref{z-inv-circle-t=0}) by $\tilde{z}_{\textrm{\tiny inv}}(L,a,b; x)$ to avoid confusion,
one finds that
\be
s\big( z_{\textrm{\tiny inv}}(x)  \big) 
\,=\,
\tilde{z}_{\textrm{\tiny inv}} \big( L, s(a), s(b) ; s(x)  \big) 
\ee
which tells us that the inversion maps at $t=0$
discussed in Sec.\,\ref{sec-inversion-gs-lorentz} and in this Appendix
can be related through conformal mappings.

%\newpage
%%%%%%%%%%%%%%%%%%%%%%%%%%%%%%%%%%%%%%%%%%%%%%%%
\section{Geodesic bit threads in global AdS$_3$}
\label{sec_app_global_AdS}
%%%%%%%%%%%%%%%%%%%%%%%%%%%%%%%%%%%%%%%%%%%%%%%%

In this appendix we show that
the inversion map discussed in the Appendix\;\ref{sec_app_circle}
can be obtained also through the geodesic bit threads in a constant time slice of global AdS$_3$.

\begin{figure}[t!]
\vspace{-.6cm}
\hspace{3.2cm}
%\begin{center}
\includegraphics[width=.6\textwidth]{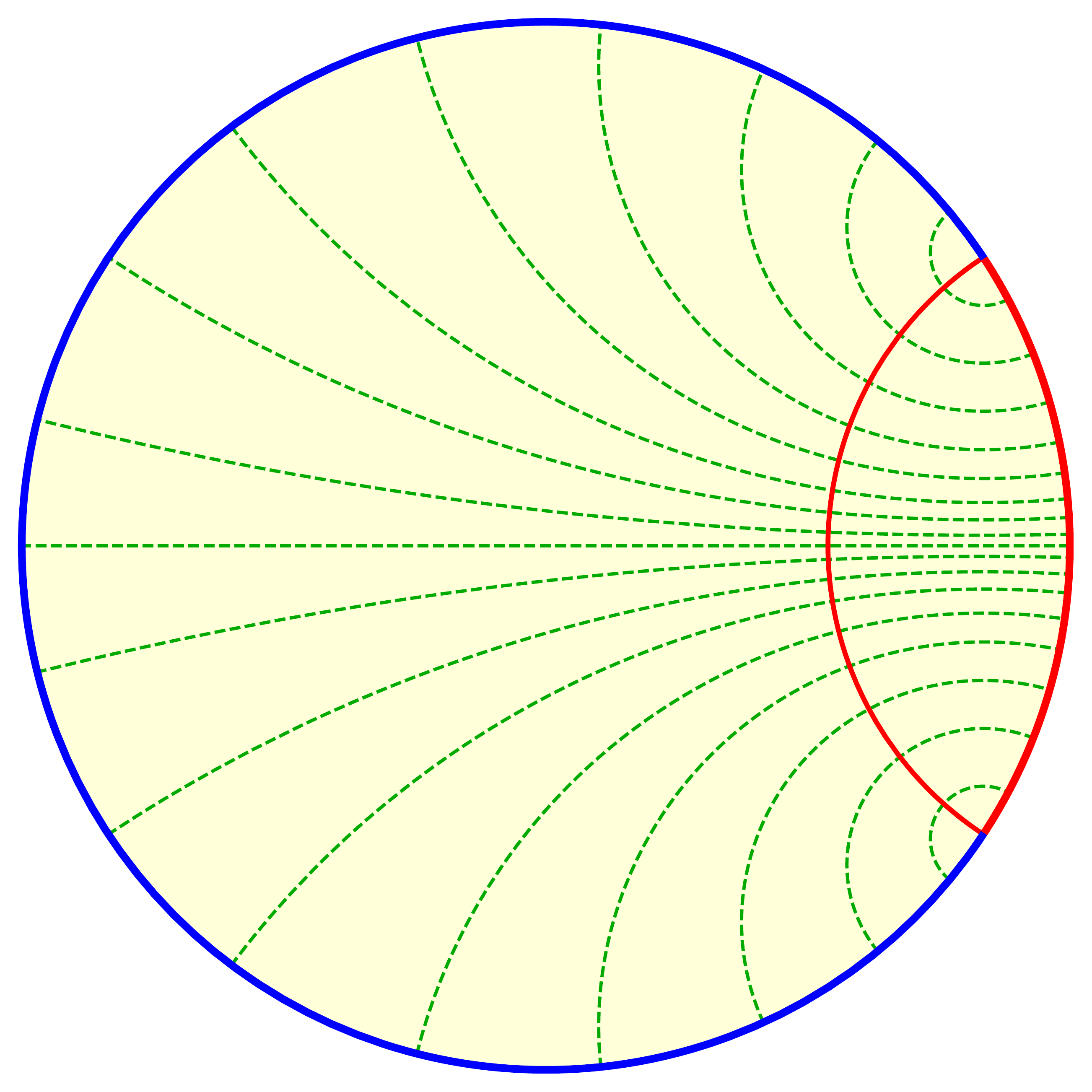}
%\end{center}
\vspace{.2cm}
\caption{
Geodesic bit threads (green dashed circular arcs) in a fixed time slice of global AdS$_3$ 
(yellow unit disk, see (\ref{metric-global-AdS3}))
for an arc of the boundary circle (red arc),
whose RT curve is the red solid arc anchored to its endpoints. 
}
\label{figure-1int-AdS-global}
\end{figure}

The metric on a fixed time slice of AdS$_3$ in global coordinates reads
\be
\label{metric-global-AdS3}
ds^2 
\,=\, 
d\rho^2 +(\sinh \rho)^2 d\theta^2
\,=\,
\frac{4}{(1 - r^2)^2}\, \big( dr^2 + r^2 d\theta^2 \big)
\;\;\;\qquad\;\;\;
\rho = \log\!\bigg(\frac{1+r}{1-r}\bigg)
\ee
where $\rho \geqslant 0$ and $\theta \in [0,2\pi)$.
The last expression in (\ref{metric-global-AdS3}) is written in terms of the polar coordinates $(r, \theta)$, with $0 \leqslant r < 1$,
and corresponds to the Poincar\'e disk (yellow unit disk in Fig.\,\ref{figure-1int-AdS-global}).
The Cayley map (\ref{cayley-map}) relates the upper half plane, 
in the complex coordinate $\mu = x + \ri \zeta$ and equipped with the hyperbolic metric (\ref{UHP-metric}),
to the Poincar\'e disk with the metric (\ref{metric-global-AdS3})
in the complex coordinate $\nu = r \,\e^{\ri \theta}$.

%\noindent
%{\bf $\bullet$ generic geodesic and RT curve.}

The geodesics in the Poincar\'e disk having their endpoints on the boundary are circular arcs
that intersect the boundary orthogonally. 
In a disk of radius $R$ in polar coordinates $(r,\theta)$, where $0 \leqslant r \leqslant R$,
the orthogonality condition for the intersection of two circles provides the following second order equation 
\be
\label{eq-geodesic-disk}
\big(r^2 + R^2 \big) \cos \alpha - 2\, r\, R \cos(\theta - \theta_0) = 0
\ee
where the intersections occur at $(R , \theta_+ )$ and $(R , \theta_-)$ in polar coordinates, 
with $\theta_\pm \equiv \theta_0 \pm \alpha$.
Solving (\ref{eq-geodesic-disk}), for the arc inside the disk one finds 
\be
\label{geoderic-global-ads}
r\,=\, R\; \frac{\cos(\theta - \theta_0) - \sqrt{[\cos(\theta - \theta_0)]^2 - (\cos \alpha)^2}}{ \cos \alpha}
\;\;\;\qquad\;\;\;
\theta \in (\theta_-\,, \theta_+ )
\ee
where $R=1$ for the Poincar\' e disk. 
%

%\noindent
%{\bf $\bullet$ RT curve}

Consider the interval $(-b, b) \subset \RR$, as done in Sec.\,\ref{sec-AdS}.
The corresponding arc on the boundary of the Poincar\'e disk (red arc on the boundary of the unit disk in Fig.\,\ref{figure-1int-AdS-global})
is given by $|\theta| \leqslant  \alpha_{\textrm{\tiny RT}}$,
where $\cos \alpha_{\textrm{\tiny RT}} = (1-b^2)/(1+b^2) = \textrm{Re}[\nu(b)]$, from (\ref{theta-p-from-cayley}).
Hence the RT curve in the Poincar\'e disk is given by 
(\ref{geoderic-global-ads}) specialised to $R=1$, $\theta_0 = 0$ and $\alpha = \alpha_{\textrm{\tiny RT}} $.
In Fig.\,\ref{figure-1int-AdS-global}, 
it corresponds to the red curve in the bulk of the Poincar\'e disk.

%\noindent
%{\bf $\bullet$ geodesic bit threads}

The geodesic bit threads in the Poincar\'e disk for this RT curve 
can be also found through the Cayley map. 
The angular coordinates $\theta_\pm $ of the endpoints are obtained from (\ref{theta-p-from-cayley}) with $p = x_s \pm R_s$.
These coordinates give the parameters $\alpha$ an $\theta_0$ (in terms of $x_A \in (-b,b)$)
to employ in (\ref{geoderic-global-ads}) with $R=1$ in order to find the geodesic bit threads.
Some of them are shown in Fig.\,\ref{figure-1int-AdS-global} (green dashed circular arcs).

%\noindent
%{\bf $\bullet$ holographic relation for the inversion}

Finally, by employing also the map $\nu \to \frac{L}{2\pi  \ri}  \log(\nu)$ introduced in the Appendix\;\ref{sec_app_circle},
we have checked that the angular distance between the endpoints of a geodesic bit thread
agrees with the CFT expression (\ref{z-inv-circle-t=0}),
as expected.

%%%%%%%%%%%%%%%%%%%%%%%%%%%%%%%%%%%%%%%%%%%%%
%\newpage 
\bibliographystyle{nb}

\bibliography{refsaction}

\end{document}